\documentclass[twocolumn, prx, aps, 10pt, floatfix, superscriptaddress, nofootinbib]{revtex4-2}
\usepackage{lmodern}
\usepackage{microtype}
\usepackage{graphicx, color, graphpap}
\usepackage{sidecap}
\usepackage{amssymb}
\usepackage{amsthm}
\usepackage{amsmath}
\usepackage{algorithm2e}
\usepackage{braket}
\usepackage{dsfont}
\usepackage{mathtools}
\usepackage[dvipsnames]{xcolor}
\usepackage[colorlinks=true,citecolor=blue,linkcolor=magenta]{hyperref}
\usepackage{enumerate}
\usepackage{enumitem}
\usepackage[utf8]{inputenc} 
\usepackage[capitalize]{cleveref}
\usepackage{orcidlink}
\usepackage{marvosym}
\usepackage{changes}
\usepackage{complexity}

\usepackage{tabularx}
\usepackage{tabularray}
\usepackage{longtable}
\usepackage{xltabular}

\usepackage{booktabs}
\usepackage[thinlines]{easytable}  

\setlist[description]{style=multiline,labelindent=0cm, leftmargin=0.3 cm}
\setlist[itemize]{style=multiline,labelindent=0cm, leftmargin=0.3 cm}
\makeatletter 
\renewcommand\onecolumngrid{
\do@columngrid{one}{\@ne}
\def\set@footnotewidth{\onecolumngrid}
\def\footnoterule{\kern-6pt\hrule width 1.5in\kern6pt}
}

\renewcommand\twocolumngrid{
        \def\footnoterule{
        \dimen@\skip\footins\divide\dimen@\thr@@
        \kern-\dimen@\hrule width.5in\kern\dimen@}
        \do@columngrid{mlt}{\tw@}
}

\newcommand{\RemoveAlgoNumber}{\renewcommand{\fnum@algocf}{\AlCapSty{\AlCapFnt\algorithmcfname}}}
\newcommand{\RevertAlgoNumber}{\algocf@resetfnum}

\makeatother

\DeclarePairedDelimiterX\norm[1]\lVert\rVert{%
  #1
}   

\newcommand{\comm}[2]{\left[ #1 , #2 \right]}

\DeclareMathOperator{\Var}{Var}
\DeclareMathOperator{\Tr}{Tr}

\renewcommand{\EE}{\mathbb{E}}
\newcommand{\CC}{\mathbb{C}}
\newcommand{\FF}{\mathbb{F}}
\newcommand{\1}{\mathds{1}} 
\DeclarePairedDelimiterX{\abs}[1]{\lvert}{\rvert}{%
 #1
}   
\DeclarePairedDelimiterX\ketbra[2]{\vert}{\vert}%
  {#1\kern0.15ex\delimsize\rangle\delimsize\langle\kern0.15ex\mathopen{}#2}
\DeclarePairedDelimiterX\sandwich[3]{\langle}{\rangle}%
  {#1\,\delimsize\vert\kern0.15ex\mathopen{}#2\kern0.15ex\delimsize\vert\kern0.15ex\mathopen{}#3}

\theoremstyle{plain}
\newtheorem{thm}{Theorem}
\newtheorem{corol}{Corollary}

\theoremstyle{definition}
\newtheorem{lemma}{Lemma}

\newtheorem{defn}{Definition}
\theoremstyle{remark}

\definecolor{FAUlightgreen}{RGB}{123, 183, 37}
\definecolor{FAUdarkgreen}{RGB}{0, 155, 119}

\definecolor{FAUred}{RGB}{141, 19, 41}
\definecolor{FAUyellow}{RGB}{253, 183, 53}

\definecolor{FAUcyan}{RGB}{0, 177, 235}
\definecolor{FAUblue}{RGB}{4, 49, 106}
\hypersetup{
pdftitle={Shot-noise reduction for lattice Hamiltonians},
pdfsubject={Quantum Computing},
pdfauthor={Timo Eckstein, Refik Mansuroglu, Stefan Wolf, Ludwig Nützel,
Stephan Tasler, Martin Kliesch, and Michael J. Hartmann},
pdfkeywords={Quantum measurement, Lattice Hamiltonian, Shot-noise reduction, Quantum computing, NISQ}
}
\begin{document}

\title{Shot-noise reduction for lattice Hamiltonians} 

\author{Timo Eckstein 
\orcidlink{0000-0002-6819-1865}}
\email{Timo.Eckstein@fau.de}
\affiliation{Department of Physics, Friedrich-Alexander-Universität Erlangen-Nürnberg, Erlangen, Germany}
\affiliation{Max-Planck Institute for the Science of Light, Erlangen, Germany}

\author{Refik Mansuroglu \orcidlink{0000-0001-7352-513X}}
\affiliation{Department of Physics, Friedrich-Alexander-Universität Erlangen-Nürnberg, Erlangen, Germany}
\affiliation{Faculty of Physics, University of Vienna, Vienna, Austria}

\author{Stefan Wolf \orcidlink{0009-0003-8297-3745}}
\affiliation{Department of Physics, Friedrich-Alexander-Universität Erlangen-Nürnberg, Erlangen, Germany}

\author{Ludwig~Nützel~\orcidlink{0000-0002-4241-1241}}
\affiliation{Department of Physics, Friedrich-Alexander-Universität Erlangen-Nürnberg, Erlangen, Germany}

\author{Stephan Tasler 
\orcidlink{0009-0000-4765-5037}}
\affiliation{Department of Physics, Friedrich-Alexander-Universität Erlangen-Nürnberg, Erlangen, Germany}

\author{Martin Kliesch \orcidlink{0000-0002-8009-0549} }
\affiliation{Institute for Quantum Inspired and Quantum Optimization, Hamburg University of Technology, Hamburg, Germany\looseness=-1}

\author{Michael J. Hartmann
\orcidlink{0000-0002-8207-3806}}
\affiliation{Department of Physics, Friedrich-Alexander-Universität Erlangen-Nürnberg, Erlangen, Germany}
\affiliation{Max-Planck Institute for the Science of Light, Erlangen, Germany}

\begin{abstract} 
Efficiently estimating energy expectation values of quantum lattice systems on quantum computers is a crucial subroutine for various quantum algorithms, which can lead to significant overhead due to the high measurement shot numbers required. 
We introduce a measurement strategy tailored to quantum lattice systems and (noisy) energy eigenstates. It is based on a geometric partitioning of the Hamiltonian into local patches, and performing the measurements in the eigenbases of those patches. The resulting energy estimator has a smaller variance than the ones of Pauli grouping schemes, which leads to a reduction of the total number of shots. 
We provide rigorous guarantees for this variance improvement for energy eigentstates, also in the presence of depolarizing noise. As one can choose the subsystem size, one can ensure that measurement circuits remain within implementable depths.
In numerical experiments, we demonstrate the shot count reduction for various 2D lattice models, including the transverse field XY and Ising models, as well as the Fermi-Hubbard model. We find sampling improvements of several orders of magnitude already for plaquettes of two by two qubits, where the required readout circuits remain very moderate in depth.
\end{abstract}

\maketitle

\section{Introduction}
\label{sec:introduction_1}

Extracting information from a quantum computer means sampling from a probability distribution defined by the prepared quantum state and the observable to be measured. 
Since the speed of individual gate and readout operations for quantum computers is much slower than for their classical counterparts, the efficiency of a quantum algorithm or quantum simulation is often limited by the number of measurements and hence executions that need to be made to obtain sufficient samples. Therefore, reducing the sampling complexity in obtaining the output of a quantum computer is a goal of utmost importance. 

Since quantum computers typically can only perform measurements in the computational basis, the observable of interest needs to be split into parts that can be measured individually after rotating their eigenbases into the computational basis. Thereby, the individual parts typically only comprise single- or two-qubit strings of Pauli operators, where the required basis rotations can be done via single-qubit gates. 

The efficiency of measurement strategies in quantum computation is a very active field of research \cite{haah2017sample, arunachalam2018optimal, takagi2022universal}, where the challenge of reducing shot counts has received significant attention, for example, in the context of applications in quantum chemistry. For corresponding measurement strategies, Pauli grouping methods \cite{gokhale2019minimizing, jena2019pauli, verteletskyi2020measurement, huggins2021efficient, gresch2025guaranteed, nutzel2025solving}, on the one hand, aim to reduce the number of parts that the observable is split into, which corresponds to minimizing the state-independent worst-case measurement error.
Classical shadows \cite{Huang2020Predicting, hadfield2022measurements, hadfield2021adaptive, elben2022randomized, brieger2025stability}, 
on the other hand, aim at estimating expectation values in an observable-independent fashion, meaning one is not restricted to knowing a priori a specific observable of interest, but also does not adapt the sampling strategy to it. Instead, one aims to use the locality of the to-be-estimated observables so that local random measurements suffice to reduce the measurement effort~\cite{crawford2021efficient, gresch2025guaranteed}.
Here, in contrast, we use the structure of both the quantum state and the observable to reduce measurement uncertainty.   

Besides quantum chemistry applications, the investigation of condensed matter quantum many-body systems is also considered a very important and promising application for quantum computers. Moreover, the Hamiltonians of such systems can often be natively explored on solid-state quantum computing platforms~\cite{chatterjee2021semiconductor, burkard2023semiconductor, arute2019quantum, blais2021circuit} with a fixed connectivity graph.
While the Hamiltonians from quantum chemistry applications have a complex non-local interaction structure, condensed matter systems often have the form of quantum lattice models, where grouping methods only have little effect on the measurement count. Hence, different strategies are required for obtaining significant improvements in the required number of measurements for lattice systems, a challenge that has received relatively little attention so far. 

In this work, we tailor measurement schemes for this broad class of quantum lattice models by splitting the observable of interest, for example, a Hamiltonian, into parts that have support on multiple adjacent lattice sites and measure these parts in their eigenbasis. We call this {\it geometric partitioning}. When applying it to a system that is prepared in an eigenstate of the total Hamiltonian one may expect that it leads to reduced sampling requirements. 

The intuition behind this expectation is the following. For a Hamiltonian with short-range interactions, the expansion of the global energy eigenstate in a basis formed by products of eigenstates of contiguous multi-qubit parts of the same Hamiltonian (geometric partitioning) should contain fewer terms than an expansion in terms of products of eigenstates of individual qubits, which one may call Pauli partitioning. The reason is that in geometric partitioning, all the Hamiltonian terms except for the interactions between patches are diagonal, whereas all qubit-qubit interactions (which are many more) are not diagonalized in Pauli partitioning. In this work, we quantify this sampling advantage in terms of the variances of the employed estimators, which allows us to derive lower bounds on the sampling improvement. We also show in numerical simulations that the improvement can often be several orders of magnitude already for measurements in the eigenbases of plaquettes of two by two qubits, where the required readout circuits remain very moderate in depth.

Our work has the following main results.
In Theorem~\ref{thm:improvement}, we show that, for systems in an energy eigenstate, geometric partitions always outperform Pauli partitions in terms of the required sample number for eigenenergy estimation and obtain lower bounds for the improvement factors. We then numerically show reductions of sampling complexity of several orders of magnitude for a number of quantum lattice models as examples. Importantly, these reductions are already seen for partitionings into very moderate-sized parts, where the measurement circuits can feasibly be implemented.

In Theorem~\ref{thm:2}, we then show that sampling improvements carry over to imperfect eigenstates, which one expects from preparation circuits on real devices. As corollaries, we also obtain upper bounds on the extra noise that the measurement circuits are allowed to generate, see Corollary \ref{lemma:noise-indep sample advantage}, and the number of gates they may be composed of, see Corollary \ref{corol: gate_budget}, while still leading to a sampling improvement. 
These findings are again corroborated by numerical studies for several models.

The remainder of the paper is organized as follows. In section \ref{sec:introduction} we explain the general concept of our approach, before deriving analytical bounds on the sampling improvement for exact eigenstates and comparing these to numerical tests in section \ref{subsec:perfect eigenstates}. Section \ref{subsec:imperfect eigenstates} generalizes the analytical results to eigenstates perturbed by global depolarizing noise, including numerical examples. Finally, section \ref{sec:discussion} provides a summary and conclusions.

\section{Concept}
\label{sec:introduction}

We first present the concept behind our approach. In our discussion, we consider the energy of a Hamiltonian $H$ and its exact or approximate eigenstates as the objects of interest. Our results, however, equally apply to other observables with a lattice structure and their exact or approximate eigenstates. 

Measuring the energy $H$ for a quantum state $\ket{\psi}$ means randomly drawing the eigenvalues $E_m$ of $H$ with probabilities defined by the amplitudes' squares $\abs{c_i}^2$ of the expansion $\ket{\psi} = \sum_i c_i \ket{E_i}$ in the eigenbasis $\{\ket{E_i}\}$ of $H$. The empirical mean estimator $\overline H_M \coloneqq M^{-1} \sum_{m=1}^M E_m$ from $M$ i.i.d.\ repetitions (or samples) then yields an approximation to the expectation value $\braket{H} = \braket{\psi | H | \psi}$ with a standard error $\sigma_H$ given by
\begin{align}
    \sigma_H^2 = \Var \left (\overline H_M \right) =\frac{1}{M} \Var_{\ket{\psi}}(H) \ .
    \label{eq: std_error}
\end{align}
This variance allows one to quantify finite sampling noise since Chebyshev's inequality upper bounds the probability to be more than a threshold $\epsilon$ away from the true expectation value, 
\begin{align}
    P\left( \left| \overline H_M - \Braket{H} \right| \geq \epsilon \right) \leq \Var \left (\overline H_M \right)/\epsilon^2 \ .
    \label{eq: Chebyshev}
\end{align}
Since measurements in quantum computation can typically only be done in the Pauli $Z$ (or computational) basis, measuring an observable that is not diagonal in this basis would generally require a diagonalization $U_H H U_H^\dagger$, which is equivalent to implementing a rotation $U_H$ that transforms the eigenbasis of $H$ into the computational basis at the end of the quantum circuit. In general, however, this is a non-trivial task, as finding and implementing the basis transformation $U_H$ is typically harder than solving the quantum computational problem in the first place. 

%
%
\begin{figure*}[t!]
    \centering
    \includegraphics[width=\linewidth]{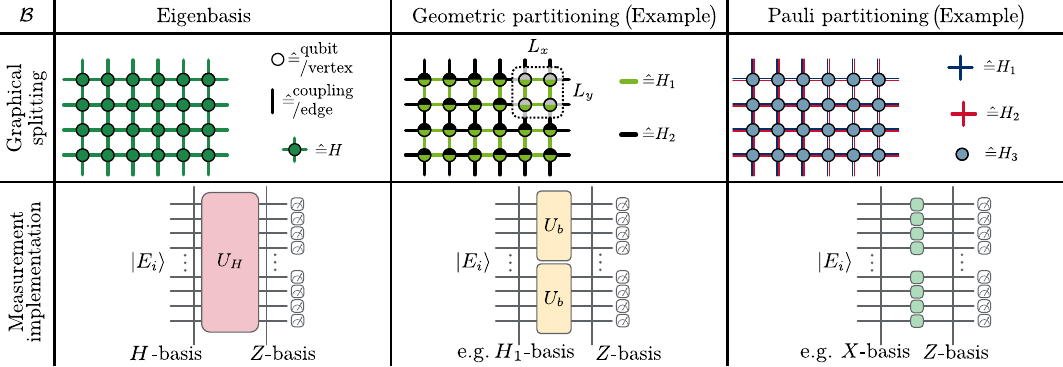}
    \caption{\textbf{Concept of geometric partitioning.} 
    For lattice Hamiltonians as in Eq.~\eqref{eq: intro_geometrically_local}, one can measure the energy in the eigenbasis (first column), which requires implementing a measurement unitary corresponding to a full diagonalization of the Hamiltonian, or in Pauli bases (third column), which only requires single-qubit unitaries, as extreme cases.     
    Here, we explore the regime between these extremes, namely measurement in the eigenbases of disjoint subsystems of $H$, which each comprise multiple lattice sites (second column). We call this a geometric partitioning. In the lattice sketch of the second column, this strategy is indicated by coloring complementary parts of the lattice in different colors (green and black). The idea is to measure all green parts in one shot and all black parts in another shot. In doing so, the single-site terms of the Hamiltonian are split into two identical parts to make the splitting symmetric, which is indicated by coloring the corresponding dots half in one and half in the other color.
    Geometric partitionings naturally take the local structure of the observable $H$ into account while keeping the basis transformations $U_{H_b} = U_b \otimes \cdots \otimes U_b$ scalably implementable due to their product structure.
    As our main analytical results, we show in \cref{thm:improvement} that these geometric partitionings of $H$ always lead to fewer required measurements
    for estimating $E_i$ from $\ket{E_i}$ compared to the Pauli partitioning baseline for the same measurement uncertainty. Furthermore, these sampling improvements also apply to imperfect eigenstates. 
    }
\label{fig: Introduction}
\end{figure*}
If the eigenbasis of $H$ cannot be obtained efficiently, it is possible to measure parts $H_b$, whose eigenbases are known, and that constitute the observable $H=\sum_b H_b$. The linearity of the expectation value then allows one to estimate $\braket{H} = \sum_b \braket{H_b}$.
The empirical mean estimator of such a partition $\mathcal{B} = \{H_1, H_2, ... \}$ of $H$ is given by
\begin{align}
    \overline H_M^{\mathcal B}
    \coloneqq \sum_{b} \left( \frac{1}{M_b} \sum_{m=1}^{M_b} E_{b, m}\right ) \ ,
\end{align}
where $E_{b, m}$ labels the sampled eigenvalues of the observable $H_b$, and $M_b$ is the number of measurements for each $H_b$ so that the overall number of measurements is $M = \sum_b M_b$. 
Since the individual samples $E_{b,m}$ are obtained in mutually independent repetitions of the experiment,  the variance of the estimate $\overline H_M^{\mathcal B}$, is the sum of the variances of the empirical mean estimators for the individual $H_b$,
\begin{align}
    \tilde{\sigma}_{\mathcal B}^2 \coloneqq  \Var \left (  \overline H_M^{\mathcal B}\right) &=\sum_{b}  \frac{\Var_{\ket{\psi}}(H_b)}{M_b} \ . 
    \label{eq: std_error_partitioned}
\end{align}
While the standard way of decomposing $H$ into measurable parts $H_b$ is to partition the Pauli-strings in $H$ into mutually commuting groups of Pauli sums, we show here that much more sampling-efficient measurement strategies can be obtained by making use of the structure of lattice Hamiltonians and their eigenstates.

While keeping much of our analysis general, we focus on translation-invariant Hamiltonians $H$ with nearest-neighbor interactions on a rectangular lattice, as can be found in numerous quantum many-body systems, with the Ising or the Hubbard models as special cases,  
\begin{equation}
\label{eq: intro_geometrically_local}
    H = \sum_{i} \mathcal{H}_i +\sum_{\left<i,j\right>} \mathcal{H}_{i,j} \, , 
\end{equation}
where $\mathcal{H}_i = \sum_{\alpha} h_i^{(\alpha)} o_i^{(\alpha)}$ and $\mathcal{H}_{i,j} = \sum_{\alpha, \beta} J_{i,j}^{(\alpha,\beta)} o_i^{(\alpha)}o_j^{(\beta)}$ 
for the one-local and two-local terms. The Latin site indices $i$ and $j$ have two components for two dimensional lattices, $i=(i_x,i_y)$ and $j=(j_x,j_y)$, whereas Greek indices label single operators $o_i^{(\alpha)}$ on individual lattice sites. These can be Pauli operators  $o_i^{(\alpha)} \in \{\mathds 1, X, Y, Z\}$ for spin models, or creation, annihilation and number operators in Fock space representation $o_i^{(\alpha)} \in \{c^\dagger_{i,\alpha}, c_{i,\alpha},n_{i,\alpha}\}$ for fermionic lattice models. The coefficents, $h_i^{(\alpha)}$ and $J_{i,j}^{(\alpha,\beta)}$ are real-valued coupling constants.

The main idea we pursue for realizing sampling improvement is to partition the geometrically local Hamiltonian $H$ into terms $H_b$, which further decompose into subsystems $H_{b,k}$ of disjoint support, 
\begin{equation}
    H=\sum_b H_b \quad \text{with} \quad H_b = \sum_{k=1}^K H_{b,k}
\end{equation}
where the $H_{b,k}$ act on multiple adjacent lattice sites. The transformations of the measurement bases, from the eigenbases of the $H_{b}$ into the computational basis, thus become direct products 
\begin{equation}
    U_{H_b} = U_{b,1} \otimes \dots \otimes U_{b,K}
\end{equation}
see also Fig.~\ref{fig: Introduction}. 
We call this a geometric partitioning and compare sample numbers $M$ for different geometric partitions with the standard partition into {fewest} groups of mutually commuting local operators, $o_i^{(\alpha)}$ and $o_i^{(\alpha)}o_j^{(\beta)}$ in Eq.~\eqref{eq: intro_geometrically_local},
under the condition of admitting the same standard error according to Eq.~\eqref{eq: std_error_partitioned}. 

Geometric partitioning becomes particularly useful if the support of the individual $H_{b,k}$ grows beyond the interaction range, such that correlations of longer and longer range can be captured well. Nonetheless, the implementation of the transformations $U_{b,k}$, which is required in this strategy, can be done efficiently in many cases for the following reasons.

{As one can choose the patch size for the $H_{b,k}$, it can be chosen constant in terms of system size $n$. For this choice, implementing $U_{H_b}$ uses only $\mathcal{O}(n)$ gates, as implementing $U_{b,k}$ has constant gate complexity, and the number of $U_{b,k}$s to implement $U_{H_b}$ scales only linearly in system size $n$. 
Moreover, the prepared state $\ket \psi = \ket{E_i}$ and local unitary $U_b$ correspond to the same lattice Hamiltonian, but only differ in size. Thus, if there is an algorithm simplifying the state preparation, as e.g. in Ref.~\cite{verstraete2009quantum}, it could be used on a smaller scale, entry by entry, to simplify the implementation of $U_b$ as well.}
In addition, for translationally invariant systems, all $U_{b,k}=U_b$ will be the same so that one needs to find only one such subsystem diagonalization circuit.

Furthermore, for specific diagonalizable 1D lattice models, like the transverse field Ising model, exact linear depth circuits to implement $U_b$ are known~\cite{verstraete2009quantum} and can thus be used for partitionings that consider patch sizes $1 \times \sqrt{n}$. 
In addition, since the state $\ket{\psi}$ will typically not be prepared exactly, an approximation of $U_b$ with a similar precision will be sufficient for our aims. Allowing an error $\epsilon$ in the implementation of $U_b$, one may thus make use of the Solovay-Kitaev algorithm~\cite{kitaev1997quantum, dawson2005solovay}, which has complexity $\mathcal{O}((\log (1/\epsilon) )^c)$, $c\approx 1.44$ ~\cite{kuperberg2023breaking}, and approximate it with a squence of easier to implement unitaries.

To show the advantages of geometric partitioning, we first seek a quantitative measure to compare a measurement strategy $\mathcal B_1$ that splits $H = \sum_{H_b \in \mathcal B_1} H_b$ with another one $\mathcal B_2$ that splits $H = \sum_{H_b \in \mathcal B_2} H_b$ by determining which one requires less measurements for achieving the same standrad error according to Eq.~\eqref{eq: std_error_partitioned}. 
To achieve this, we first need to resolve the issue that the standard error of a partitioned observable in Eq.~\eqref{eq: std_error_partitioned} not only depends on the different partitionings but also on the specific measurement budget allocation $M_b$. It turns out that comparing partitioning strategies for the optimal measurement budget allocation, which minimizes the sampling error, makes the comparison independent of the distribution of the $M_b$~\cite{crawford2021efficient}: 
\begin{align}
   \sigma^2_{\mathcal{B}} = \min_{M_{\mathcal{B}} = \sum M_b} \tilde{\sigma}^2_{\mathcal{B}} = \frac{1}{ M_{\mathcal{B}}} \left[ \sum_{H_b \in \mathcal B}  \sqrt{ \Var_{\ket \psi} \left(  H_b\right)} \right ]^2 \ .
   \label{eq: std_error_min}
\end{align}
One can prove Eq. \eqref{eq: std_error_min} using the positivity of the variance and minimizing $\sigma_{\mathcal{B}}^2$ using Lagrange multipliers to find that all $\sqrt{\Var_{\ket \psi} \left(  H_b\right)}/{M_b}$ are equal~\cite{crawford2021efficient}, see also Appendix~\ref{appendix: compare measurement strategies}, where we give the proof for completeness.

As a direct consequence of Eq.~\eqref{eq: std_error_min}, partitioning $H$ always increases the sampling error independently of the state $\ket \psi$ and partitioning $\mathcal B$ (see also Appendix~\ref{appendix: compare measurement strategies}):
\begin{align}
    \frac{1}{M} \Var_{\ket{\psi}}(H)
    \leq \sum_{H_b \in \mathcal B}  \frac{\Var_{\ket{\psi}}(H_b)}{M_b} \ .
    \label{eq: std_error_ineq}
\end{align}

Further, Eq.~\eqref{eq: std_error_min} motivates the definition of the following cost function to compare measurement strategies, being different splittings of $H$, that is not only independent of the specific measurement budget allocation $M_b$ but also independent of the specific number of measurements $M$.

\begin{defn}[Relative sampling complexity (pure state version)]
  \label{def:Sampling improvement}
Let $\mathcal B_1$ and $\mathcal B_2$ be partitionings of $H$. We call the ratio of the number of samples $M_{\mathcal B_1}/M_{\mathcal B_2}$ to obtain the same standard error $\sigma^2_{\mathcal{B}_1} = \sigma^2_{\mathcal{B}_2}$ (see Eq.~\eqref{eq: std_error_min}), the \emph{relative sampling complexity}: 
\begin{align}
   &\mathcal G_{\ket \psi} \left(\mathcal B_1, \mathcal B_2 \right)
   \coloneqq\frac{M_{\mathcal B_1}}{M_{\mathcal B_2}}
   =  \left[ \frac{\sum\limits_{H_b \in \mathcal B_1}  \sqrt{ \Var_{\ket \psi} \left(  H_b\right)}  }{  \sum\limits_{H_b^\prime \in \mathcal B_2}  \sqrt{ \Var_{\ket \psi} \left(  H_b^\prime \right)}  }\right]^2 . 
   \label{eq: intro_cost_function}
\end{align}
\end{defn}
Note that the second equality holds thanks to Eq. \eqref{eq: std_error_min}. 
For $\mathcal G_{\ket \psi} \left(\mathcal B_1, \mathcal B_2 \right) > 1$, we thus get a sampling improvement for partitioning~$\mathcal B_2$ compared to partitioning~$\mathcal B_1$, and the value of $\mathcal G_{\ket \psi} \left(\mathcal B_1, \mathcal B_2 \right)$ tells us how much less often we have to measure within partitioning $\mathcal B_2$ compared to $\mathcal B_1$ for the same standard error, see Eq.~\eqref{eq: std_error_partitioned}, under optimal sampling budget allocation. 
Our results can also be interpreted as quantifying which strategy leads to a smaller standard error for the same number of measurements $M$.
Equally, one can view $\mathcal G_{\ket \psi} \left(\mathcal B_1, \mathcal B_2 \right) $ as the relative efficiency of unbiased estimators $\overline H_M^{\mathcal B_1}$ and $\overline H_M^{\mathcal B_2}$, as it is also equal to  $\sigma^2_{\mathcal{B}_1} / \sigma^2_{\mathcal{B}_2}$ for $M_{\mathcal B_1}=M_{\mathcal B_2} = M$.

\section{Sampling improvements for exact eigenstates} 
\label{subsec:perfect eigenstates}

Many quantum algorithms aim to prepare eigenstates of Hamiltonians, including ground states.
In this section, we thus consider the task of estimating energies $E_i$ of eigenstates, where $H \ket{E_i} = E_i \ket{E_i}$. Our aim is to make statements about the sampling improvement factor $\mathcal G_{\ket{E_i}} \left( \mathcal B_{\rm Pauli}, \mathcal B_{L_x, L_y} \right)$, where $\mathcal B_{\rm Pauli}$ labels the partitioning of $H$ into {fewest} groups of mutually commuting local operators, which often coincide for lattice Hamiltonians with partitioning into different Pauli operators. {$\mathcal B_{L_x, L_y}$ labels geometric partitioning, where each partition is obtained by splitting of $H$ into disjoint patches of size $L_x \times L_y$}, compare Fig.~\ref{fig: Introduction}.
For explicit definitions see Appendix ~\ref{app:def_partitionings}.
For all considered partitionings of $H$, we have the following lemma.

\begin{lemma}[Variance (in-)equalities]
\label{lemma:variance_in_equalities}
For a state vector $\ket \psi$ and Hamiltonian partitioning $H = \sum_{H_b \in \mathcal B} H_b$ we have
\begin{align}
    \label{eq:linear upper bound}
    \abs{\mathcal B} \sum_{H_b \in \mathcal{B}} \Var_{\ket{\psi}} (H_b) &\geq \left( \sum_{H_b \in \mathcal{B}} \sqrt{\Var_{\ket{\psi}} (H_b)} \right)^2 \\
    &\geq  \sum_{H_b \in \mathcal{B}} (2b-1)\Var_{\ket{\psi}} (H_b) \ ,
    \label{eq:linear lower bound}
\end{align}
where $b \in \{1,2,..., |\mathcal B|\}$.
The variance terms are, without loss of generality, taken to be ordered non-increasingly as $\Var_{\ket{\psi}} (H_1)\geq \Var_{\ket{\psi}} (H_2)\geq \dots$.
\begin{proof}[Proof]
    Eq.~\eqref{eq:linear lower bound} is a purely algebraic observation. Let us denote $ \Var_{\ket{\psi}} (H_b) \coloneqq a_b \geq a_{b+1}\geq 0$ and $|\mathcal B| =B$. From $0 \leq (a_b-a_k)^2 = (a_b+a_k)^2-4a_ba_k$ follows $\sqrt{a_ba_k} \leq \frac{1}{2}(a_b+a_k)$ and hence:
\begin{align}
   \left( \sum_{b=1}^B \sqrt{a_b} \right)^2
   &\leq   \sum_{b=1}^B a_b  + \sum_{b > k = 1}^B (a_b + a_k) = B \sum_{b=1}^B a_b \nonumber \ .
\end{align}
Similarly, from $\sqrt{a_b a_k} \geq \min(a_b,a_k)$ folllows
\begin{align}
   \left( \sum_{b=1}^B \sqrt{a_b} \right)^2 
   &\geq  \sum_{b=1}^B a_b  + \sum_{b  = 2}^B 2(b-1) a_b = \sum_{b=1}^B (2b-1) a_b \nonumber \ .
\end{align}
\end{proof}
\end{lemma}
As an intermediate consequence for any partitioning scheme into 2 parts, and energy estimation of an eigenstate, we know that we optimally have to measure both parts equally often, as both variances are the same. 

\begin{corol}[Variance equality for eigenstates and two partitionings]
\label{corol: Ei_2_partitions}
Let $H$ be partitioned into two parts $H_1$ and $H_2$ and $\ket \psi = \ket{E_i}$ be an eigenstate of $H$, then:  
\begin{align}
    \Var_{\ket{E_i}} (H_1) &= \Var_{\ket{E_i}} (H_2) \ . 
    \label{eq: var_2_partition}
\end{align}
\begin{proof}[Proof]
For bipartions of eigenstates we obtain from $\Var_{\ket{E_i}}(H)=0$:  
\begin{align}
   &\Var_{\ket{E_i}} (H_1) + \Var_{\ket{E_i}} (H_2)  =  2 |{\rm CoV}_{\ket{E_i}}(H_1,H_2)| \nonumber \\
   &\leq 2 \sqrt{\Var_{\ket{E_i}} (H_1) \Var_{\ket{E_i}} (H_2) }
   \label{eq: var_2__inequality_partition}
\end{align}
 where we used that the covariance is an inner product and applied the Cauchy-Schwarz inequality. From $\sqrt{ab} \leq \frac{1}{2}(a+b)$, Eq.~\eqref{eq:linear upper bound}, we obtain that Eq. \eqref{eq: var_2__inequality_partition} needs to be an equality, and hence Eq. \eqref{eq: var_2_partition}.
\end{proof}
\end{corol}

We note that \cref{corol: Ei_2_partitions} also implies that the commutator-based lower bound (commonly named uncertainty principle~\cite{robertson1929uncertainty}) vanishes since ${\rm CoV}_{\ket{E_i}}(H_1,H_2)=  {\rm CoV}_{\ket{E_i}}(H_2,H_1)$, and thus
$\left| \Braket{E_i | \left[H_1, H_2 \right] | E_i}\right| = 0$.

Furthermore, we emphasize that even if a Pauli partitioning $\mathcal B_{\rm{Pauli}}$ requires more than two non-commuting Pauli groupings, we can still restrict our considerations to the case $H = H_1 + H_2$.
To see this, let us assume we need to partition the mutually commuting grouping measurements into 3 parts. As Eq.~\eqref{eq: std_error_ineq} is independent of the state, we know that measuring finer partitions always increases the combined variance (e.g. insert $H_2$ instead of $H$ on the l.h.s.). Hence, for finding the lower bound, it suffices to consider a two-partition, where we can measure $H_1$ and put all other terms into $H_2$.

We consider geometric partitionings that split $H$ into two equal parts, $H = H_1 + H_2$ with $|\mathcal{B}| = 2$,
\begin{align}
H_1 &\coloneqq \frac{1}{2} \left( H - H_{\rm cut}+H_{\rm cut}^\prime \right) \ , \\
H_2 &\coloneqq \frac{1}{2} \left( H + H_{\rm cut}-H_{\rm cut}^\prime \right) \ ,
\end{align}
where $H_{\rm cut}$ and $H_{\rm cut}^\prime$ are sums of the 2-local terms in $H$, which are cut into their local parts in $H_1$ and $H_2$, respectively.
In particular, we consider cutting the lattice into one-dimensional slices of thickness $L$, two-dimensional patches of size $L_x \times L_y$ and 2-local dimers of lattice sites.
As an example, a partitioning into slices of thickness $L \geq 2$ is explicitly given by,  
\begin{align}
H_{\rm cut} &= \sum_{m=1}^{\lfloor n_x/L\rfloor} \sum_{\ell=1}^{n_y}  \mathcal{H}_{(mL,\ell),(mL+1,\ell)} \ , \\
H_{\rm cut}^\prime &= \sum_{m=1}^{\lfloor n_x/L\rfloor} \sum_{\ell=1}^{n_y} 
\mathcal{H}_{(mL-1,\ell),(mL,\ell)}  \ ,
\end{align}
with the $\mathcal{H}_{(i_x, i_y),(j_x,j_y)}$ as introduced in Eq.~\eqref{eq: intro_geometrically_local}. 
We provide detailed descriptions of all considered partitionings in Appendix~\ref{app:def_partitionings} and Appendix~\ref{appendix:list_all_partitions}.
Furthermore, the correlation
\begin{align}
    \operatorname{CoR}_{\ket{E_i}} (H_{\rm cut}, H_{\rm cut}^\prime) = \frac{{\rm CoV}_{\ket{E_i}} (H_{\rm cut}, H_{\rm cut}^\prime)}{ \sqrt{\Var_{\ket{E_i}} (H_{\rm cut}) \Var_{\ket{E_i}} (H_{\rm cut}^\prime)}}
\end{align}
will play a significant role in the bounds we derive below.

With these definitions, we can now quantify the ratios of required measurement numbers, $\mathcal G_{\ket \psi} \left(\mathcal B_1, \mathcal B_2 \right)$, for the measurement strategies discussed in Fig.~\ref{fig: Introduction}.
We prove the following lower bounds on the {relative} sampling {complexity} (Def.~\ref{def:Sampling improvement}) improvement.

\begin{thm}[Rel.\ sampling complexity improvement lower bound]
\label{thm:improvement}
Let $H$ be a translation-invariant $2$-local Hamiltonian on an $n_x\times n_y$ rectangular lattice as specified by Eq.~\eqref{eq: intro_geometrically_local}. 
Let $\mathcal{B}_{\rm Pauli}$ be a Pauli partitioning 
and $\mathcal{B}_{L_x,L_y}$ be a geometric partitioning. 
Then, for any non-degenerate eigenstate $\ket{E_i}$ of $H$,
\setcounter{equation}{\theequation + 1}
\begin{align}
   &\mathcal G_{\ket{E_i}} \left( \mathcal{B}_{\rm Pauli}, \mathcal B_{L_x, L_y} \right) 
   \geq \left\{ \begin{array}{l l l}
  4 L \frac{1 }{1 - \operatorname{CoR}_{\ket{E_i}} (H_{\rm cut}, H_{\rm cut}^\prime)}  & \ \rm{(\theequation a)} \\ 
  4 \frac{L_xL_y}{L_x+L_y} \frac{1 }{1 - {\rm CoR}_{\ket{E_i}} (H_{\rm cut}, H_{\rm cut}^\prime)} & \ \rm{(\theequation b)}
  \\
  \frac{4}{3} & \ \rm{(\theequation c)}  
  \end{array} \right  . \nonumber
\end{align}
where the cases denote different types of geometric partitionings.
(a) corresponds to cutting $H$ in one direction so that $(L_x,L_y) = (n_x,L)$ or equivalently $(L_x,L_y) = (L,n_y)$ with $L$ being the thickness of the 1D cuts. 
(b) corresponds to cutting $H$ into rectangular patches of size $L_x \times L_y$ with
$n_{x}/2 \geq L_{x} \geq 2$ and $n_{y}/2 \geq L_{y} \geq 2$.
(c) corresponds to the equal splitting of $H$ into four 2-local partitions, $(L_x,L_y) = \text{2-local}$. 
Further, the correlation
is always non-negative, 
$ \operatorname{CoR}_{\ket{E_i}} (H_{\rm cut}, H_{\rm cut}^\prime) \geq 0$, and thus $\frac{1 }{1 - {\rm CoR}_{\ket{E_i}} (H_{\rm cut}, H_{\rm cut}^\prime)} \geq 1$. 
\end{thm}

\begin{proof}[Proof (sketch)]
    In all three cases, the proof is a composition of the variance of the Hamiltonians in $\mathcal{B}_{\rm Pauli}$ and $\mathcal{B}_{L_x, L_y}$ into the ones of the unit cell interaction $V$. The multi-linearity of the covariance and the invariance of the ground state enable an identification of terms yielding the respective factor for each case. See Appendix \ref{app:def_partitionings} for technical details. For explicit examples of Pauli and geometric partitionings see Appendix~\ref{appendix:list_all_partitions}.
\end{proof}
We have formulated Theorem \ref{thm:improvement} as a sample improvement when measuring the Hamiltonian on non-degenerate eigenstates. Note that case (a) is not a special case of (b), as the latter cuts in both directions, s.t. for example, setting $L_x=n_x$ will yield a weaker bound. It readily generalizes to other measurement schemes $\mathcal{B}$. A generalization from 2-local to $k$-local interactions is straightforward, albeit it comes with a weakening factor on the sample improvement, or respectively a scaling factor of necessary cluster thicknesses. 

\begin{figure*}[t!]
    \centering
    \includegraphics[width=\linewidth]{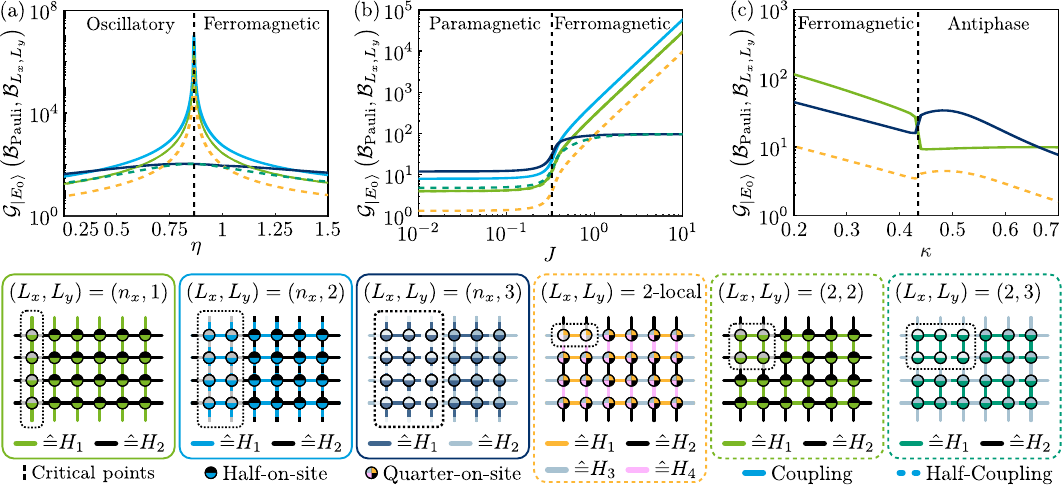}
    \caption{\textbf{Numerical examples: TFXYM, TFIM and TF-BNNNI.} These are examples for the application of Theorem~\ref{thm:improvement}. All examples are on a $4 \times 6$ rectangular lattice with periodic boundary conditions. Critical points of phase transitions are indicated with dashed vertical lines and the different geometric splittings are color-coded  (compare graphical legend as well as Tables~\ref{table: explicit_partitions_TFXYM}~-~\ref{table: explicit_partitions_TFBNNNI} in Appendix~\ref{appendix:list_all_partitions}). In the legend, the dotted boxes show the size of the unit cell, for which one needs to find a $U_m$ in a geometric partitioning, compare also Fig.~\ref{fig: Introduction}. Lattice vertices that are colored with two colors indicate that the geometric partitioning splits the single-site terms symmetrically into two parts. For 2-local partitionings, the single-site terms are symmetrically split into 4 parts. 
    %
    %
    (a)~2D~Transverse field XY-model (TFXYM): For $\eta \rightarrow \eta_c$ (here: $\eta_c=\sqrt{3}/2$) the improvement increases and diverges around the phase transition, where correlation lengths increase (see Theorem \ref{thm:improvement}). 
    %
    %
    (b) 2D Transverse Field Ising Model (TFIM). For $J/h \ll 1$ this is an example where lower bounds are approximately realized. In contrast, for $J/h \gg 1$, we see improvements noticeably surpassing the lower bounds, with scalings of $16L$ up to $\mathcal O((J/h)^2)$. For details see Lemma~\ref{lemma:TFIM improvements}. Here, the Pauli cost is multiplied by 100 to fit the common scale.
    %
    %
    (c) 2D Transverse Field Biaxial Next Nearest Neighbor Ising (BNNNI) Model. $\kappa$  is the NNN coupling, where $\kappa=0$ corresponds to the TFIM. As the interaction range for this model is that of a 3-local Hamiltonian, fewer geometric partitionings are available for the considered lattice sizes.  
     }
\label{fig: Examples}
\end{figure*}
The lower bounds (\theequation a) and (\theequation b) in Theorem \ref{thm:improvement} become arbitrarily large for ${\rm CoR}(H_{\rm cut}, H_{\rm cut}^\prime) \rightarrow 1$, or equivalently $\Var_{\ket{E_i}} (H_{L_x,L_y}^{(1)})=\Var_{\ket{E_i}} (H_{L_x,L_y}^{(2)}) \rightarrow 0$. 
This behavior is expected for our strategy of comparing the required measurements to achieve the same precision. For $\sigma_{\mathcal B_{L_x,L_y}} \to 0$ one would need a diverging number of measurements via Pauli partitioning with $\sigma_{\mathcal B_{\rm Pauli}}  \neq 0$ to achieve the same standard error, c.f. Eq. \eqref{eq: std_error_min}.

We conclude that for pure eigenstates, geometric partitioning will always yield a sampling advantage compared to standard Pauli partitioning, provided it can be efficiently implemented. The implementability of the measurement circuits $U_{H_b}$ is mainly determined by the extra noise they generate due to imperfect implementations of the involved gates. To answer this question and analyze the impact of errors in the preparation of the measured state, we will cover in the following how the above results generalize to imperfect state preparation and imperfect implementation of $U_{H_b}$.

Next, we provide numerical examples illustrating our analytical results. Readers who are primarily interested in the generalization of our analytical results to scenarios that take imperfections in the state preparation and readout circuits into account may directly continue with section \ref{subsec:imperfect eigenstates}. 
All examples considered in this section are on a two-dimensional $(n_x,n_y)=(4,6)$ rectangular lattice with periodic boundaries, apart from the (spinful) Fermi-Hubbard model, which also uses 24 qubits on a $3\times 4$ lattice.
The considered examples are the 2D Transverse Field XY-Model (Sec.~\ref{subsubsec: TFXYM}), the 2D Transverse Field Ising-Model (Sec.~\ref{subsubsec:TFIM}), and the 2D Transverse Field Biaxial Next-Nearest Neighbor Ising Model (Sec.~\ref{subsubsec:BNNNI}), which are illustrated in Fig.~\ref{fig: Examples}. In Fig.~\ref{fig: Examples1b}, we show the Hard-Core Bose Hubbard Model (Sec.~\ref{subsubsec:HBH}), the spinless (Sec.~\ref{subsubsec: Spinless Hubbard}) and the spinful Fermi-Hubbard model (Sec.~\ref{subsubsec: Spinful Hubbard}).

\subsection{Two-dimensional Transverse Field XY-Model} 
\label{subsubsec: TFXYM}
Theorem \ref{thm:improvement} indicates that measuring via geometric partitionings can become (almost) as good as measuring in the eigenbasis. The relative sampling complexity improvement lower bound comparing geometric partitioning with mutually commuting local operator grouping, which mostly coincides with Pauli partitioning for simple lattice models, can even diverge for ${\rm CoR}(H_{\rm cut}, H_{\rm cut}^\prime) \rightarrow 1$. This means that a specific eigenstate correlates parts of the Hamiltonian $H_{\rm cut}$ and $H_{\rm cut}^{'}$ very efficiently. Plausible candidates are highly entangled states, for instance, around zero-temperature (= ground state) phase transitions, where band gaps typically close.
If $H_{\rm cut}$ and $H_{\rm cut}^{'}$ have at least one lattice site in common, then their graph distance is 0. This applies, for example, for $L=1$ or $(L_x,L_y)=(2,2)$. Then e.g. the Hastings-Koma lemma~\cite{hastings2006spectral} does not yield an upper bound on the covariance making these specific geometric partitionings particularly well-suited candidates for divergent relative sampling complexity improvements. 

Indeed, the 2D Transverse Field XY-Model~\cite{henkel1984statistical}, provides such an example, which is described by the following Hamiltonian:
\begin{align}
    H_\textrm{TFXYM} = &- \frac{1}{2} \sum_{\left< i, j\right>} \left( (1+\eta) X_iX_j + (1-\eta) Y_iY_j \right) \nonumber \\
    &- h \sum_i Z_i \ ,
   \label{eq: results_examples_XY_Hamiltonian}
\end{align}
with anisotropy parameter $\eta$. Note that for $\eta=1$, we would obtain a Transverse Field Ising model with $J=h=1$, which we show as a second example (see Section~\ref{subsubsec:TFIM}). $H_\textrm{TFXYM}$ has three phases: a paramagnetic, a ferromagnetic, and an oscillatory one. This 'oscillatory' phase is a type of ferromagnetic phase with oscillating band gap~\cite{nishiyama2019multicritical}.
The phase boundary between ferromagnetic and oscillatory phase is at $\eta^2 + (h/2)^2 = 1$~\cite{henkel1984statistical}. 

Hence, we fix $\eta$ or $h$ to scan the other parameter over the phase transition. In Fig.~\ref{fig: Examples} (a) we set $h=1$ and scan $\eta$. Indeed, we find around the phase transition, $\eta \approx \sqrt{3}/{2}$, as predicted by Theorem~\ref{thm:improvement} for $L=1$ (light green), $L=2$ (cyan) as well as $(L_x,L_y)=(2,2)$ (dashed light green, nearly not distinguishable from $L=1$) a divergence of the relative sampling improvement compared to Pauli partitioning. In addition, this divergence is also realized for the geometric 2-local partitioning (dashed yellow), providing an example that also the easier-to-implement (c)-type measurement strategies in Theorem~\ref{thm:improvement} can exhibit diverging relative sampling complexity improvements, even though we were not able to show this analytically. 
Further, the TFXYM is an example for the extension of Theorem~\ref{thm:improvement}, which is proven for Pauli-2-partitions, onto $\geq2$-Pauli-partitions due to Eq.~\eqref{eq: std_error_ineq}. 

\subsection{Two-dimensional Transverse field Ising-Model} 
\label{subsubsec:TFIM}
Another class of cases with large sampling improvement are examples where $\Var_{\ket{E_i}} (H_1)=\Var_{\ket{E_i}} (H_2)$ scale more beneficial than Pauli partitioning.
One such instance is a standard test bed of quantum spin models, the 2D Transverse Field Ising Model~\cite{ising1924beitrag, onsager1944crystal, schultz1964two}:
\begin{align}
    H_\textrm{TFIM} = -J \sum_{\left< i, j\right>} Z_iZ_j - h \sum_i X_i \ ,
\end{align}
which has a phase transition at $h_c \approx 3.044 J$~\cite{blote2002cluster}, where it exhibits a ferromagnetic phase for $h<h_c$ and a paramagnetic phase for $h>h_c$. We show the following Lemma:
\begin{lemma}[Perturbative improvement estimations]
\label{lemma:TFIM improvements}
The sampling complexity improvement of estimating the ground state energy $E_0$ of the 2D TFIM on a rectangular lattice of geometric partitioning compared to Pauli sampling is given by:
\setcounter{equation}{\theequation + 1}
\begin{align}
   &\mathcal G_{\ket{E_0}} \left( \mathcal B_{\rm Pauli}, \mathcal B_{L_x, L_y} \right) \nonumber \\
   &= \left\{ \begin{array}{l l l}
  4L + \mathcal O ( J/h ) & \quad  J/h \ll 1  &\qquad \rm{(\theequation a)} 
  \\
  \mathcal O ( (J/h)^2)  & \quad h/J \ll 1, L \leq 2 &\qquad \rm{(\theequation b)} 
  \\
  32L + \mathcal O ( (h/J)^2) & \quad  h/J \ll 1, L > 2 \ &\qquad \rm{(\theequation c)}  
  \end{array} \right  . \nonumber
  \label{eq: result_perturbative_estimate}
\end{align}
%
%
\begin{proof}[Proof (Sketch)]
    The Lemma follows from tedious, but straightforward second perturbation theory calculations of the respective variances. See Appendix~\ref{appendix:TFIM}.
\end{proof}
\end{lemma}

\begin{figure}[t!]
    \centering
    \includegraphics[width=\linewidth]{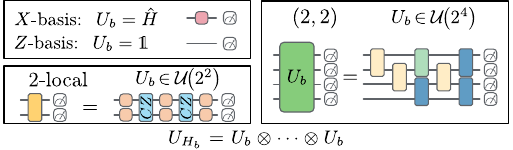}
     \vspace{-0.5cm}
    \caption{{\textbf{Circuit decomposition example for the TFIM.} 
    Here, we give decompositions of the to-be-implemented local measurement unitaries $U_b$. For Pauli measurements, we can either just measure (Z) or need to apply Hadamard gates on each qubit (X). For 2-local geometric partitioning, we can decompose $U_b$ into 1-qubit gates and 2 CZ gates. For 4-qubit $2\times 2$ unitaries, we would in general need to use exact~\cite{shende2004minimal,shende2005synthesis} or approximate~\cite{dawson2005solovay,khatri2019quantum,madden2022best} compiling techniques. For the $2\times 2$ TFIM, which is the same as a 4-qubit 1D TFIM with periodic boundaries, an efficient circuit using only seven 2-qubit unitaries is known~\cite{verstraete2009quantum}.
    }}
    \vspace{-0.5cm}
    \label{fig: tfim_circuits}
\end{figure}

This means that for $J/h \ll 1$ the lower bounds (a) in Theorem~\ref{thm:improvement} are saturated, and hence tight. As shown in Fig.~\ref{fig: Examples} (b), in all three cases: 1D cuts, 2D cuts and 2-local partitions, the numerical results saturate the lower bounds, indicating that we cannot hope for larger lower bounds within the given setting.

Furthermore, for $J/h \gg 1$ we find within the numerical example for $L\leq 2$, $(L_x, L_y) = (2,2)$ as well as 2-local geometric partitioning a model parameter dependent improvement of order $\mathcal O((J/h)^2)$, which matches Lemma~\ref{lemma:TFIM improvements} for $L\leq 2$. Even though our Theorem~\ref{thm:improvement} gives a stronger lower bound for larger geometric partitions, this is an example of the possibility that smaller and thus easier to implement geometric partitions can yield a higher relative sampling complexity advantage. 
The respective decompositions of the local measurement unitaries $U_b$ are given in Fig.~\ref{fig: tfim_circuits}. The 2-local measurement can be decomposed into just 2 CZ-gates and some 1-qubit gates. The $2 \times 2$ partition, where Theorem~\ref{thm:improvement} guarantees at least a 4-fold reduction of the measurement effort, is equal to a 4 qubit chain with periodic boundary conditions. For this case, a particular efficient decomposition requiring only 7 2-qubit unitaries is known~\cite{verstraete2009quantum}.
\subsection{Two-dimensional Transverse field biaxial next-nearest-neighbor Ising model} 
\label{subsubsec:BNNNI}
To arrive at a model, which we need to partition similarly to an (axial) 3-local one, we can add an axial next nearest-neighbor $Z_iZ_j$ interaction to the TFIM. Then, we obtain the Transverse field biaxial next-nearest-neighbor Ising (TF-BNNNI) model on a two-dimensional rectangular lattice~\cite{hornreich1979lifshitz, eckstein2024large}:
\begin{align}
H_\text{TF-BNNNI}= J &\left( - \sum_{\langle i,j \rangle} Z_i  Z_j + \kappa \sum_{\langle\langle i,j \rangle\rangle}  Z_i  Z_j\right) \\
&- h  \sum_i  X_i \nonumber \, .
\end{align}
The TF-BNNNI model exhibits three distinct phases, compare e.g. Ref.~\cite{eckstein2024large} Fig. 2. For $J=1$ (to fix the energy scale), $h=0$, and $\kappa < \kappa_c = \frac{1}{2}$, the nearest neighbor coupling dominates and the ground state is ferromagnetically ordered. If $\kappa$ becomes larger than 0.5, it becomes energetically favorable that $2\times2$ plaquettes are aligned and neighboring plaquettes are flipped with respect to each other. Thus, the BNNNI model can be viewed as a basic spin model for magnetic frustration. 

\begin{figure*}[t!]
    \centering
    \includegraphics[width=\linewidth]{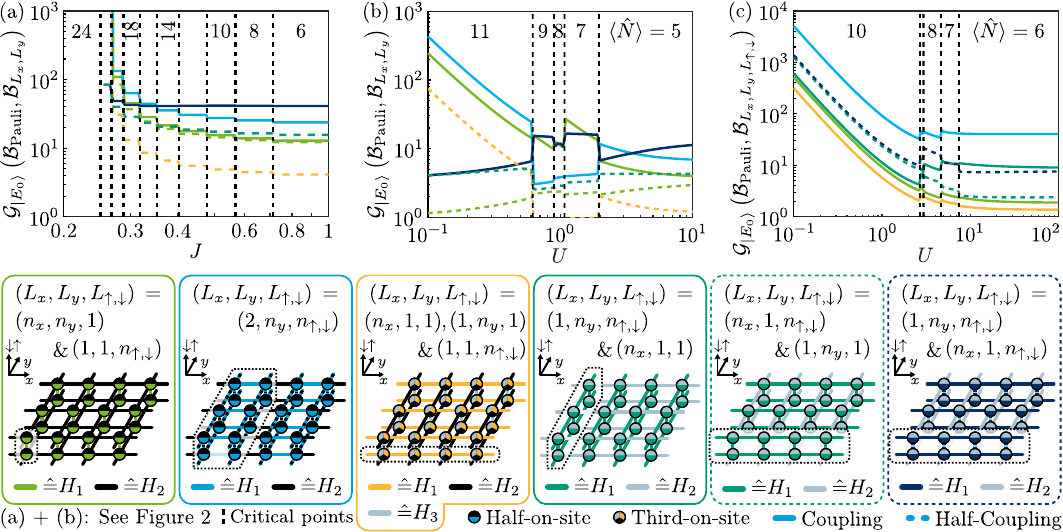}
    \caption{\textbf{Numerical examples: Hubbard-typed models.} The spinless examples (a) and (b) are on a $4 \times 6$ rectangular lattice with periodic boundary conditions. Their partitionings and colors, including solid and dashed lines, correspond to the ones in Fig.~\ref{fig: Examples}.
    Dashed vertical lines indicate transitions between symmetry sectors of different particle number, and the different geometric splitting are color-coded. 
    %
    %
    (a) Hard-core Bose Hubbard (HCBH) model. This model is similar to the TFXYM where the XY-couplings are tuned symmetrically. The numbers show the amount of bosons in the system $\braket{E_0 |\hat N | E_0}$. 
    %
    %
    (b) Spinless Fermi Hubbard model. This model resembles the HCBH model, but with fermions instead of bosons on the lattice sides. Here, as a baseline the partitioning separating the Hamiltonian into mutually commuting groups of 2-local fermionic Hermitian operators is used.  
    %
    %
    (c) (Spinful) Fermi Hubbard model. In comparison with its spinless version, here the repulsive nearest-neighbor coupling is replaced with a repulsive on-site spin-spin coupling. Thus, it can be viewed as two spinless Hubbard models, coupled by their spin degree of freedom. Thus, it also consists of 24 lattice sides, but on a $4\times 3 \times 2$ graph  (compare graphical legend as well as Tables~\ref{table: explicit_partitions_TFXYM},~\ref{table: explicit_partitions_spinless_Hubbard} and~\ref{table: explicit_partitions_spinfull_Hubbard} in Appendix~\ref{appendix:list_all_partitions}).
     }
\label{fig: Examples1b}
\end{figure*}

Still, without a transverse field, the model remains diagonal or classical, and the eigenstate sampling task  trivial as all eigenstates are computational basis states. In the limit $h \rightarrow \infty$ the ground state becomes a disordered $X$ eigenstate. For our demonstration of how geometric partitioning works for a (quasi-)3-local model, we chose $h=1$, and then $\kappa_c \approx 0.435$, compare Figure~\ref{fig: Examples} (c). Compared to the 2-local next-nearest-neighbor models, we need to pick larger geometric partitions. The $L=1$-partitioning (light green) still works as the NNN coupling is axial and both partitions are rotated with respect to each other. In contrast, the $L=2$-cutting would not allow to include the NNN-terms, which is why the only other one-dimensional cut we consider is $L=3$ (dark blue, leading to 3 parts). Similarly, the four 2-local partitions become six axial 3-local partitions (dotted yellow), or for larger systems four axial 5-local partitions.
For similar reasons, we do not show partitionings into two-dimensional patches, which would only be useful for larger lattices of $5\times 10$ or $10 \times 10$ sites.

\subsection{Two-dimensional hard-core Bose-Hubbard model} 
\label{subsubsec:HBH}
If we parameterize the TFXYM, see Eq. \eqref{eq: results_examples_XY_Hamiltonian}, with a symmetric interaction strength $J$,
\begin{align}
    H_\textrm{HCBH} = &- \frac{J}{2} \sum_{\left< i, j\right>} \left(  X_iX_j + Y_iY_j \right) +  \frac{h}{2}  \sum_i Z_i \ ,
   \label{eq: Spin_HCBH}
\end{align}
we obtain a Hamiltonian equivalent to the hard-core limit of the Bose-Hubbard model~\cite{murg2007variational,jordan2009numerical}
\begin{align}
    H_\textrm{HCBH} = &- J \sum_{\left< i, j\right>} \left(  a_i^\dagger a_j + a_i a_j^\dagger \right) +  h \sum_i n_i \ ,
   \label{eq: Fock_op_HCBH}
\end{align}
with $a_i$, $a_i^\dagger$ and $n_i = a_i^\dagger a_i$ being bosonic annihilation, creation and number operators at lattice side $i$, respectively. $\left< i, j\right>$ indicates nearest neighbors on a rectangular lattice. Compared to Bose-Hubbard model~\cite{gersch1963quantum}, in the hard-core limit the on-site repulsion term $U/2 \sum_i n_i(n_i-1)$ is "frozen" out as $U \rightarrow \infty$. Hence, the system is described by Eq.~\eqref{eq: Fock_op_HCBH} with the constraint that a lattice site can maximally be occupied by up to one boson due to the infinite on-site repulsion. 
This means that each lattice site is only a two-level system, and the Hilbert space of the model is only $2^n$-dimensional. Further, the number operator $\sum_i n_i$ (or respectively $\sum_i Z_i$) commutes with the Hamiltonian, and thus gives an excitation number conservation symmetry, separating the Hilbert space into $n+1$ $n \choose m$-large sectors ($m\in \{0, ..,n\}$) \cite{yanay2020two}. 
Hence, for $J \lessapprox h/4$, we are in the one-dimensional all (non-)occupied sector, where the ground state is simply a $Z$ eigenstate. 

The numerical results for this model are shown in Figure~\ref{fig: Examples1b} (a), where for $J \lessapprox h/4$ the ground state is a $Z$ eigenstate. In turn, the configuration space becomes the largest, and hence the model most complex, if about half the sites are occupied ($m=n/2$), which is the case for the $4\times6$ hard-core Bose-Hubbard model for $0.4 h \lessapprox J \lessapprox 0.47 h$. Still, we find also in this most challenging regime sampling improvements of geometric partitioning $\mathcal G_{\ket{ E_i}} \left( \mathcal B_{\rm Pauli}, \mathcal B_{L_x,L_y} \right)$ ranging from ca. 5 to~40. 

\subsection{Two-dimensional Spinless Hubbard Model} 
\label{subsubsec: Spinless Hubbard}
Next, we demonstrate that our geometric partitioning improves sampling efficiency for fermionic lattice models as well. For these models, we geometrically partition the Hamiltonian in the fermionic Fock basis and then perform a local fermion-to-qubit mapping for each patch separately (together with the transformations $U_{b}$).
We start with the Hubbard model for spinless interacting fermions, that is similar to the hard-core Bose Hubbard model from a lattice perspective, just that fermions live on the lattice sites,
\begin{align}
  H_\textrm{Spinless Hubbard} = &- t \sum_{\left< i, j\right>} \left(  c_{i}^\dagger c_{j} + c_{j}^\dagger c_{i}  \right) \nonumber \\& +U \sum_{\left< i, j\right>} n_{i} n_{j}  -  \mu \sum_{i} n_{i} \ .
   \label{eq: Fock_op_spinless_Hubbard}
\end{align}
Here $c_i$, $c_i^\dagger$ and $n_i = c_i^\dagger c_i$ label annihilation, creation and number operators for Fermions. The $t$-term implements fermionic hopping between neighboring sites on the rectangular lattice, where we set $t=1$ to fix the energy scale. The $U$-term (for $U>0$) penalizes the nearest neighbor occupation~\cite{zhang2003stripes}. Thus, for zero on-site energy $\mu = 0$, the model remains below half-filling ($\Braket{\sum_i n_i} < n/2$). Similarly to the hard-core Boson Hubbard model, the number operator $\hat N= \sum_i n_i$ commutes with the Hamiltonian, is hence a symmetry, and as a consequence sub-divides the Hilbert space into $n \choose m$, $m\in \{0,..,n\}$ large subspaces. 

From this one would conclude that the largest sub-Hilbert space occurs for $m=n/2$, at half filling, but at this point the system has a further conserved quantity due to a particle-hole symmetry. Thus, we identify as a challenging scenario systems close to half-filling, but not exactly half-filled $n/2 \approx m\neq n/2$. For the $4\times 6$ lattice model we consider here, this is realized for $\mu=0$ and $U \lessapprox 0.6$, where we have $\braket{\hat N} = 11$ (compare Figure~\ref{fig: Examples1b} (b)). 
For this reason, we set in the following $\mu = 0$. Further, note that this does not make the (non-geometric) sampling problem easier as the number operators can be measured together anyway, and hence the number of {standard} partitions is unaffected{, compare Appendix~\ref{appendix:list_all_partitions}}. Similarly to the spirit of the baseline within the bosonic basis, we group here as a comparison the operators in $H_\textrm{Spinless Hubbard}$ in groups of mutually commuting operators. For $U\neq 0$ there are 5 groups, 4 groups being from the hopping part of the Hamiltonian with disjoint support, and one group measuring the number operators. Under Jordan-Wigner mapping, this corresponds to the Pauli grouping, which is why we still call it Pauli partitioning in Figure~\ref{fig: Examples}. 

Further, we find for $U \approx 1$ ($\braket{\hat N} = 8$), that the ground state of the model is degenerate or respectively gapless. Even though our Theorem~\ref{thm:improvement} does then not strictly apply any longer, we find in this numerical test that the sampling improvement for geometric partitioning persists.

\subsection{Two-dimensional Spinful Hubbard Model} 
\label{subsubsec: Spinful Hubbard}
As a final example for geometric partitioning to reduce the measurement effort when estimating energies of eigenstates, we show the Hubbard model. This model, which was originally proposed as an effective model to study electron correlations, for example how transitions from conductors to insulators occur~\cite{hubbard1963electron}, is of central importance in condensed matter physics.
It resembles the previously shown spinless fermion model, but features an on-site spin-spin (repulsive) interaction, instead of neighbor repulsion: 
\begin{align}
  H_\textrm{Hubbard} = &- t \sum_{\left< i, j\right>, \sigma} \left(  c_{i, \sigma}^\dagger c_{j, \sigma} + c_{j, \sigma}^\dagger c_{i, \sigma}  \right) \nonumber \\& +U \sum_i n_{i, \uparrow} n_{i, \downarrow}  -  \mu \sum_{i, \sigma} n_{i, \sigma} \ .
   \label{eq: Fock_op_Hubbard}
\end{align}
One can view this (spinful) Hubbard model as two spinless ones (w.o. neighbor repulsion), that are repulsively coupled only at the same lattice sites $i$, forming a biplanar graph with the two two-dimensional planes being the up and down spin sectors. Thus, one can consider the Hubbard model as an example of geometric partitioning on a three-dimensional lattice. Testing a real three-dimensional example, including cubic patches, remains challenging as one would need 4 lattice sites in each dimensional direction or 64 overall. 

We keep $t=1$ and $\mu = 0$, and expect again to remain close to but below half-filling for all $U$. Indeed, for $U \lessapprox 2$ we have $\braket{\hat N} = 10$, which is still a larger Hilbert space than half-filling due to the particle-hole symmetry at half-filling that halves the dimension of the Hilbert subspace~\cite{lieb1989two}. The baseline we compare geometric partitioning with works analogously to the spinless case, with four partitions to measure the hopping terms and one to measure the on-site repulsion.

For the spinful Hubbard model, all three particle sectors in the middle in Figure~\ref{fig: Examples1b} (c), $8 \leq \braket{\hat N} \leq 10$, and $2.5\lessapprox U \lessapprox 7.5$, are gapless. We see that the improvement noticeably leaves the smooth trajectory it has before and after, but nonetheless and despite these cases not being captured within Theorem~\ref{thm:improvement}, we find the sampling improvement by geometric partitioning persists.
\section{Sampling improvements under global depolarization noise} 
\label{subsec:imperfect eigenstates}

In real-world sampling situations, where we aim to prepare eigenstates, we can typically not avoid small perturbations. 
These can originate from approximation errors in the state preparation $U_{\ket{E_i}}$ and measurement circuits $U_{H_b}$ as well as imperfect operations on the quantum hardware. While a detailed study of noise effects would require a hardware specific noise model, the noise in local gates can often be well approximated by local depolarization noise~\cite{dur2005standard, gonzalez2022error}. Further, local depolarization converges already for $\Omega(\sqrt n)$ deep 2D brick wall circuits to a global depolarization noise~\cite{gonzalez2022error}. We thus consider a noise channel for global depolarization, which also provides an instance and hardware agnostic conservative estimate for approximation errors in the unitaries $U_{\ket{E_i}}$ and $U_{H_b}$,
\begin{align}
    \Delta_{\epsilon} (\Ket\psi \Bra \psi) = (1-\epsilon) \Ket\psi \Bra \psi + \epsilon \frac{\1_d}{d} \ .
    \label{eq: global_depolarization_noise_channel}
\end{align}
$\Delta_{\epsilon}$ conserves the initial state with probability $(1-\epsilon)$, and maps to the maximally mixed state $\1_d/d$ with probability $\epsilon$, which can be thought of as one of the $4^n -1$ possible Pauli errors occurring with the same probability.

As it is a unital noise channel ($\Delta_{1-\epsilon}(\1_d/d) = \1_d/d$), we can model the results for the noisy states $\Delta_{\epsilon} (\Ket\psi \Bra \psi)$ as averages over pure states,
\begin{align}
    \ket{\tilde\psi (\epsilon)}\coloneqq \sqrt{1-\epsilon} \ket{\psi} + \sqrt{\epsilon} \ket\xi.
    \label{eq: tilde_psi}
\end{align}
where the error states $\ket \xi$ are effectively Haar random states. As we consider expectation values only up to second moments here, it suffices to demand that averages over the $\ket \xi$ equal averaging over the full Hilbert space up to second moments. More technically, this means that $\ket \xi$ is distributed according to a spherical $2$-design in $\CC^d$, compare also Appendix~\ref{appendix:Lemma_2}. 

If the main error contribution to the entire quantum circuit is due to the state preparation circuit $U_{\ket{E_i}}$, we have the same noise strength $\epsilon$ for both Pauli and geometric partitioning. 
We can however account for noise due to the additional gates, that geometric partitioning will use to rotate into the local eigenbases via $U_{L_x, L_y}$, by considering different noise levels for Pauli and geometric partitioning. We thus introduce the ratio $\gamma \in [0,1]$ of both noise levels
\begin{equation} \label{eq:gamma_def}
    \gamma \coloneqq \frac{\epsilon_{\rm Pauli}}{\epsilon_{L_x,L_y}}.
\end{equation}
Assuming that the noise is dominated by gate errors due to imperfect implementation of the gates, we can, for each case, estimate the global depolarization noise strength $\epsilon$ via the average gate error $\epsilon_{\rm gate}$ as well as the number of gates of the entire circuit. Since 1-qubit gate errors are typically one to two orders of magnitude smaller than 2-qubit gate errors in state-of-the-art hardware platforms~\cite{quantinuum-hardware-specifications}, we only count 2-qubit gates and denote their number as ${\mathcal{N}_{2q}}$. We can thus estimate~\cite{gonzalez2022error}, 
$(1 - \epsilon) \approx (1- \epsilon_{\rm gate})^{\mathcal{N}_{2q}}$. 
Further, circuit approximation errors can be factored in by increasing the effective gate count. For instance, if we have a state preparation approximation error $\epsilon_{U_{\ket{E_i}}}= 2 \%$ and a 2-qubit gate error $\epsilon_{\rm gate} = 0.1\%$, we can take this into account by increasing $\mathcal{N}_{2q}(U_{\ket{E_i}})$ by $20$.
We can thus approximate $\gamma$ by
\begin{align}
    \gamma &\approx \frac{1-(1-\epsilon_{\rm gate})^{\mathcal{N}_{2q} \left( U_{\ket{E_i}} \right)}}{1-(1-\epsilon_{\rm gate})^{\mathcal{N}_{2q} \left( U_{L_x,L_y} \right) + \mathcal{N}_{2q} \left( U_{\ket{E_i}} \right)}} \ ,
    \label{eq:gamma_gates} 
\end{align}
where $\mathcal{N}_{2q} (U)$ is the number of 2-qubit gates in the circuit $U$ with $U_{\ket{E_i}}\ket 0^{\otimes n}=\ket{E_i}$ being the preparation circuit preparing $\ket{E_i}$ from $\ket 0^{\otimes n}$ and $ U_{L_x,L_y}$ the readout circuit for the geometric partitioning. As the rotation in local Pauli bases can be done with 1-qubit gates, we use $\mathcal{N}_{2q}\left( U_{\rm Pauli} \right ) = 0$. For $\mathcal{N}_{2q}\epsilon_{\rm gate}\ll 1$ one can further approximate
\begin{align}
    \gamma &
    \approx \frac{\mathcal{N}_{2q}(U_{\ket{E_i}})}{\mathcal{N}_{2q}(U_{\ket{E_i}}) + \mathcal{N}_{2q}\left( U_{L_x,L_y} \right )} 
    \gtrapprox  1 - \frac{\mathcal{N}_{2q}\left( U_{L_x,L_y} \right )}{\mathcal{N}_{2q}(U_{\ket{E_i}}) } \ .
\end{align}
$\gamma = 1/2$ would mean that readout via geometric partitioning uses as many 2-qubit gates as the state preparation circuit. Hence for small enough patches one typically has $\gamma > 1/2$. Our numerical results show that such patch sizes nonetheless lead to very significant sampling improvements. 

To take depolarizing noise into account, we need to consider the statistical error average $\EE$ over the corresponding ensemble of states. Nevertheless, a similar upper bound to Eq.~\eqref{eq: std_error_ineq} holds true:
\begin{align}
&\frac{1}{ M} \EE \left[ \Var_{\ket{ \tilde \psi}} \left(  H\right) \right ] 
    \leq \frac{1}{ M} \EE \left[ \left( \sum_{H_b \in \mathcal B}  \sqrt{   \Var_{\ket{ \tilde \psi}} \left(  H_b\right) } \right)^2 \right ] \nonumber \\
   &\qquad \qquad  \leq \frac{1}{ M} \left[ \sum_{H_b \in \mathcal B}  \sqrt{ \EE \left[  \Var_{\ket{ \tilde \psi}} \left(  H_b\right) \right ] } \right ]^2 {=: \overline \sigma^2_{\mathcal{B}}} \ ,
   \label{eq: average_sampling_error}
\end{align}
where, we applied a second time the Cauchy-Schwarz inequality onto the expectation value over $\ket{\tilde \psi}$, to  obtain the second inequality, which would be an equality if
the different random variables $\sqrt{   \Var_{\ket{\tilde \psi}}\left(  H_b\right) }$ were stochastically independent.
See Appendix.~\ref{appendix: compare measurement strategies} for the explicit calculation.
The above inequality allows us to expand the previously defined relative sampling complexity (Definition~\ref{def:Sampling improvement}) from pure to mixed states:
\begin{defn}[Relative Sampling complexity (Mixed state version)]
  \label{def:improvement imperfect states}
Let $\epsilon_1$ and $\epsilon_2$ be the global depolarization noise strength for measuring the quantum state $\ket{\tilde \psi (\epsilon)}$ (Eq.~\eqref{eq: tilde_psi}) in the partitionings $\mathcal B_1$ and $\mathcal B_2$ of $H$.
Then we define the ratio of the number of samples $M_{\mathcal B_1}/M_{\mathcal B_2}$ to obtain $\overline \sigma^2_{\mathcal{B}_1} = \overline \sigma^2_{\mathcal{B}_2}$, c.f. Eq.~\eqref{eq: average_sampling_error}, as the relative sampling complexity: 
\begin{align}
   &\overline{\mathcal G}_{\ket{\psi}} \left( \mathcal B_1^{\epsilon_1}, \mathcal B_2^{\epsilon_2} \right)
   \coloneqq 
   \frac{M_{\mathcal B_1}}{M_{\mathcal B_2}}
   = \left[ 
   \frac{  \sum\limits_{H_b \in \mathcal B_1 } \sqrt{ \mathbb E \left [ \Var_{\ket{ \tilde \psi (\epsilon_1)}} \left(  H_b\right)\right ] }  }
   {  \sum\limits_{H_b^\prime \in \mathcal B_2}  \sqrt{ \mathbb E \left [ \Var_{\ket{ \tilde \psi (\epsilon_2)}}  \left(  H_b^\prime \right) \right ]}  }\right]^2 \ .
   \label{eq: improvement_imperfect_eigenstate}
\end{align}
where the second equality holds because of Eq. \eqref{eq: average_sampling_error}.
\end{defn}

One might wonder whether Eq. \eqref{eq: improvement_imperfect_eigenstate} is a reliable measure because  $\overline{\sigma}_{\mathcal{B}}$ is only an upper bound to the average of variances, see Eq. \eqref{eq: average_sampling_error}. To confirm that $\overline{\sigma}_{\mathcal{B}}$ provides an accurate estimate for the error in an energy measurement by sampling from a quantum state, we compare it to the error of a simulated sampling process in section \ref{subsubsec:Sampling}. 

Note that $\overline{\mathcal G}_{\ket{\psi}} \left( \mathcal B_1^{\epsilon_1}, \mathcal B_2^{\epsilon_2} \right) \to \mathcal G_{\ket{\psi}} \left( \mathcal B_1, \mathcal B_2 \right)$ for $\epsilon \to 0$. Moreover, even though we here focus on global depolarization noise, one could similarly study a relative sampling complexity w.r.t. to two quantum channels $\Lambda_1$ and $\Lambda_2$ instead of the same channel $\Delta_\epsilon$ with different noise strengths $\epsilon_1$ and $\epsilon_2$.
The following Lemma allows us to resolve $\mathbb E \left [ \Var_{\ket{\tilde \psi}}  \left(  H_b \right) \right ]$:

\begin{lemma}[Variance of isotropically perturbed states]
\label{thm:imperfect eigenstates} 
Let $\ket{\psi}\in \CC^d$ be some  fixed reference state vector, 
$\ket\xi$ a random vector distributed according to a spherical $2$-design in $\CC^d$,
$O$ an observable, and $\epsilon\in [0,1]$.
Then, for $\ket{\tilde\psi}\coloneqq \sqrt{1-\epsilon} \ket{\psi} + \sqrt{\epsilon} \ket\xi$, 
\begin{subequations}
\begin{align}
    \EE \left[ \Var_{\ket{\tilde \psi}}(O) \right ]
    &= 
    (1-\epsilon) \Var_{\ket{\psi}}(O) \qquad \label{eq: noisy_var_V}
    \\
    &\quad 
    + \epsilon(1-\epsilon)\sandwich{\psi}{O}{\psi}^2  \label{eq: noisy_var_E} 
    \\&\quad 
    + \epsilon  \norm{O}^2_F/d + \mathcal{O}(1/d) \ .
    \label{eq: noisy_var_H}
\end{align}
\end{subequations}
where $\norm{\cdot}_F/d$ denotes the Frobenius norm renormalized by a dimension factor $d=2^n$ for system size $n$.

\begin{proof}[Proof (Sketch)]
We use known results from representation theory, more concretely the Schur-Weyl duality, and the so-called swap-trick to compute the Hilbert space averages.
For the explicit calculation see Appendix~\ref{appendix:Lemma_2}.
\end{proof}
\end{lemma}

{Note that the setting in Lemma~\ref{thm:imperfect eigenstates} is equivalent to the pure state $\ket \psi$ undergoing global depolarization noise of strength $\epsilon$. Further,
$\EE \left[ \Var_{\ket{\tilde \psi}}(O) \right ] \geq \Var_{\ket{\psi}}(O)$, if $\Var_{\ket{\psi}}(O) \leq \norm{O}^2_F/d + \mathcal{O}(1/d)$. This means that white noise increases the variance unless the noise-free state-specific sampling error is worse than state-independent average case sampling.}

Unlike for perfect eigenstates, we {have for $\epsilon \neq 0$ a non-zero shot noise $\mathbb E \left [\Var_{\ket{\tilde E_i} } \left( H \right) \right ] =  \epsilon (1- \epsilon) E_i^2 +  \epsilon  \norm{H}^2_F/d + \mathcal{O} ( 2^{-n}) \neq 0$, even for measurements in the eigenbasis of $H$}. 
We can thus interpret Eq. \eqref{eq: noisy_var_V} as a contribution to $ \EE \left[ \Var_{\ket{\tilde \psi}}(O) \right ]$ coming from the quantum fluctuations of $O$ in the state $\ket{\psi}$, whereas the contributions in Eqs. \eqref{eq: noisy_var_E} and \eqref{eq: noisy_var_H} are due to the noise. Indeed, in our further results, the following ratios between these contributions,
\begin{align}
    \label{eq:chi_def}
    \epsilon_{I\leftrightarrow II} &\coloneqq \frac{ 4 \Var_{\ket{E_i}} \left( H_{L_x, L_y}^{(1)} \right)}{E_i^2 + \norm{H}^2_F/d } \\
    \epsilon_{II\leftrightarrow III} &\coloneqq \frac{ 4 \Var_{\ket{E_i}} \left( H_{\rm Pauli}^{(1)} \right)}{E_i^2 + \norm{H}^2_F/d }
\end{align}
play a central role.

\begin{figure}[h!]
    \centering
    \includegraphics[width=\linewidth]{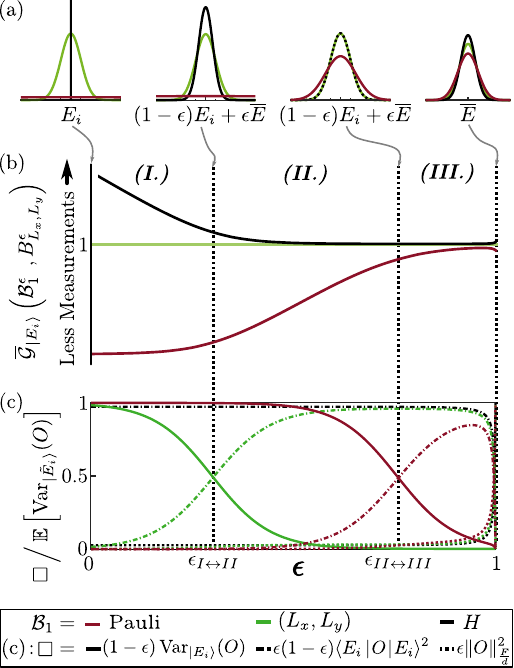}
     \vspace{-0.5cm}
    \caption{\textbf{Sampling improvement: from perfect eigenstates to Haar-random states:} 
    Here, we consider states of the form $\ket{\tilde E_i} = \sqrt{1-\epsilon}\ket{E_i} + \sqrt \epsilon \ket \xi $ with $\ket \xi$ Haar random up to second moments for for a 2D Transverse field Ising model. Yet all graphs will qualitatively look similar for different models. All three subplots share the same color coding: red for Pauli, green for geometric, and black for eigenbasis partitioning. Subplots (b) and (c) share the same x-axis ($\epsilon$). 
    In (a) we show the probability density corresponding to the empirical mean estimator $\overline O_M$ for different measurement strategies (color coded). The different snapshots depict (from left to right) $\epsilon = 0$, $\epsilon \approx \epsilon_{I\leftrightarrow II}$, 
    $\epsilon \approx \epsilon_{II\leftrightarrow III}$ and $\epsilon \approx 1$. 
    In (b) we provide a qualitative graph of the relative sampling complexity as defined in Def.~\ref{def:improvement imperfect states}.
    In agreement with Corollary.~\ref{corol:eps-delta-threashold}, regime (II.) exists and has at least the size $\mathcal G_{\ket{E_i}} \left( \mathcal B_{\rm Pauli}, \mathcal B_{L_x, L_y} \right) \approx \epsilon_{II\leftrightarrow III} / \epsilon_{I\leftrightarrow II}$, if the noise mainly originates from imperfect state preparation, $\gamma \approx 1$, compare Eq.~\eqref{eq:gamma_gates}.
      In (c) we plot the relative weight of the terms in Eqs. \eqref{eq: noisy_var_V}, \eqref{eq: noisy_var_E} and \eqref{eq: noisy_var_H} of Lemma~\ref{thm:imperfect eigenstates}, which are encoded by line type (solid, dash-dotted and dotted), see legend. 
    This shows that the transition from low to intermediate random regime (I.) $\rightarrow$ (II.) happens when the measurement uncertainty of geometric partitioning (green lines of the three types) becomes dominated by the noise-induced terms \eqref{eq: noisy_var_E} and \eqref{eq: noisy_var_H}.
    Similarly, the transition from (II.) $\rightarrow$ (III.) corresponds to the same phenomena for Pauli partitioning (red lines). 
    For reference, the noise-induced uncertainty of sampling in the eigenbasis is shown in black, which almost coincides with geometric partitioning in the middle of regime (II.) as well as regime (III.), as we expect from Theorem~\ref{thm:2}.
    }
\label{fig: Noisy_Results}
\vspace{0.5 cm}
\end{figure}

Thus, $\mathcal B_1 = \mathcal B_H$ provides a nontrivial finite upper bound for the sampling improvement (see also Fig.~\ref{fig: Noisy_Results} (c)). This means that, in the case of $\epsilon \neq 0$, there will be both an upper and lower bound, compare also Appendix~\ref{appendix: For_Sec_II_C}.

\begin{thm}[Rel. Sampling complexity improvement under global depolarization noise]
\label{thm:2}
Let $\overline{\mathcal G}_{\ket{E_i}} \left( \mathcal B_1^{\epsilon_1}, \mathcal B_2^{\epsilon_2} \right)$ be the relative sampling complexity with noise strength $\epsilon_1$ for partition $\mathcal B_1$, and  $\epsilon_2$ for $\mathcal B_2$ (Def.~\ref{def:improvement imperfect states}). Let the noisy state under global depolarization noise be given by $\ket{\tilde E_i(\epsilon)} = \sqrt{1-\epsilon}\ket{E_i} + \sqrt{\epsilon} \ket \xi$, with $\ket{E_i}$ an eigenstate of $H$. 
Then we have the following regimes: 
\begin{enumerate}[label=\Roman*.]
\item For low randomness, $\epsilon \leq \epsilon_{I\leftrightarrow II}$ the sampling improvement is at worst a factor 2 less than for $\epsilon=0$, 
\begin{align} 
    &\overline{\mathcal G}_{\ket{ E_i}} \left(  \mathcal B_{\rm Pauli}^{\gamma \epsilon}, \mathcal B_{L_x, L_y}^\epsilon\right) \geq \frac{1}{2}
    \mathcal G_{\ket{ E_i}} \left(  \mathcal B_{\rm Pauli}, \mathcal B_{L_x, L_y}\right) .
    \label{eq:thm2_1}
\end{align} \vspace{- 5 mm}
independently of $\gamma$, see Eq. \eqref{eq:gamma_def}.
\item For intermediate randomness, we have
\begin{align}
\label{eq:thm2_2a}
    \overline{\mathcal G}_{\ket{ E_i}} \left(  \mathcal B_{L_x, L_y}^\epsilon, \mathcal B_{H}^\epsilon\right) \leq 2&\Rightarrow 
    \epsilon \geq \epsilon_{I\leftrightarrow II} \\
    \overline{\mathcal G}_{\ket{ E_i}} \left(  \mathcal B_{\rm Pauli}^{\gamma \epsilon}, \mathcal B_{H}^\epsilon\right) \geq 2&\Rightarrow 
    \epsilon \leq  \frac{1}{2-\gamma} \epsilon_{II\leftrightarrow III} ,
    \label{eq:thm2_2b}
\end{align}
\item For a randomness dominated regime with $\epsilon  \geq \epsilon_{II\leftrightarrow III} /(C-2\gamma)$, 
where $C>2$ is a constant, the sampling efforts become dominated by noise, and
\begin{align}
    \overline{\mathcal G}_{\ket{ E_i}} \left(  \mathcal B_{\rm Pauli}^{\gamma \epsilon}, \mathcal B_{H}^\epsilon\right) &\leq C \ ,
\end{align}
for any bi-partition of $H$ ($H=H_1+H_2$), in particular also for $\mathcal{B}_{L_x,L_y}^\epsilon$.
\end{enumerate}
\begin{proof}[Proof (Sketch)]
The proof in full detail is given in Appendix~\ref{appendix:Theorem_2}.
The basic idea is to use Lemma~\ref{thm:imperfect eigenstates} to resolve $\mathbb E \left [ \Var_{\ket{\tilde E_i(\epsilon)}}  \left(  H_b \right) \right ]$ on the r.h.s. of Eq.~\eqref{eq: improvement_imperfect_eigenstate}. Then one uses Lemma~\ref{lemma:variance_in_equalities} and Corollary~\ref{corol: Ei_2_partitions} to linearize $(\sum \sqrt .)^2$ both in the numerator and denominator, as well as: 
\begin{align}
    \left[ \sum_{k}  
    \sqrt{ A_k+B_k} \right ]^2 \geq \left[\sum_{k} \sqrt{ A_k} \right ]^2  + \left[\sum_{k} \sqrt{ B_k} \right ]^2 \ .
    \label{eq: inequality_sqrt_sum}
\end{align}
To handle, the resulting fraction, we use iteratively the following algebraic inequality:
\begin{align}
    \frac{A}{A^\prime} \geq \frac{B}{B^\prime} \Rightarrow \frac{A}{A^\prime} \geq \frac{A+B}{A^\prime+B^\prime}\geq \frac{B}{B^\prime} \ ,
    \label{eq: frac_inequality}
\end{align}
for $A, A^\prime, B, B^\prime \geq 0$.
To see that Eq.~\eqref{eq: frac_inequality} holds add $AA^\prime$ or $BB^\prime$ to $AB^\prime \geq BA^\prime$ and resolve for $A/A^\prime$ or $B/B^\prime$. Eventually, to bound the resulting fractions, we use insights from our noise-free analysis, e.g. Theorem~\ref{thm:improvement}. 
\end{proof}
\end{thm}

Note that regime (I.) does not depend on the noise level of Pauli partitioning $\gamma\epsilon$. Similarly, the high-noise threshold of regime (II.), $\propto 1/(2-\gamma)$, for $\gamma = 0$ is only decreased by a factor of 2 compared to $\gamma=1$; if we were interested only in geometric partitioning performing better than Pauli partitioning even only a factor of $4/3$. We have the following Corollary to determine whether geometric partitioning yields lower sampling noise than Pauli partitioning for any noise strength $\epsilon$:
\begin{corol}[Criteria for sampling advantage of geometric partitioning under global depolarization]
\label{lemma:noise-indep sample advantage}
Let all objects be defined as in Theorem~\ref{thm:2}. Then we have
\begin{align}
 \overline{\mathcal G}_{\ket{ E_i}} \left( \mathcal B_{\rm Pauli}^{\gamma \epsilon}, \mathcal B_{L_x, L_y}^\epsilon \right) \geq 1 \ ,
\end{align}
if 
\begin{align}
\label{eq:G_epsilon=1}
\gamma \left( \frac{\sum\limits_{H_b \in \mathcal B_{\rm Pauli}} \norm{H_b}_F }{\sum\limits_{H_b^\prime \in \mathcal B_{L_x, L_y} } \norm{H_b^\prime}_F }\right)^2 &\geq 1 \ ,\\
\qquad {\rm and} \qquad \gamma (1 + \alpha_{\rm Pauli}^2) &\geq 1
\label{eq:G_epsilon=1b}
\end{align}
with $\alpha_{\rm Pauli}  \coloneqq \left| \Braket{ H_{\rm Pauli}^{(1)} }/E_i -1/2 \right | \geq 0$.\\
or if
\begin{align}
    \epsilon \leq  c_{\epsilon} \epsilon_{II\leftrightarrow III}
     \label{eq:thm2_2c}
\end{align}
with $c_{\epsilon} = \max(\frac{1}{2-\gamma},  \,  \left[ 1 - {1}/{\mathcal G_{\ket{ E_i}} \left(  \mathcal B_{\rm Pauli}, \mathcal B_{L_x, L_y}\right)} \right ])$
\begin{proof}[Proof (Sketch)]
Similarly to the methods for Theorem~\ref{thm:2}, we use Eq.~\eqref{eq: inequality_sqrt_sum} to linearized the noisy sampling error of Pauli partition (Eq.~\eqref{eq: average_sampling_error}), and then Eq.~\eqref{eq: frac_inequality} iteratively to obtain:
    \begin{align}
 \overline{\mathcal G}_{\ket{\tilde E_i}} \left( \mathcal B_{\rm Pauli}, \mathcal B_{L_x, L_y} \right)  
 \geq \min \left( \frac{A^\prime}{A}, \frac{B^\prime}{B}, \frac{C^\prime}{C}  \right) \ ,
 \end{align}
 with $A$, $B$, $C$ according to the three lines in Lemma~\ref{thm:imperfect eigenstates}. 
 $A^\prime/A\geq\mathcal G_{\ket{ E_i}} \left(  \mathcal B_{\rm Pauli}, \mathcal B_{L_x, L_y}\right)$ is larger than 1 due to Theorem~\ref{thm:improvement}, 
 $B^\prime/B \geq \gamma (1+\alpha_{\rm Pauli}^2)$ as geometric partitioning equally distributes the energy ($\alpha_{L_x,L_y}^2=0$) and 
 $C^\prime/C \geq \gamma \left({\sum_{H_b \in \mathcal B_{\rm Pauli}} \norm{H_b}_F } \right / \left . {\sum_{H_b^\prime \in \mathcal B_{L_x, L_y} } \norm{H_b^\prime}_F } \right)^2$. 
 The first part of, Eq.~\eqref{eq:thm2_2c} corresponds to Eq.~\eqref{eq:thm2_2b} and the second part one obtains in a similar calculation using $\overline{\mathcal G}_{\ket{ E_i}} \left(  \mathcal B_{\rm Pauli}^{\gamma \epsilon}, \mathcal B_{L_x, L_y}^\epsilon\right) 
\geq \overline{\mathcal G}_{\ket{ E_i}} \left(  \mathcal B_{\rm Pauli}^{0 \epsilon}, \mathcal B_{L_x, L_y}^\epsilon\right)$. 
 The proof in full detail is given in Appendix~\ref{appendix:Lemma_4}.
\end{proof}
\end{corol}

For $\gamma \approx 1$, Eq.~\eqref{eq:G_epsilon=1b} is typically fulfilled as Pauli partitioning in general does not distribute the energy equally. Then, it is sufficient to compute Eq.~\eqref{eq:G_epsilon=1}, which can be calculated classically at scale for sparse observables such as lattice Hamiltonians. As we know the Pauli sum representation of our sparse observables $H_b$, and we have for Paulistring $P_i$ and Kronecker delta $\delta_{i,j} $: $\Tr(P_iP_j)/d=\delta_{i,j}$, calculating the Frobenius norm boils down to summing the squared Pauli sum coefficients $\norm{H_b}^2_F/d = \sum_{c_jP_j \in H_b} c_j^2$. Indeed, $\sum_{H_b \in \mathcal B} \norm{H_b}_F/d$ gives precisely the state-independent average sampling error for partitioning $H$ according to splitting $\mathcal B$, which we would have obtained if we had averaged over the Hilbert space, instead of deriving state-tailored sampling bounds~\cite{crawford2021efficient}.  
Therefore, Eq.~\eqref{eq:G_epsilon=1} tells us, for $\gamma \approx 1$ on average what improvement we can expect by geometric partitioning compared to Pauli partitioning for an arbitrary state.
Thus, even if we cannot take advantage of any specific state structure, we can still improve our measurement scheme on the observable partitioning level by optimizing the partitioning w.r.t. Eq.~\eqref{eq:G_epsilon=1}. 

Complementary to Corollary \ref{lemma:noise-indep sample advantage}, one can require a targeted relative sampling improvement and ask how large the gate error $\epsilon_{\rm gate}$ and the number of gates $\mathcal{N}_{2q}(U_{\ket{E_i}})$ and $\mathcal{N}_{2q}\left( U_{L_x,L_y} \right )$ can be to guarantee such an improvement. This is answered by the following corollary:
\begin{corol}[Gate budget for noisy geometric partitioning]
\label{corol: gate_budget}
Let all objects be defined as in Theorem~\ref{thm:2} and prior. Let $G$ be a desired relative sampling improvement. Then
\begin{align}
 \overline{\mathcal G}_{\ket{ E_i}} \left( \mathcal B_{\rm Pauli}^{\gamma \epsilon}, \mathcal B_{L_x, L_y}^\epsilon \right) \geq G \ ,
\end{align}
provided that 
\begin{align} \label{eq: max-meas-circuit}
\mathcal{N}_{2q}\left( U_{L_x,L_y} \right ) \lessapprox \frac{\log\left( 1 - c_{\epsilon}^{(G)}\epsilon_{II \leftrightarrow III}\right)}{\log(1- \epsilon_{\rm gate})} - \mathcal{N}_{2q}(U_{\ket{E_i}})
\end{align}
with $c_{\epsilon}^{(G)} = \max(\frac{1}{2G-\gamma},  \,  \left[ 1 - {G}/{\mathcal G_{\ket{ E_i}} \left(  \mathcal B_{\rm Pauli}, \mathcal B_{L_x, L_y}\right)} \right ])$
\begin{proof}[Proof (sketch)]
The first $c_{\epsilon}$ follows by replacing "$2$" in Eq.~\eqref{eq:thm2_2b} with "$2G$"
and plugging in Eq.~\eqref{eq:gamma_gates}. $G \leq {\mathcal G}_{\ket{ E_i}} \left( \mathcal B_{\rm Pauli}, \mathcal B_{L_x, L_y} \right) /2$ ensures that we are still in regime (II.) 
The second $c_{\epsilon}$ similarly takes Eq.~\eqref{eq:thm2_2c} and replaces "$\geq 1$" with "$\geq G$". 
For more details see Appendix~\ref{appendix:Corollary_4}.
\end{proof}
\end{corol}

One can see that for very good gate fidelities, $\epsilon_{\rm gate} \to 0$, the upper bound in Eq. \eqref{eq: max-meas-circuit} diverges. This is consistent with the intuition of having an arbitrary gate budget for measurement in this case. In turn, for very noisy circuits, $\epsilon_{\rm gate} \to 1$, the bound would become negative, excluding, as expected, any further gates for the measurement circuit. One can see from Eq. \eqref{eq: max-meas-circuit} that, in order to have a positive bound for $\mathcal{N}_{2q}\left( U_{L_x,L_y} \right)$, one must have
\begin{align}
    \mathcal{N}_{2q}(U_{\ket{E_i}}) \lessapprox \frac{\log\left( 1 - c_{\epsilon}^{(G)}\epsilon_{II \leftrightarrow III}\right)}{\log(1- \epsilon_{\rm gate})}.
\end{align}
State preparation circuits of larger size will not allow for a subsequent measurement circuit.
For $G=1$, we thus re-obtain the intuitive expectation that the gate error $\epsilon_{\rm gate}$ needs to be smaller than the threshold $\epsilon_{II \leftrightarrow III}$ for $\mathcal{N}_{2q}(U_{\ket{E_i}}) \gtrapprox 1$.

In Theorem \ref{thm:2}, one can think of regime (I.) as effectively noise-free. In regime (II.), measuring in geometric partitioning is effectively as good (up to a factor 2) as the optimal strategy, measuring in the eigenbasis. Finally, regime (III.) is too noisy for state-tailored shot noise improvements since the main sampling error will be $\propto \epsilon(1-\epsilon)E_i^2 + \epsilon \norm{H}^2_F/d$, and hence independent of the chosen measurement strategy.  

The regimes of randomness predicted by Theorem \ref{thm:2} are illustrated in Figure~\ref{fig: Noisy_Results} for the example of a 2D Transverse Field Ising model. The data clearly shows the transitions between the regimes as predicted by Theorem \ref{thm:2}. To see that indeed all three regimes are realized, note that for $\epsilon \rightarrow 0$ all but the variance-type terms vanish, and respectively for $\epsilon \rightarrow 1$, all but the norm-type terms vanish. 
The following Corollary, lower bounds the size of regime (II.), and hence guarantees its existence:

\begin{figure*}[t!]
    \centering
    \includegraphics[width=\linewidth]{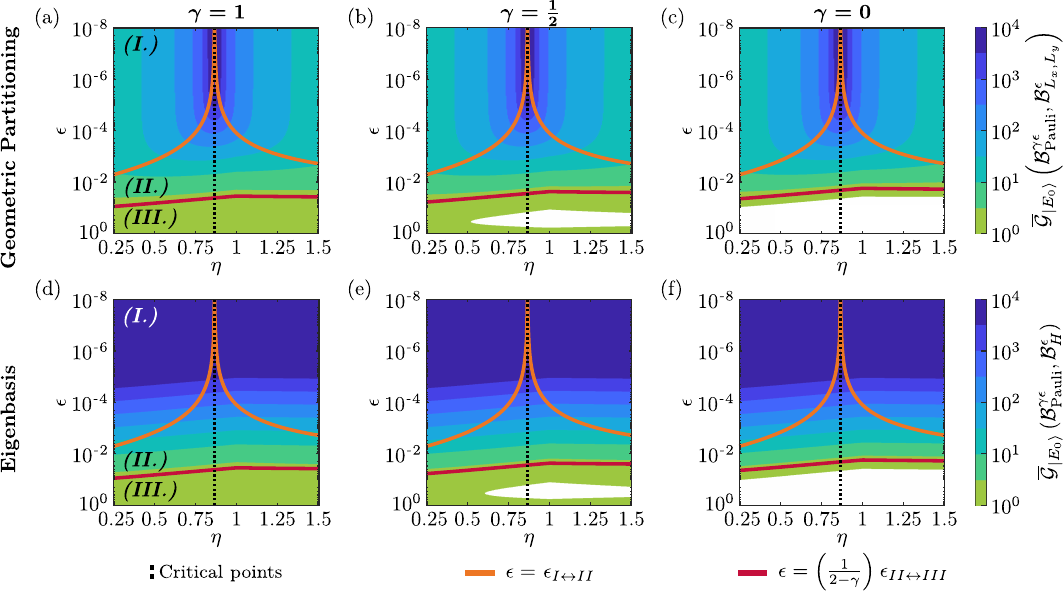}
    \caption{\textbf{Shot-noise reduction for the TFXYM under noise.} 
   Here we present numerical results that confirm Theorem~\ref{thm:2} for the 2D XY-model with Z field on a $4\times6$ rectangular lattice with periodic boundary conditions (see also Sec.~\ref{subsubsec: TFXYM} and Fig.~\ref{fig: Examples} (a) for the $\epsilon = 0$ case). 
   The vertical dashed lines mark, as in Fig.~\ref{fig: Examples} (a), the critical point at $\eta_c = \sqrt 3/2$.
   We pick a geometric partitioning with $(L_x,L_y)=(2,2)$, and plot relative sampling complexities $\overline{\mathcal G}_{\ket{E_0}} \left( \mathcal B_{\rm{Pauli}}^{\gamma \epsilon}, \mathcal B_{L_x, L_y}^\epsilon \right )$ and $\overline{\mathcal G}_{\ket{E_0}} \left( \mathcal B_{\rm{Pauli}}^{\gamma \epsilon}, \mathcal B_{H}^\epsilon \right )$, which are both color coded, for varying depolarization noise $\epsilon$ and system parameter $\eta$. 
   The orange line marks $\epsilon = \epsilon_{I \leftrightarrow II}$ and the red line $\epsilon = \frac{1}{2-\gamma}\epsilon_{II \leftrightarrow III}$, see legend.
   We cut the improvement scale at $10^4$ and a white area indicates that Pauli partitioning is better.
   The upper row shows the sampling improvement of geometric partitioning compared to Pauli partitioning, and the lower row compares hypothetical sampling in the eigenbasis with Pauli partitioning. 
   The columns show different relative noise strength $\gamma$ to take into account potentially additional gates when using geometric partitioning, or the eigenbasis.
   One can see that regime (I.) is effectively noiseless as there is no noticeable $\epsilon$-dependence. Similarly, geometric partitioning (top row) is essentially as good as sampling in the true eigenbasis (bottom row) in regimes (II.) and (III.), as predicted by Theorem~\ref{thm:2}.
   In agreement with Corollary~\ref{lemma:noise-indep sample advantage}, we find that geometric partitioning is always better than geometric partitioning in the high-sampling advantage regimes (I.) and (II.).
   Moreover, the size of regime (II.) is given by $\frac{1}{2-\gamma}{\mathcal G}_{\ket{E_0}} \left( \mathcal B_{\textrm{Pauli}}, \mathcal B_{L_x, L_y} \right )$, as stated in Corollary~\ref{corol:eps-delta-threashold}. Therefore, a large noiseless sampling complexity improvement will lead to a large noise range where geometric partitioning performs effectively as well as measuring in the true eigenbasis and is guaranteed to be better than Pauli partitioning.  
     }
\label{fig: Examples2}
\end{figure*}

\begin{corol}[Existence and intermediate noise regime]
\label{corol:eps-delta-threashold}
Let all objects be defined as in Theorem~\ref{thm:2}. Then there exist noise thresholds $\epsilon^{\rm th}_{L_x, L_y}, \epsilon^{\rm th}_{\rm Pauli}> 0$ s.t.:
\begin{align}
\forall \epsilon \geq \epsilon^{\rm th}_{L_x, L_y}:& \quad \overline{\mathcal G}_{\ket{ E_i}} \left( \mathcal B_{L_x,L_y}^\epsilon, \mathcal B_H \right) 
   \leq 2 \ ,\\
   \forall \epsilon \leq \epsilon^{\rm th}_{\rm Pauli}:& \quad  \overline{\mathcal G}_{\ket {E_i}} \left( \mathcal B_{\rm Pauli}^{\gamma \epsilon}, \mathcal B_H \right) \geq 2 \ ,
   \label{eq: corollary_4_1}
\end{align}
and the ratio of those noise thresholds is given by:
\begin{align}
 &\frac{\epsilon^{\rm th}_{\rm Pauli}}{\epsilon^{\rm th}_{L_x,L_y}} \geq \left ( \frac{1}{2 - \gamma}  \right ) \mathcal G_{\ket{E_i}} \left( \mathcal B_{\rm Pauli}, \mathcal B_{L_x, L_y} \right)   \ .
 \label{eq: corollary_4_2}
 \end{align} 
For all but 2-local partitioning, the r.h.s. of Eq.~\eqref{eq: corollary_4_2} $\geq 2$ due to Theorem~\ref{thm:improvement}.

\begin{proof}[Proof]
This is a direct consequence of Theorem~\ref{thm:2}, as have found there explicit approximate thresholds of the form $\epsilon_{I\leftrightarrow II} \geq \epsilon^{\rm th}_{L_x, L_y}$ and $\epsilon^{\rm th}_{\rm Pauli} \geq \frac{1}{2-\gamma} \epsilon_{II\leftrightarrow III}$, from which the statement follows.
\end{proof}
\end{corol}

In the limit of large depolarization noise ($\epsilon \rightarrow 1$), one thus recovers the same estimates as via averaging over Hilbert space, compare Appendix~\ref{appendix: compare measurement strategies}. Our results show that sampling improvements can be more significant for small noise regimes, as the structure of the observable and its eigenstate can be exploited.

\subsection{Sampling improvements under noise: numerical analysis}
\label{subsec: example imperfect eigenstates}
\begin{figure*}[t!]
    \centering
    \includegraphics[width=\linewidth]{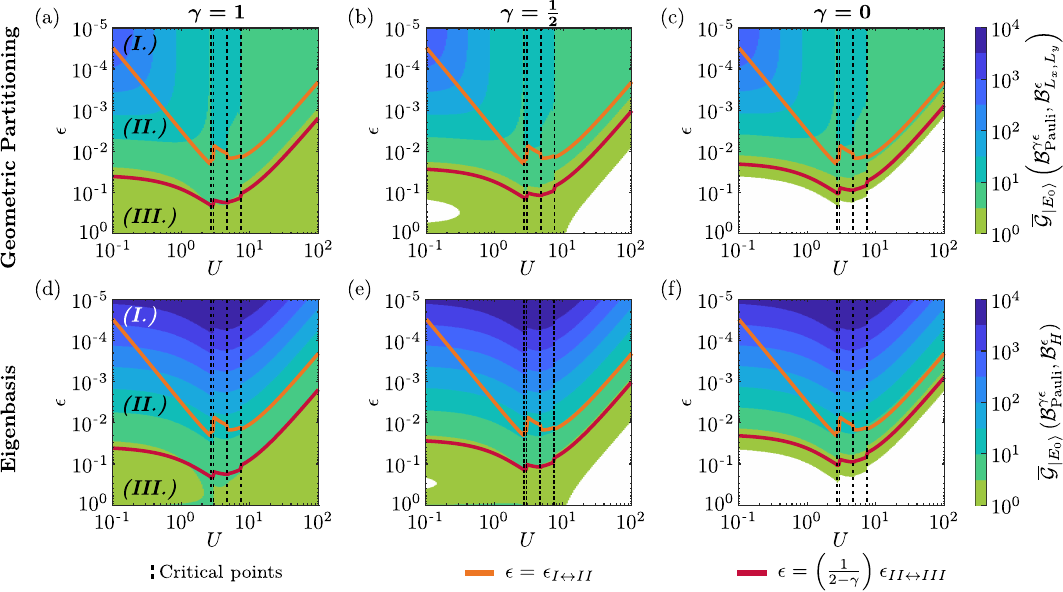}
    \caption{\textbf{Shot-noise reduction for the Fermi-Hubbard under noise.} 
   Here we show numerical results for the spinful Fermi-Hubbard model on a $3\times 4$ lattice under noise, which corresponds to 24 qubits on a biplanar graph (The corresponding noise-free results are shown in Fig.~\ref{fig: Examples1b} (c)). 
   This figure is similarly structured as Fig.~\ref{fig: Examples2}. Vertical dashed lines show, as in Fig.~\ref{fig: Examples1b} (c), the critical points, which are given by transitions of the average occupation number from $\braket{\hat N} = 10$ to $\braket{\hat N} = 6$ in integer steps. 
   As the geometric partition we choose 1D cuts with $L=1$ (which was seen to perform as well as $(L_x,L_y)= (2,2)$ in all other examples) and plot relative sampling complexities $\overline{\mathcal G}_{\ket{E_0}} \left( \mathcal B_{\rm{Pauli}}^{\gamma \epsilon}, \mathcal B_{L_x, L_y}^\epsilon \right )$ and $\overline{\mathcal G}_{\ket{E_0}} \left( \mathcal B_{\rm{Pauli}}^{\gamma \epsilon}, \mathcal B_{H}^\epsilon \right )$ for varying depolarization noise $\epsilon$ and system parameter $U$. 
   Similarly to the noisy TFXYM, we find that geometric partitioning yields lower shot noise for any $\epsilon$ for $\gamma=1$, 
   while $\gamma \neq 1$ only affects the high-noise regime (III.), where geometric partitioning already performs comparably to measuring in the eigenbasis. White areas correspond to regions where Pauli partitioning is better than geometric partitioning or eigenbasis measurements.
     }
\label{fig: Examples2b}
\end{figure*}
Here, we give numerical examples of energy estimations of eigenstates under global depolarization noise. Our main objective is to explore the tightness of the analytical bounds, mostly Theorem~\ref{thm:2}.  Compared to the noiseless examples in Sec.~\ref{subsubsec: TFXYM} - \ref{subsubsec: Spinful Hubbard}, we thus have two more variable parameters in our numerical study, the noise strength $\epsilon$ and relative noise strength $\gamma$, see Eq. \eqref{eq:gamma_def}. 
Even though, in real world scenarios, measuring in geometric partitioning will have lower noise than measuring in the eigenbasis due to shallower measurement circuits, we use here the same noise strength for both as a very conservative approximation to illustrate when geometric partitioning effectively works as well as the noisy eigenbasis.
We focus our analysis here on one bosonic model, the TFXYM, see Fig.~\ref{fig: Examples2},  and one fermionic model, the Fermi-Hubbard model, see Fig.~\ref{fig: Examples2b}.

In Figs. \ref{fig: Examples2} and \ref{fig: Examples2b}, all subfigures are structured in the same way. Along the horizontal axis, we vary a model parameter, the anisotropy $\eta$ for the TFXYM and the interaction strength $U$ for the Fermi-Hubbard model. Along the vertical axis, we vary the noise level in logarithmic scale. The color coding shows, in steps of half an order of magnitude, the relative sampling complexity $\overline{\mathcal G}_{\ket{E_0}}$ for the compared measurement strategies. 

For exploring the transition between regimes (I.) and (II.) we compare three noise thresholds. The first is our approximation to the transition noise strength, $\epsilon_{I \leftrightarrow II}$ as defined in Eq. \eqref{eq:chi_def}, the second is the noise level for which the sampling improvement is reduced by a factor two compared to the noiseless case and the third is the threshold above which geometric partitioning needs twice as many measurements as measuring in the eigenbasis.
All three match very closely throughout most of the parameter ranges in Fig.~\ref{fig: Examples2} and Fig.~\ref{fig: Examples2b},
which is one of our main numerical findings, confirming that our analytical approximations are indeed tight. 
Thus, we draw in Figs. \ref{fig: Examples2} and \ref{fig: Examples2b} only the line $\epsilon = \epsilon_{I \leftrightarrow II}$ in orange. The $\epsilon$-values, where $\overline{\mathcal G}_{\ket{E_0}} \left( \mathcal B_{L_x, L_y^{\epsilon}}, \mathcal B_{H}^\epsilon \right ) = 2$ and $\overline{\mathcal G}_{\ket{E_0}} \left( \mathcal B_{L_x, L_y^{\epsilon}}, \mathcal B_{H}^\epsilon \right ) = {\mathcal G}_{\ket{E_0}} \left( \mathcal B_{\textrm{Pauli}}, \mathcal B_{L_x, L_y} \right )/2$ have relative deviations from the $\epsilon = \epsilon_{I \leftrightarrow II}$ values of at most 17\% for the TFXYM, see Fig. \ref{fig: Examples2}, and at most 9.4\% for $U \lessapprox 4.7$ for the Fermi-Hubbard model, see Fig. \ref{fig: Examples2b}. For large $U$-values of  the Fermi-Hubbard model (see Fig.~\ref{fig: Examples2b}) the $\epsilon$-value for the threshold $\overline{\mathcal G}_{\ket{E_0}} \left( \mathcal B_{\rm{Pauli}}^{\gamma \epsilon}, \mathcal B_{L_x, L_y}^\epsilon \right ) = \mathcal G_{\ket{E_0}} \left( \mathcal B_{\rm{Pauli}}, \mathcal B_{L_x, L_y} \right )/2$ is shifted towards larger $\epsilon$ by at most a factor 2.4  for $\gamma = 1$ and by a factor 1.5 for $\gamma = 1/2$, which is consistent with Theorem~\ref{thm:2}.  

In regime (I.) we further expect the shot noise reductions to differ at most by a factor two from the improvements in the noiseless case. Consequently,
we expect $\overline{\mathcal G}_{\ket{E_0}} \left( \mathcal B_{\rm{Pauli}}^{\gamma \epsilon}, \mathcal B_{L_x, L_y}^\epsilon \right )$ to barely change with $\epsilon$ and to be independent of the relative noise strength of Pauli partitioning $\gamma$ in this regime. This is confirmed by Fig.~\ref{fig: Examples2}~(a)~-~(c) and Fig.~\ref{fig: Examples2b}~(a)~-~(c).
For regimes (II.) and (III.), we expect that geometric partition performs up to a factor of two as good as measuring in the eigenbasis of $H$. Indeed, when we compare plots (a) to (d), (b) to (e), and (c) to (f) of Fig.~\ref{fig: Examples2} and Fig.~\ref{fig: Examples2b}, these regions appear very comparable. In agreement with Theorem~\ref {thm:2}, we only find for the high noise regime (III.) and lower relative noise ($\gamma = 1/2$ and $\gamma = 0$), that Pauli partitioning becomes better than geometric partitioning (white areas). Yet here, measuring via Pauli splitting is typically also better than measuring in the eigenbasis. 

In real-world settings and for small enough patch sizes, we expect the majority of the gates to be used for state preparation, and hence $\gamma \approx 1$. In such cases, we expect geometric partitioning to perform better than Pauli partitioning for any noise level $\epsilon$ due to Corollary~\ref{lemma:noise-indep sample advantage}. This can nicely be observed in Fig.~\ref{fig: Examples2} (a) and Fig.~\ref{fig: Examples2b} (a).
These results are in agreement with the observation that Eqs.~\eqref{eq:G_epsilon=1} and \eqref{eq:G_epsilon=1b} are fulfilled, as geometric splitting distributes the energy equally among the parts and also spreads the Pauli weights equally, so that typically also $\sum \norm{H_b}_F$ is smaller for geometric than Pauli partitioning.

The case with $\gamma = 1/2$ corresponds to the same effective number of 2-qubit gates for both state preparation and geometric partitioning as shown in Eq.~\eqref{eq:gamma_gates}. 
We also show $\gamma = 0$, which compares noise-free perfect state preparation and Pauli measurement against noisy geometric partitioning not as a realistic application scenario, but to emphasize the validity of our analysis. 
Another observation in both noisy numerical examples is, that for $U/t,  \eta =\mathcal{O}(1)$, the transition  (II.) $\leftrightarrow$ (III.) typically lies at (large) noise strength of $\epsilon \approx$ 2 - 10 \%. At the same time, the strongly correlated and thus particularly interesting regimes are typically also located in these parameter regimes.  

When comparing the noisy sampling improvements here with their noiseless counterparts in Fig~\ref{fig: Examples} (a) and Fig~\ref{fig: Examples}~(c), one can see, particularly for the TFXYM, a close relation between noisy and noiseless results as the large improvements carry over to the imperfect setting. Moreover, the size of regime (II.) is approximately equal to the value of $\mathcal G_{\ket{E_0}} \left( \mathcal B_{\rm{Pauli}}, \mathcal B_{L_x, L_y} \right )$, as stated in Corollary~\ref{corol:eps-delta-threashold}. This means that for higher improvement of geometric partitioning in the noiseless case, is, the smaller the noise level $\epsilon$ needs to be for geometric partitioning performing recognizable worse than measuring in the eigenbasis. Note that this is independent of the relative noise strength of Pauli partitioning $\gamma$. 

\subsection{Simulated sampling}
\label{subsubsec:Sampling}
\begin{figure}[t!]
    \centering
    \includegraphics[width=\linewidth]{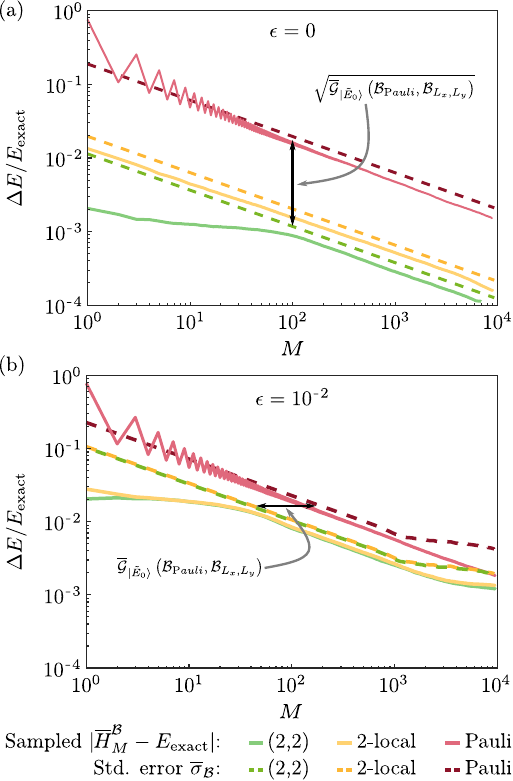}
     \vspace{-0.5cm}
    \caption{\textbf{Numerical example: Sampled energy estimations.} 
    The TFIM with $J/h=1$ is the same system as the TFXYM in Fig.~\ref{fig: Examples2} with $\eta=1$, where we simulate sampling from a perfect $\ket \psi = \ket{E_0}$ (a), and a noisy one with global depolarization noise $\epsilon = 1 \%$ (b). On real hardware, the noise could have originated from noisy hardware gates, imperfect implementation of the state preparation or of the measurement unitaries $U_b$.
    In light red, yellow, and green, we show the relative error of the sampled energy $\overline H_M^\mathcal{B}$ for Pauli, 2-local, and $2\times 2$ geometric partitioning. To make a fair comparison, we report the mean energy error over a random ordering of the simulated samples. In darker shades of the same colors, we give the corresponding relative standard errors $\overline \sigma_{\mathcal B}/E_{\rm exact}$ (compare also Eq.~\eqref{eq: average_sampling_error}), which match well with the sampled energy estimates.
    For few samples, Pauli partitioning suffers from its integer spectrum, while the sampled energies with geometric partitioning are more fine-grained and closer to the exact energy. Consistently with Fig~\ref{fig: Examples2} (c), we find that even the noisy geometric partitionings in (b) yield lower shot-noise than the noise free Pauli bases measurements in (a). 
    }
    \vspace{-0.5cm}
    \label{fig: sampled_example}
\end{figure}

In this section, we give further evidence that the measures for the accuracy of sampling strategies that we use, see Eqs.~\eqref{eq: std_error_min} and~\eqref{eq: average_sampling_error}, are good approximations to the actual error occurring in the sampling process. These results thus support that the starting point for our comparison of sampling efficiency is well-founded. 
To provide this evidence, we show simulations of actual sampling. This means we do not compute the considered variances from the full wave-function but instead simulate actual measurements by applying the measurement unitaries $U_{H_b}$ and sampling bit-strings from which we compute the corresponding energy values. For the analysis of noisy situations, we add random contributions to the state according to Eq. \eqref{eq: tilde_psi} before sampling the bit-strings. For these simulations, we compute the relative energy error from the sampling process
\begin{equation}
    \left|\frac{\overline H_M^\mathcal{B} - E_{\text{exact}}}{E_{\text{exact}}}\right|
\end{equation}
and compare it to $\sigma_{\mathcal{B}}/|E_{\text{exact}}|$ for the noiseless and $\overline{\sigma}_{\mathcal{B}}/|E_{\text{exact}}|$ for the noisy case.

We do this comparison for the TFXYM and TFIM ground state $\ket{E_0}$ at $J=\eta=h=1$ on a $4\times 6$ rectangular lattice as examples.
Figs.~\ref{fig: sampled_example} (a) and (b), show the relative energy error under  simulated samples in light red for Pauli sampling, light yellow for 2-local sampling, and light green for a $(2,2)$ geometric partition as well as the corresponding values of   $\sigma_{\mathcal{B}}/|E_{\text{exact}}|$ for the noiseless case in panel (a) and $\overline{\sigma}_{\mathcal{B}}/|E_{\text{exact}}|$ for the noisy case in panel~(b).

A first observation is, that the light and darker lines of the same color are indeed very close to each other. As expected, $\sigma_{\mathcal{B}}/|E_{\text{exact}}|$ and $\overline{\sigma}_{\mathcal{B}}/|E_{\text{exact}}|$ are slightly above the corresponding observed mean sampling error (light colors), see Eqs.~\eqref{eq: std_error_min} and \eqref{eq: std_error_ineq}.
For few samples, $M \lessapprox 100$, there are discrete jumps, also above the standard error for Pauli partitioning. This reflects the coarse discretization of the Pauli spectrum, where obtainable one-shot expectation values are integer. On the other hand, for few samples, geometric partitioning performs even better than its standard error. This is consistent with a further observation, namely that only rather few eigenstates of each patch of a geometric partitioning have non-vanishing occupation, leading to a better discrete representation of $E_{\text{exact}}$ with very few samples. We conjecture this beneficial few-sample effect also to occur at other examples, making geometric partitioning even more beneficial than quantified by $ \overline{\mathcal G}_{\ket{\tilde E_0}} \left( \mathcal B_{\rm Pauli}, \mathcal B_{L_x, L_y} \right)$.

Besides showing the appropriateness of the quantities $\sigma_{\mathcal{B}}$ and $\overline{\sigma}_{\mathcal{B}}$ for quantifying sampling errors, Fig. \ref{fig: sampled_example} again confirms the reduction of sampling complexity via geometric partitioning as discussed in sections \ref{subsubsec:TFIM} and \ref{subsec: example imperfect eigenstates}. In particular, Fig.~\ref{fig: sampled_example} (b), where we take errors in the state preparation and measurement circuit implementation into account, confirms that there is no fundamental change in the accuracy of our analysis, while the improvement $\mathcal G$ (Def.~\ref{def:Sampling improvement}) is reduced to $\overline{\mathcal G}$ (Def.~\ref{def:improvement imperfect states}), as predicted in Theorem~\ref{thm:2} and shown in Fig.~\ref{fig: Examples2} and Fig.~\ref{fig: Examples2b}. In good agreement with Theorem~\ref{thm:2}, even for a noise level of 1 \% geometric partitioning, green and yellow in Fig.~\ref{fig: sampled_example} (b), still yield a lower shot-noise than perfect state preparation and noiseless sampling in the Pauli bases, red in Fig.~\ref{fig: sampled_example} (a). 

\newpage
\section{Discussion}
\label{sec:discussion}
The main idea we pursued in this work was to make use of the geometric locality of quantum systems to improve measurement strategies compared to sampling in Pauli bases. Conceptually, we think of these geometrically local subsystems to be diagonalizable on the quantum hardware, meaning that the transformations mapping the eigenstates of the subsystems to the computational basis can be found. 

In our first main result (Sec.~\ref{subsec:perfect eigenstates}),
we have proven that indeed geometric partitioning always yields an advantage over Pauli partitioning for energy estimates of a system in an energy eigenstate, where we identified three cases. These are geometric partition coming from a one-dimensional cuts, two-dimensional patches or even 2-local partitions. Typically, smaller guaranteed improvements coincide with easier-to-implement measurement circuits, as these act on less qubits/lattice sites.  There is thus something for (almost) any available measurement gate budget, starting from as little as one additional 2-qubit gate per patch when using the 2-local geometric partitioning strategy. Twice as large partitioning patches with e.g. $(L_x,L_y)=(2,2)$ then require to implement a 4-qubit measurement unitary, which appears still feasible also in near-term applications. 
{Throughout our numerical examples, including different Ising and Hubbard models, we find that typically already 4-qubit $(L_x,L_y)=(2,2)$ partitioning suffices to obtain orders of magnitude relative sampling complexity improvements. In such case, the gate complexity to implement measurement circuit $U_{H_b}$ scales with the number of blocks $U_b$, and hence only linear in the number of qubits $n$} 

With our second main result (Sec.~\ref{subsec:imperfect eigenstates}), we cover the more general case of imperfect eigenstates, perturbed by global depolarization noise, which can be viewed as a proxy for both imperfect state preparation and quantum hardware noise. 
A future extension of our work may include tailoring readout circuits not only based on state symmetries as we did here, but also based on a quantum hardware-specific noise model. 
We identified three noise regimes. In the low noise one $(I.)$, the noiseless sampling improvement is maintained up to at worst a factor 2. In the intermediate $(II.)$ and high-noise $(III.)$ regime, measuring in the geometric basis effectively becomes as good as measuring in the eigenbasis, again up to a factor 2. 
Importantly, we provide upper bounds on the number of additional gates that can be used for the measurement circuits and the amount of additional noise they may introduce while leading to a sampling advantage for geometric partitioning, see corollaries \ref{lemma:noise-indep sample advantage} and \ref{corol: gate_budget}. 

Typically, more correlated states require larger measurement efforts~\cite{anshu2016concentration}. Hence, our approach, which measures geometrically extended operators, yields very significant improvements for highly correlated states, which are typically of interest. 
We observed this behavior featuring improvements $\gtrapprox 10^7$ for TFXYM around its oscillatory to ferromagnetic phase transition (Fig.~\ref{fig: Examples} (a)). 
We focused on proving sampling improvements for geometric local models, both in bosonic and fermionic bases. We have shown in total six examples, where the geometric partitioning strategy led to a sampling improvement compared to the (Pauli) baseline, as our analysis suggested. Further, we demonstrated with the Transverse Field Biaxial Next-Nearest Neighbor Ising model, how this idea extends beyond 2-local models to a quasi-3-local one. 

To go forward an interesting future direction might be potentially lifting the requirement of geometric locality, and extending the idea to further types of symmetries beyond translational invariance. 
Similarly, we found examples of improvements beyond our lower bound, e.g. within the ferromagnetic phase $J/h \gg 1$ of the TFIM. This motivates to investigate further more model-specific sampling improvement estimations or to exploit perturbation theory results to construct better-performing sampling basis as we provided with our Lemma~\ref{lemma:TFIM improvements}. 

We took the stand here that we are given a state and aim to estimate the energy expectation value. In future work, one could flip this view around and rather use geometrically local bases to compress ansätze for approximated state preparation. For instance, for ground state preparation, one could expect the low-energy subspace of the geometrically local bases to be mostly relevant for the ground state of the entire system. Provided the inverses of our measurement unitaries not only lead to vanishing local energy variances but also low local energy expectation values, this strategy would support approximations with reduced effective Hilbert spaces.

\begin{acknowledgments}
This work was German Research Foundation (DFG) – Project-ID 429529648 – TRR 306 QuCoLiMa, the German Federal Ministry of Education and Research (BMBF) contract number 13N16067 “EQUAHUMO”, and Munich Quantum Valley, which is supported by the Bavarian state government with funds from the Hightech Agenda Bayern Plus. This work  used resources of the Erlangen National High Performance Computing Center.
T.E. acknowledges support from the International Max-Planck Research School for Physics of Light. 
M.K.\ is co-funded by 
the ERDF of the European Union and by Fonds of the Hamburg Ministry of Science, Research, Equalities and Districts (BWFGB)
and funded by the Fujitsu Germany GmbH as part of the endowed professorship ``Quantum Inspired and Quantum Optimization.''
\end{acknowledgments}
\section*{Data Availability}
The data for this paper are available from the corresponding author upon reasonable request.
\newpage


\bibliography{references}

@article{haah2017sample,
  title={Sample-optimal tomography of quantum states},
  author={Haah, Jeongwan and Harrow, Aram W and Ji, Zhengfeng and Wu, Xiaodi and Yu, Nengkun},
  journal={IEEE Transactions on Information Theory},
  volume={63},
  number={9},
  pages={5628--5641},
  year={2017},
  publisher={IEEE},
  doi={10.1145/2897518.2897585},
  url={https://doi.org/10.1145/2897518.2897585}
}

@article{arunachalam2018optimal,
  title={Optimal quantum sample complexity of learning algorithms},
  author={Arunachalam, Srinivasan and De Wolf, Ronald},
  journal={Journal of Machine Learning Research},
  volume={19},
  number={71},
  pages={1--36},
  year={2018},
  url={http://jmlr.org/papers/v19/18-195.html}
}

@article{takagi2022universal,
  title={Universal sampling lower bounds for quantum error mitigation},
  author={Takagi, Ryuji and Tajima, Hiroyasu and Gu, Mile},
  journal={Physical Review Letters},
  volume={131},
  number={21},
  pages={210602},
  year={2023},
  publisher={APS},
url={https://journals.aps.org/prl/abstract/10.1103/PhysRevLett.131.210602},
doi={10.1103/PhysRevLett.131.210602}
}

@article{gokhale2019minimizing,
  title={Minimizing state preparations in variational quantum eigensolver by partitioning into commuting families},
  author={Gokhale, Pranav and Angiuli, Olivia and Ding, Yongshan and Gui, Kaiwen and Tomesh, Teague and Suchara, Martin and Martonosi, Margaret and Chong, Frederic T},
  journal={arXiv preprint arXiv:1907.13623},
  year={2019},
  url={https://doi.org/10.48550/arXiv.1907.13623},
  doi={10.48550/arXiv.1907.13623}
}

@article{jena2019pauli,
  title={Pauli partitioning with respect to gate sets},
  author={Jena, Andrew and Genin, Scott and Mosca, Michele},
  journal={arXiv preprint arXiv:1907.07859},
  year={2019},
  url={https://doi.org/10.48550/arXiv.1907.07859},
  doi={10.48550/arXiv.1907.07859}
}

@article{verteletskyi2020measurement,
  title={Measurement optimization in the variational quantum eigensolver using a minimum clique cover},
  author={Verteletskyi, Vladyslav and Yen, Tzu-Ching and Izmaylov, Artur F},
  journal={The Journal of chemical physics},
  volume={152},
  number={12},
  pages={124114},
  year={2020},
  publisher={AIP Publishing LLC},
  doi = {10.1063/1.5141458},
  url = {https://doi.org/10.1063/1.5141458}
}

@article{huggins2021efficient,
  title={Efficient and noise resilient measurements for quantum chemistry on near-term quantum computers},
  author={Huggins, William J and McClean, Jarrod R and Rubin, Nicholas C and Jiang, Zhang and Wiebe, Nathan and Whaley, K Birgitta and Babbush, Ryan},
  journal={npj Quantum Information},
  volume={7},
  number={1},
  pages={1--9},
  year={2021},
  publisher={Nature Publishing Group},
  url={https://www.nature.com/articles/s41534-020-00341-7}
}

@article{gresch2025guaranteed,
  title={Guaranteed efficient energy estimation of quantum many-body Hamiltonians using ShadowGrouping},
  author={Gresch, Alexander and Kliesch, Martin},
  journal={Nature communications},
  volume={16},
  number={1},
  pages={689},
  year={2025},
  publisher={Nature Publishing Group UK London},
  doi={10.1038/s41467-024-54859-x},
  url={https://doi.org/10.1038/s41467-024-54859-x}
}

@article{nutzel2025solving,
  title={Solving an industrially relevant quantum chemistry problem on quantum hardware},
  author={N{\"u}tzel, Ludwig and Gresch, Alexander and Hehn, Lukas and Marti, Lucas and Freund, Robert and Steiner, Alex and Marciniak, Christian D and Eckstein, Timo and Stockinger, Nina and Wolf, Stefan and others},
  journal={Quantum Science and Technology},
  volume={10},
  number={1},
  pages={015066},
  year={2025},
  publisher={IOP Publishing},
  doi={10.1088/2058-9565/ad9ed3},
  url={https://doi.org/10.1088/2058-9565/ad9ed3}
}

@article{huang2020predicting,
  title={Predicting many properties of a quantum system from very few measurements},
  author={Huang, Hsin-Yuan and Kueng, Richard and Preskill, John},
  journal={Nature Physics},
  volume={16},
  number={10},
  pages={1050--1057},
  year={2020},
  publisher={Nature Publishing Group},
  doi={10.1038/s41567-020-0932-7},
  url={https://www.nature.com/articles/s41567-020-0932-7}
}

@article{hadfield2022measurements,
  title={Measurements of quantum Hamiltonians with locally-biased classical shadows},
  author={Hadfield, Charles and Bravyi, Sergey and Raymond, Rudy and Mezzacapo, Antonio},
  journal={Communications in Mathematical Physics},
  volume={391},
  number={3},
  pages={951--967},
  year={2022},
  publisher={Springer},
  doi = {10.1007/s00220-022-04343-8},
  url = {https://doi.org/10.1007/s00220-022-04343-8}
}

@article{hadfield2021adaptive,
  title={Adaptive Pauli shadows for energy estimation},
  author={Hadfield, Charles},
  journal={arXiv preprint arXiv:2105.12207},
  year={2021},
  doi = {10.48550/arXiv.2105.12207},
  url = {https://doi.org/10.48550/arXiv.2105.12207}
}

@article{elben2022randomized,
  title={The randomized measurement toolbox},
  author={Elben, Andreas and Flammia, Steven T and Huang, Hsin-Yuan and Kueng, Richard and Preskill, John and Vermersch, Beno{\^\i}t and Zoller, Peter},
  journal={Nature Review Physics},
  year={2022},
  doi={10.1038/s42254-022-00535-2},
  url={https://www.nature.com/articles/s42254-022-00535-2}
}

@article{brieger2025stability,
  title={Stability of classical shadows under gate-dependent noise},
  author={Brieger, Raphael and Heinrich, Markus and Roth, Ingo and Kliesch, Martin},
  journal={Physical review letters},
  volume={134},
  number={9},
  pages={090801},
  year={2025},
  publisher={APS},
  doi={0.1103/PhysRevLett.134.090801},
  url={https://doi.org/10.1103/PhysRevLett.134.090801}
}

@article{crawford2021efficient,
  doi = {10.22331/q-2021-01-20-385},
  url = {https://doi.org/10.22331/q-2021-01-20-385},
  title = {Efficient quantum measurement of {P}auli operators in the presence of finite sampling error},
  author = {Crawford, Ophelia and Straaten, Barnaby van and Wang, Daochen and Parks, Thomas and Campbell, Earl and Brierley, Stephen},
  journal = {{Quantum}},
  issn = {2521-327X},
  publisher = {{Verein zur F{\"{o}}rderung des Open Access Publizierens in den Quantenwissenschaften}},
  volume = {5},
  pages = {385},
  month = jan,
  year = {2021}
}

@article{chatterjee2021semiconductor,
  title={Semiconductor qubits in practice},
  author={Chatterjee, Anasua and Stevenson, Paul and De Franceschi, Silvano and Morello, Andrea and de Leon, Nathalie P and Kuemmeth, Ferdinand},
  journal={Nature Reviews Physics},
  volume={3},
  number={3},
  pages={157--177},
  year={2021},
  publisher={Nature Publishing Group UK London},
  doi = {10.1038/s42254-021-00283-9},
  url = {https://doi.org/10.1038/s42254-021-00283-9}
}

@article{burkard2023semiconductor,
  title={Semiconductor spin qubits},
  author={Burkard, Guido and Ladd, Thaddeus D and Pan, Andrew and Nichol, John M and Petta, Jason R},
  journal={Reviews of Modern Physics},
  volume={95},
  number={2},
  pages={025003},
  year={2023},
  publisher={APS},
  doi = {10.1103/RevModPhys.95.025003},
  url = {https://doi.org/10.1103/RevModPhys.95.025003}
}

@article{arute2019quantum,
author={Arute, Frank
and Arya, Kunal
and Babbush, Ryan
and Bacon, Dave
and Bardin, Joseph C.
and Barends, Rami
and Biswas, Rupak
and Boixo, Sergio
and Brandao, Fernando G. S. L.
and Buell, David A.
and Burkett, Brian
and Chen, Yu
and Chen, Zijun
and Chiaro, Ben
and Collins, Roberto
and Courtney, William
and Dunsworth, Andrew
and Farhi, Edward
and Foxen, Brooks
and Fowler, Austin
and Gidney, Craig
and Giustina, Marissa
and Graff, Rob
and Guerin, Keith
and Habegger, Steve
and Harrigan, Matthew P.
and Hartmann, Michael J.
and Ho, Alan
and Hoffmann, Markus
and Huang, Trent
and Humble, Travis S.
and Isakov, Sergei V.
and Jeffrey, Evan
and Jiang, Zhang
and Kafri, Dvir
and Kechedzhi, Kostyantyn
and Kelly, Julian
and Klimov, Paul V.
and Knysh, Sergey
and Korotkov, Alexander
and Kostritsa, Fedor
and Landhuis, David
and Lindmark, Mike
and Lucero, Erik
and Lyakh, Dmitry
and Mandr{\`a}, Salvatore
and McClean, Jarrod R.
and McEwen, Matthew
and Megrant, Anthony
and Mi, Xiao
and Michielsen, Kristel
and Mohseni, Masoud
and Mutus, Josh
and Naaman, Ofer
and Neeley, Matthew
and Neill, Charles
and Niu, Murphy Yuezhen
and Ostby, Eric
and Petukhov, Andre
and Platt, John C.
and Quintana, Chris
and Rieffel, Eleanor G.
and Roushan, Pedram
and Rubin, Nicholas C.
and Sank, Daniel
and Satzinger, Kevin J.
and Smelyanskiy, Vadim
and Sung, Kevin J.
and Trevithick, Matthew D.
and Vainsencher, Amit
and Villalonga, Benjamin
and White, Theodore
and Yao, Z. Jamie
and Yeh, Ping
and Zalcman, Adam
and Neven, Hartmut
and Martinis, John M.},
title={Quantum supremacy using a programmable superconducting processor},
journal={Nature},
year={2019},
month={Oct},
day={01},
volume={574},
number={7779},
pages={505-510},
issn={1476-4687},
doi={10.1038/s41586-019-1666-5},
url={https://doi.org/10.1038/s41586-019-1666-5}
}

@article{blais2021circuit,
  title={Circuit quantum electrodynamics},
  author={Blais, Alexandre and Grimsmo, Arne L and Girvin, Steven M and Wallraff, Andreas},
  journal={Reviews of Modern Physics},
  volume={93},
  number={2},
  pages={025005},
  year={2021},
  publisher={APS},
  doi = {10.1103/RevModPhys.93.025005},
  url = {https://doi.org/10.1103/RevModPhys.93.025005}
}

@article{verstraete2009quantum,
  title={Quantum circuits for strongly correlated quantum systems},
  author={Verstraete, Frank and Cirac, J Ignacio and Latorre, Jos{\'e} I},
  journal={Physical Review A},
  volume={79},
  number={3},
  pages={032316},
  year={2009},
  publisher={APS},
  url = {https://doi.org/10.1103/PhysRevA.79.032316},
  doi = {10.1103/PhysRevA.79.032316}
}

@article{kitaev1997quantum,
doi = {10.1070/RM1997v052n06ABEH002155},
url = {https://dx.doi.org/10.1070/RM1997v052n06ABEH002155},
year = {1997},
month = {dec},
publisher = {},
volume = {52},
number = {6},
pages = {1191},
author = {A Yu Kitaev},
title = {Quantum computations: algorithms and error correction},
journal = {Russian Mathematical Surveys}
}

@article{dawson2005solovay,
  title={The Solovay-Kitaev algorithm},
  author={Dawson, Christopher M and Nielsen, Michael A},
  journal={Quantum Information \& Computation},
  volume={6},
  number={1},
  pages={81--95},
  year={2006},
  publisher={Rinton Press, Incorporated Paramus, NJ},
  doi={10.5555/2011679.2011685},
  url={https://doi.org/10.5555/2011679.2011685}
}

@article{kuperberg2023breaking,
  title={Breaking the cubic barrier in the Solovay-Kitaev algorithm},
  author={Kuperberg, Greg},
  journal={arXiv preprint arXiv:2306.13158},
  year={2023},
  doi={10.48550/arXiv.2306.13158},
  url={https://doi.org/10.48550/arXiv.2306.13158}
}

@article{robertson1929uncertainty,
  title={The uncertainty principle},
  author={Robertson, Howard Percy},
  journal={Physical Review},
  volume={34},
  number={1},
  pages={163},
  year={1929},
  publisher={APS},
  doi = {10.1103/PhysRev.34.163},
  url = {https://doi.org/10.1103/PhysRev.34.163}
}

@article{dur2005standard,
  title={Standard forms of noisy quantum operations via depolarization},
  author={D{\"u}r, Wolfgang and Hein, Marc and Cirac, J Ignacio and Briegel, H-J},
  journal={Physical Review A—Atomic, Molecular, and Optical Physics},
  volume={72},
  number={5},
  pages={052326},
  year={2005},
  publisher={APS},
  doi={10.1103/PhysRevA.72.052326},
  url={https://doi.org/10.1103/PhysRevA.72.052326}
}

@article{gonzalez2022error,
  title={Error propagation in nisq devices for solving classical optimization problems},
  author={Gonz{\'a}lez-Garc{\'\i}a, Guillermo and Trivedi, Rahul and Cirac, J Ignacio},
  journal={PRX Quantum},
  volume={3},
  number={4},
  pages={040326},
  year={2022},
  publisher={APS},
  doi={10.1103/PRXQuantum.3.040326},
  url={https://doi.org/10.1103/PRXQuantum.3.040326}
}

@misc{quantinuum-hardware-specifications,
author = "{Quantinuum Ltd.}",
title = {{Quantinuum Hardware Specifications}},
url = {https://www.github.com/CQCL/quantinuum-hardware-specifications}
}

@article{hastings2006spectral,
  title={Spectral gap and exponential decay of correlations},
  author={Hastings, Matthew B and Koma, Tohru},
  journal={Communications in mathematical physics},
  volume={265},
  pages={781--804},
  year={2006},
  publisher={Springer},
  url={https://doi.org/10.1007/s00220-006-0030-4},
  doi={10.1007/s00220-006-0030-4}
}

@article{henkel1984statistical,
  title={Statistical mechanics of the 2D quantum XY model in a transverse field},
  author={Henkel, Malte},
  journal={Journal of Physics A: Mathematical and General},
  volume={17},
  number={14},
  pages={L795},
  year={1984},
  publisher={IOP Publishing},
  doi = {10.1088/0305-4470/17/14/013},
  url = {https://link.aps.org/doi/10.1088/0305-4470/17/14/013},
}

@article{nishiyama2019multicritical,
  title={Multicritical behavior of the fidelity susceptibility for the 2D quantum transverse-field XY model},
  author={Nishiyama, Yoshihiro},
  journal={The European Physical Journal B},
  volume={92},
  pages={1--7},
  year={2019},
  publisher={Springer},
  doi = {10.1140/epjb/e2019-100269-8},
  url = {https://doi.org/10.1140/epjb/e2019-100269-8}
}

@book{ising1924beitrag,
  title={Beitrag zur Theorie des Ferro-und Paramagnestismus},
  author={Ising, Ernst},
  year={1924},
  publisher={Hamburgische Universität}
}

@article{onsager1944crystal,
  title={Crystal statistics. I. A two-dimensional model with an order-disorder transition},
  author={Onsager, Lars},
  journal={Physical Review},
  volume={65},
  number={3-4},
  pages={117},
  year={1944},
  publisher={APS},
  doi={10.1103/PhysRev.65.117},
  url={https://doi.org/10.1103/PhysRev.65.117}
}

@article{schultz1964two,
  title={Two-dimensional Ising model as a soluble problem of many fermions},
  author={Schultz, Theodore D and Mattis, Daniel C and Lieb, Elliott H},
  journal={Reviews of Modern Physics},
  volume={36},
  number={3},
  pages={856},
  year={1964},
  publisher={APS},
  doi={10.1103/RevModPhys.36.856},
  url={https://doi.org/10.1103/RevModPhys.36.856}
}

@article{blote2002cluster,
  title = {Cluster Monte Carlo simulation of the transverse {I}sing model},
  author = {Bl\"ote, Henk W. J. and Deng, Youjin},
  journal = {Phys. Rev. E},
  volume = {66},
  issue = {6},
  pages = {066110},
  numpages = {8},
  year = {2002},
  month = {Dec},
  publisher = {American Physical Society},
  doi = {10.1103/PhysRevE.66.066110},
  url={https://doi.org/10.1103/PhysRevE.66.066110}
}

@article{shende2004minimal,
  title = {Minimal universal two-qubit controlled-NOT-based circuits},
  author = {Shende, Vivek V. and Markov, Igor L. and Bullock, Stephen S.},
  journal = {Physical Review A},
  volume = {69},
  issue = {6},
  pages = {062321},
  numpages = {8},
  year = {2004},
  publisher = {American Physical Society},
  doi = {10.1103/PhysRevA.69.062321},
  url = {https://link.aps.org/doi/10.1103/PhysRevA.69.062321}
}

@inproceedings{shende2005synthesis,
  title={Synthesis of quantum logic circuits},
  author={Shende, Vivek V and Bullock, Stephen S and Markov, Igor L},
  booktitle={Proceedings of the 2005 Asia and South Pacific Design Automation Conference},
  pages={272--275},
  year={2005},
  doi={10.1109/TCAD.2005.855930},
  url={https://doi.org/10.48550/10.1109/TCAD.2005.855930}
}

@article{khatri2019quantum,
  title={Quantum-assisted quantum compiling},
  author={Khatri, Sumeet and LaRose, Ryan and Poremba, Alexander and Cincio, Lukasz and Sornborger, Andrew T and Coles, Patrick J},
  journal={Quantum},
  volume={3},
  pages={140},
  year={2019},
  publisher={Verein zur F{\"o}rderung des Open Access Publizierens in den Quantenwissenschaften},
  url={https://quantum-journal.org/papers/q-2019-05-13-140/},
  doi={10.22331/q-2019-05-13-140}
}

@article{madden2022best,
  title={Best approximate quantum compiling problems},
  author={Madden, Liam and Simonetto, Andrea},
  journal={ACM Transactions on Quantum Computing},
  volume={3},
  number={2},
  pages={1--29},
  year={2022},
  publisher={ACM New York, NY},
  doi={10.1145/3505181},
  url={https://doi.org/10.1145/3505181}
}

@article{hornreich1979lifshitz,
  title={Lifshitz points in Ising systems},
  author={Hornreich, RM and Liebmann, R and Schuster, Heinz Georg and Selke, Walter},
  journal={Zeitschrift f{\"u}r Physik B Condensed Matter},
  volume={35},
  number={1},
  pages={91--97},
  year={1979},
  publisher={Springer},
  doi={10.1007/BF01322086},
  url={https://doi.org/10.1007/BF01322086}
}

@article{eckstein2024large,
  title={Large-scale simulations of Floquet physics on near-term quantum computers},
  author={Eckstein, Timo and Mansuroglu, Refik and Czarnik, Piotr and Zhu, Jian-Xin and Hartmann, Michael J and Cincio, Lukasz and Sornborger, Andrew T and Holmes, Zo{\"e}},
  journal={npj Quantum Information},
  volume={10},
  number={1},
  pages={84},
  year={2024},
  publisher={Nature Publishing Group UK London},
  doi = {10.1038/s41534-024-00866-1},
  url = {https://doi.org/10.1038/s41534-024-00866-1}
}

@article{murg2007variational,
  title={Variational study of hard-core bosons in a two-dimensional optical lattice using projected entangled pair states},
  author={Murg, Valentin and Verstraete, Frank and Cirac, J Ignacio},
  journal={Physical Review A—Atomic, Molecular, and Optical Physics},
  volume={75},
  number={3},
  pages={033605},
  year={2007},
  publisher={APS},
  doi = {10.1103/PhysRevA.75.033605},
  url = {https://doi.org/10.1103/PhysRevA.75.033605}
}

@article{jordan2009numerical,
  title={Numerical study of the hard-core Bose-Hubbard model on an infinite square lattice},
  author={Jordan, Jacob and Or{\'u}s, Rom{\'a}n and Vidal, Guifr{\'e}},
  journal={Physical Review B—Condensed Matter and Materials Physics},
  volume={79},
  number={17},
  pages={174515},
  year={2009},
  publisher={APS},
  doi = {10.1103/PhysRevB.79.174515},
  url = {https://doi.org/10.1103/PhysRevB.79.174515}
}

@article{gersch1963quantum,
  title={Quantum cell model for bosons},
  author={Gersch, Harold A and Knollman, Gilbert C},
  journal={Physical Review},
  volume={129},
  number={2},
  pages={959},
  year={1963},
  publisher={APS},
  doi={10.1103/PhysRev.129.959},
  url={https://doi.org/10.1103/PhysRev.129.959}
}

@article{yanay2020two,
  title={Two-dimensional hard-core Bose--Hubbard model with superconducting qubits},
  author={Yanay, Yariv and Braum{\"u}ller, Jochen and Gustavsson, Simon and Oliver, William D and Tahan, Charles},
  journal={npj Quantum Information},
  volume={6},
  number={1},
  pages={58},
  year={2020},
  publisher={Nature Publishing Group UK London},
  doi={10.1038/s41534-020-0269-1},
  url={https://doi.org/10.1038/s41534-020-0269-1}
}

@article{zhang2003stripes,
  title={Stripes and holes in a two-dimensional model of spinless fermions or hardcore bosons},
  author={Zhang, NG and Henley, CL},
  journal={Physical Review B},
  volume={68},
  number={1},
  pages={014506},
  year={2003},
  publisher={APS},
  url={https://doi.org/10.1103/PhysRevB.68.014506},
  doi={10.1103/PhysRevB.68.014506}
}

@article{hubbard1963electron,
  title={Electron correlations in narrow energy bands},
  author={Hubbard, John},
  journal={Proceedings of the Royal Society of London. Series A. Mathematical and Physical Sciences},
  volume={276},
  number={1365},
  pages={238--257},
  year={1963},
  publisher={The Royal Society London},
  url={https://doi.org/10.1098/rspa.1963.0204}
}

@article{lieb1989two,
  title={Two theorems on the Hubbard model},
  author={Lieb, Elliott H},
  journal={Physical review letters},
  volume={62},
  number={10},
  pages={1201},
  year={1989},
  publisher={APS},
  url={https://doi.org/10.1103/PhysRevLett.62.1201},
  doi={10.1103/PhysRevLett.62.1201}
}

@article{anshu2016concentration,
  title={Concentration bounds for quantum states with finite correlation length on quantum spin lattice systems},
  author={Anshu, Anurag},
  journal={New Journal of Physics},
  volume={18},
  number={8},
  pages={083011},
  year={2016},
  publisher={IOP Publishing},
  doi = {10.1088/1367-2630/18/8/083011},
  url = {https://doi.org/10.1088/1367-2630/18/8/083011},
}

@article{romero2005structure,
  title={Structure of the sets of mutually unbiased bases for N qubits},
  author={Romero, JL and Bj{\"o}rk, Gunnar and Klimov, AB and S{\'a}nchez-Soto, LL},
  journal={Physical Review A—Atomic, Molecular, and Optical Physics},
  volume={72},
  number={6},
  pages={062310},
  year={2005},
  publisher={APS},
  doi={10.1103/PhysRevA.72.062310},
  url={https://doi.org/10.1103/PhysRevA.72.062310}
}

@article{kliesch2021theory,
  title={Theory of quantum system certification},
  author={Kliesch, Martin and Roth, Ingo},
  journal={PRX quantum},
  volume={2},
  number={1},
  pages={010201},
  year={2021},
  publisher={APS},
  doi = {10.1103/PRXQuantum.2.010201},
  url = {https://doi.org/10.1103/PRXQuantum.2.010201},
}

\newpage
\ \\
\newpage

\onecolumngrid

\appendix
\section{How to compare measurement strategies (Proof of Eq.~\eqref{eq: std_error_min})}
\label{appendix: compare measurement strategies}
A measurement strategy in quantum mechanics has the aim of determining an estimator for the true observable expectation value $\Braket{ \psi | H |  \psi}$ as accurate as possible with a given number of measurements $M$. Then standard deviation of an unbiased estimator is given by:
\begin{align}
   \sigma_{\rm min. }^2 = \frac{1}{ M} \Delta H^2  = \frac{1}{ M} {\Var_{\ket \psi} \left( H \right) } \ .
\end{align}
Realizing this sampling error requires the ability to measure observable $H$ at once. However, finding the eigenbasis, in which one would need to measure, generally cannot be done in polynomial time, as we need to diagonalize the entire exponentially large Hilbert space to map $\ket \psi$ into the eigenbasis of $H$, in the worst case. Thus, in typical non-trivial sampling tasks, one needs to split $H$ into parts $H_b$ of which the transformations into their respective eigenbasis are known, such that $H  = \sum_{H_b \in \mathcal B} H_b$ with partitioning $\mathcal{ B }$. Then the measurement error can be upper bounded by (step by step calculation of Eq.~\eqref{eq: intro_cost_function}):
\begin{align}
   &\frac{1}{ M} \Delta H^2  = \frac{1}{ M} \left[ \sum_{H_b \in \mathcal B}  \Var_{\ket \psi} \left( H_b\right) + \sum_{\substack{H_b, H_b^\prime \in \mathcal B  \\ H_b \neq H_b^\prime }} {\rm CoV}_{\ket \psi} \left( H_b, H_b^\prime \right) \right]
   \leq \frac{1}{ M} \left[ \sum_{H_b \in \mathcal B}  \Var_{\ket \psi} \left( H_b\right) + \sum_{\substack{H_b, H_b^\prime \in \mathcal B  \\ H_b \neq H_b^\prime }} \left|  {\rm CoV}_{\ket \psi} \left( H_b, H_b^\prime \right) \right | \right] \nonumber \\
   &\leq \frac{1}{ M}  \left[  \sum_{H_b \in \mathcal B}  \Var_{\ket \psi} \left( H_b\right) + \sum_{\substack{H_b, H_b^\prime \in \mathcal B  \\ H_b \neq H_b^\prime }}  \sqrt{ \Var_{\ket \psi} \left(H_b\right) \Var_{\ket \psi} \left(H_b^\prime \right) } \right ] 
   = \frac{1}{ M} \left[  \sum_{H_b \in \mathcal B} \sqrt{ \Var_{\ket \psi} \left( H_b\right) } \right ]^2 
   \leq \sum_{H_b \in \mathcal B}  \frac{ \Var_{\ket \psi} \left( H_b\right) }{M_b} \ ,
   \label{eq: subsystem_error}
\end{align}
where $M = \sum_{b=1}^{|\mathcal B|} M_b$. The first inequality adds absolute values to the covariances and the second inequality is the Cauchy–Schwarz inequality. Using Lagrangian multipliers, one finds that the last inequality becomes an equality precisely for the optimal distribution of sampling budget $M$. Let us write this down explicitly. 
The Lagrangian is given by:
\begin{align}
\mathcal{L} (\vec M, \lambda) \coloneqq \sum_{b=1}^{|\mathcal B|}  \frac{ \Var_{\ket \psi} \left( H_b\right) }{M_b} - \lambda \left( M- \sum_{b=1}^{|\mathcal B|} M_b\right) \ ,
\end{align}
where we named $\vec M$ being a vector of all $M_b$. Then the stationary conditions yields:
\begin{align}
\label{eq: part_Lagragian_Mb}
0 &\stackrel{!}{=} \frac{\mathcal{L} (\vec M, \lambda) }{\partial M_b} = - \frac{\Var_{\ket \psi} \left( H_b\right)}{M_b^2} + \lambda \qquad \forall b \in \{1,2,..., B \} \\
0 &\stackrel{!}{=} \frac{\mathcal{L} (\vec M, \lambda) }{\partial \lambda} = M - \sum_{b=1}^B M_b \ ,
\end{align}
where $\stackrel{!}{=} $ means that we demand this equality to be fulfilled for the stationary conditions.
As $\Var_{\ket \psi} \left( H_b\right) \geq 0$, we know that $\lambda = \Var_{\ket \psi} \left( H_b\right)/M_b^2 \geq 0$, and in order to minimize the standard error, $M_b$ needs to be chosen such that all $\sqrt{\Var_{\ket \psi} \left( H_b\right)}/M_b$ become equal. This allows us to express $\lambda$ in terms of M and $\Var_{\ket \psi} \left( H_b\right)$: 
\begin{align}
M&= \sum_{b=1}^{|\mathcal B|} M_b = \sum_{b=1}^{|\mathcal B|} \sqrt{\frac{\Var_{\ket \psi} \left( H_b\right)}{\lambda}} \quad \Rightarrow  \quad  \sqrt \lambda = \frac{1}{M} \sum_{b=1}^{|\mathcal B|} \sqrt{\Var_{\ket \psi} \left( H_b\right)} \ .
\label{eq: sqrt_lambda}
\end{align}
Then Eq.~\eqref{eq: sqrt_lambda} and Eq.~\eqref{eq: part_Lagragian_Mb} solve the constraint minimization problem:
\begin{align}
\min_{M = \sum M_b} \left(\sum_{H_b \in \mathcal B}  \frac{ \Var_{\ket \psi} \left( H_b\right) }{M_b} \right) 
= \sum_{H_b \in \mathcal B} \sqrt{\Var_{\ket \psi} \left( H_b\right)} \sqrt \lambda 
= \frac{1}{ M} \left[  \sum_{H_b \in \mathcal B} \sqrt{ \Var_{\ket \psi} \left( H_b\right) } \right ]^2 \ .
\end{align}
$\Var_{\ket \psi} \left( H_b\right)$ are the sampling errors we will experience when measuring $H$ in the partitioning $\mathcal B$. 
The partitioning $\mathcal B$ is by no means unique. Rather it gives us a way to compare sampling budget allocation agnostic sampling strategies by comparing $\left[  \sum_{H_b \in \mathcal B} \sqrt{ \Var_{\ket \psi} \left( H_b\right) } \right ]^2$, motivation our relative sampling complexity cost function in Definition~\ref{def:Sampling improvement}.

In a more general setting, where we are interested to upper bound the (average) measurement error, not only w.r.t. a single state $\ket \psi$, but w.r.t. to an assemble of states $\ket{\tilde \psi}$, we need to consider the expectation value over $\ket{\tilde \psi}$:
\begin{align}
   &\frac{1}{ M}  \overline{\Delta  H}^2  = \frac{1}{ M} \EE \left[  \Var_{\ket{\tilde \psi}} \left(  H\right) \right] \ .
\end{align}
Applying once more the Cauchy-Schwarz inequality, this time on the expectation value over  $\ket{\tilde \psi}$, as well as using linearity of the expectation value, we can obtain a similar upper bound to Eq.~\eqref{eq: subsystem_error}:
\begin{align}
   \frac{1}{ M}  \overline{\Delta  H}^2  &= \frac{1}{ M} \EE \left[  \Var_{\ket{\tilde \psi}} \left(  H\right) \right] 
   \leq \frac{1}{ M} \EE \left[   \left[ \sum_{H_b \in \mathcal B}  \sqrt{ \Var_{\ket \psi} \left(  H_b\right)} \right ]^2 \right]  \nonumber \\
   &= \frac{1}{ M} \EE \left[  \sum_{H_b \in \mathcal B}   \Var_{\ket \psi} \left(  H_b\right)  \right] +  \frac{1}{ M}
   \EE \left[ \sum_{H_b^\prime \neq H_b \in \mathcal B}  \sqrt{ \Var_{\ket \psi} \left(  H_b\right)} \sqrt{ \Var_{\ket \psi} \left(  H_b^\prime \right)} \right]  \nonumber\\
   &\leq \frac{1}{ M}  \sum_{H_b \in \mathcal B} \EE \left[    \Var_{\ket \psi} \left(  H_b\right)  \right] +  \frac{1}{ M}  \sum_{H_b^\prime \neq H_b \in \mathcal B}  \sqrt{
   \EE \left[\Var_{\ket \psi} \left(  H_b\right) \right] } \sqrt{
   \EE \left[\Var_{\ket \psi} \left(  H_b^\prime \right) \right] } \nonumber \\
   &= \frac{1}{ M} \left[ \sum_{H_b \in \mathcal B}  \sqrt{ \mathbb E \left [ \Var_{\ket{\tilde \psi}} \left(  H_b\right)\right ] } \right]^2 = \sigma_{\mathcal B}^2\ .
   \label{eq: subsystem_error_assemble}
\end{align}

To obtain the state-independent average sampling error, one can average Eq.~\eqref{eq: subsystem_error_assemble} over all states $\ket{\tilde \psi}$. Lemma~\ref{thm:imperfect eigenstates} covers this case for $\epsilon = 1$, $\ket{\tilde \psi}= \ket \xi$, and $ \ket \xi$ Haar-random up to second moment:
\begin{align}
   \frac{1}{ M}  \overline{\Delta  H}^2 
   &\leq \frac{1}{ M} \left[ \sum_{H_b \in \mathcal B}  \sqrt{ \mathbb E \left [ \Var_{\ket{\xi}} \left(  H_b\right)\right ] } \right]^2 
   = \frac{1}{ M} \left[ \sum_{H_b \in \mathcal B}  \sqrt{ \norm{H_b}^2_F/d } \right]^2 + \mathcal{O} \left( \frac{1}{d}\right)
   \coloneqq \sigma_{\rm av. }^2
   \label{eq: average_measurement_error}
\end{align}

Note that shadow norms are closely related to the Frobenius norm, also called Hilbert-Schmidt norm~\cite{Huang2020Predicting}.
Similarly, one can derive a worst-case sampling error from Eq.~\ref{eq: subsystem_error}, by partitioning $H = \sum_b \omega_b P_b$ into Pauli weights $\omega_b$ and Pauli strings $P_b$. Then the each variance $\Var(P_b) = \braket{P_b^2} - \braket{P_b}^2 = 1- \braket{P_b}^2$ is upper bound by 1, and we get:
\begin{align}
   \frac{1}{ M}  \Delta  H^2 
   &\leq \frac{1}{ M} \left[ \sum_{H_b \in \mathcal B}  \sqrt{  \Var_{\ket{\psi}} \left(  H_b\right) } \right]^2 
   = \frac{1}{ M} \left[ \sum_{H = \sum_b \omega_b P_b}  \sqrt{ \omega_b^2 \Var_{\ket{\psi}} \left(  P_b\right) } \right]^2 
   \leq \frac{1}{ M} \left[ \sum_{H = \sum_b \omega_b P_b}  \left | \omega_b \right |\right]^2 
   \coloneqq \sigma_{\rm max. }^2
   \label{eq: maximal_measurement_error}
\end{align}

\section{Noiseless Sampling Improvements: Additional material for Sec.~\ref{subsec:perfect eigenstates}}
\subsection{Definitions of employed partitionings}
\label{app:def_partitionings}

{
Here, we write out explicitly all used partitionings. To recall Eq.~\eqref{eq: intro_geometrically_local}, the considered Hamiltonians have following form:
\begin{align}
 H = &\sum_{i, \alpha} h_i^{(\alpha)} o_i^{(\alpha)} +\sum_{\left<i,j\right>, \alpha, \beta} J_{i,j}^{(\alpha,\beta)} o_i^{(\alpha)}o_j^{(\beta)} \ ,
\end{align}
with $o_i^{(\alpha)}$ Pauli or annihilation/creation operators. }

\begin{defn}[Mutually commuting operator/Pauli partitioning]
  \label{def:Pauli partitionings}
Let $H$ be a lattice Hamiltonian as defined in Eq.~\eqref{eq: intro_geometrically_local}, which we denote for simplicity by:
\begin{align}
 H = &\sum_{\boldsymbol \iota=1}^{\boldsymbol I} O_{\boldsymbol \iota} \ ,
\end{align}
with finite multi-index $\boldsymbol \iota$ mapped to the natural numbers, and $O_{\boldsymbol \iota} \in  \{h_i^{(\alpha)} o_i^{(\alpha)}, J_{i,j}^{(\alpha,\beta)} o_i^{(\alpha)}o_j^{(\beta)} \}$. As $H$ is a 2-local Hamiltonian, $\boldsymbol I = \mathcal O(n)$. 
We define the mutually commuting operator partitioning by the following algorithm:\\ 
{
\RemoveAlgoNumber
\RestyleAlgo{ruled}
\begin{algorithm}
\caption{for construction of a mutually commuting operator partitioning}
$H_1 \gets O_{1}$\;
\For{$\boldsymbol \iota\leftarrow 2$ \KwTo $\boldsymbol I$}{
$b \gets 1$\;
\While{$\exists O^\prime \in H_b: [O^\prime, O_{\boldsymbol \iota} ] \neq 0$}{
$b \gets b + 1$\;
}
$H_b \gets H_b + O_{\boldsymbol \iota} $\;
}
\end{algorithm} 
}

where $b \gets b + 1$ means that variable $b$ is assigned value $b+1$.
We call the resulting partitioning $\mathcal B_{\rm Pauli} = \{ H_b\}$ interchangeably Pauli partitioning, as it coincides for most local lattice models with naively separating $H$ by local Pauli operator type. 
\end{defn}

{
This way, each $H_b$ can be measured at once. The maximal number of Pauli partitions  $\left | \mathcal B_{\rm Pauli} \right |$ is given by the number of mutually unbiased bases of a $k$-local observable $2^k + 1$, with locality $k = 2$, $\left | \mathcal B_{\rm Pauli} \right | \leq 5$~\cite{romero2005structure}. 

Similarly, we define geometric partitioning explicitly as follows:

\begin{defn}[Explicit geometric partitionings]
  \label{def:Explicit geometric partitionings}
Let $H$ be a lattice Hamiltonian as defined in Eq.~\eqref{eq: intro_geometrically_local}. We define its geometric partitioning, $\mathcal{B}_{L_x,L_y}$, $H_1 + H_2 = H$ by:
\begin{align}
H_1 &\coloneqq \frac{1}{2} \left( H - H_{\rm cut}+H_{\rm cut}^\prime \right) \ , \\
H_2 &\coloneqq \frac{1}{2} \left( H + H_{\rm cut}-H_{\rm cut}^\prime \right) \ ,
\end{align}
with $H_{\rm cut}$ being the 2-local terms of $H$ cut in $H_1$ and added to $H_2$, and vice versa $H_{\rm cut}^\prime$ to obtain subsystem patches $H_{b,k}$ with disjoint support. 
We use the notation
\begin{align}
        \mathcal{H}_i = \sum_{\alpha} h_i^{(\alpha)} o_i^{(\alpha)} \quad \text{and} \quad \mathcal{H}_{i,j} = \sum_{\alpha, \beta} J_{i,j}^{(\alpha,\beta)} o_i^{(\alpha)}o_j^{(\beta)} \ ,
\end{align}
for the one-local and two-local terms in Eq. \eqref{eq: intro_geometrically_local}, where the site indices $i$ and $j$ have two components for two dimensional lattices, $i=(i_x,i_y)$ and $j=(j_x,j_y)$, and consider the following three specific partitionings.

\paragraph{1D partitioning} with 1D strip thickness $L$:
\begin{align}
L=1: &
\qquad 
H_1 = \sum_{m=1}^{\lfloor n_x/2\rfloor} \sum_{\ell=1}^{n_y} 
\mathcal{H}_{(2m,\ell),(2m+1,\ell)} + \frac{1}{2}\sum_{i} \mathcal{H}_i \ ,
\qquad
H_2 = \sum_{m=1}^{n_x} \sum_{\ell=1}^{\lfloor n_y/2\rfloor} 
\mathcal{H}_{(m,2\ell),(m,2\ell+1)} + \frac{1}{2}\sum_{i} \mathcal{H}_i \ ,
\\
L\geq2: & 
\qquad H_{\rm cut} = \sum_{m=1}^{\lfloor n_x/L\rfloor} \sum_{\ell=1}^{n_y}  \mathcal{H}_{(mL,\ell),(mL+1,\ell)} \ , 
\qquad \qquad \qquad
H_{\rm cut}^\prime = \sum_{m=1}^{\lfloor n_x/L\rfloor} \sum_{\ell=1}^{n_y} 
\mathcal{H}_{(mL-1,\ell),(mL,\ell)}  \ ,
\end{align}
where $L=1$ is a special case with $H_{\rm cut}$ consisting of all 2-local terms in $x$-direction and respectively  $H_{\rm cut}^\prime$ containing all 2-local terms in $y$-direction.

\paragraph{2D partitioning}
with patch size $L_x \times L_y$:
\begin{align}
H_{\rm cut} &= H_{\rm cut}^{x} + H_{\rm cut}^{y},
\qquad
 H_{\rm cut}^{x} = 
\sum_{m=1}^{\lfloor n_x/L_x\rfloor} \sum_{\ell=1}^{n_y}  
\mathcal{H}_{(mL_x,\ell),(mL_x+1,\ell)},
\qquad 
H_{\rm cut}^{y} = 
\sum_{m=1}^{n_x} 
\sum_{\ell=1}^{\lfloor n_y/L_y\rfloor}  
\mathcal{H}_{(m,\ell L_y),(m,\ell L_y +1)} \ , \\
H_{\rm cut}^\prime &= H_{\rm cut}^{x'} + H_{\rm cut}^{y'},
\qquad
H_{\rm cut}^{x'} =
\sum_{m=1}^{\lfloor n_x/L_x\rfloor} \sum_{\ell=1}^{n_y}  
\mathcal{H}_{(mL_x-1,\ell),(mL_x,\ell)},
\qquad
H_{\rm cut}^{y'} =\sum_{m=1}^{n_x} 
\sum_{\ell=1}^{\lfloor n_y/L_y\rfloor}  
\mathcal{H}_{(m,\ell L_y-1),(m,\ell L_y)} \ .
\end{align}

\paragraph{2-local partitioning}
Further, we can partition $H$ into $H_b$'s with disjoint blocks of size 2, which requires four parts $H=H_1+H_2+H_3+H_4$:
\begin{align}
H_1 &= \sum_{m=1}^{\lfloor n_x/2\rfloor} \sum_{\ell=1}^{n_y} 
\mathcal{H}_{(2m,\ell),(2m+1,\ell)} + \frac{1}{4}\sum_{i} \mathcal{H}_i \ ,
\\
H_2 &= \sum_{m=1}^{\lfloor n_x/2\rfloor} \sum_{\ell=1}^{n_y} 
\mathcal{H}_{(2m-1,\ell),(2m,\ell)} + \frac{1}{4}\sum_{i} \mathcal{H}_i \ , \\
H_3 &= 
\sum_{m=1}^{n_x} 
\sum_{\ell=1}^{\lfloor n_y/2\rfloor}  
\mathcal{H}_{(m,2\ell ),(m,2\ell  +1)} 
+ \frac{1}{4}\sum_{i} \mathcal{H}_i \ ,
\\
H_4 &= 
\sum_{m=1}^{n_x} 
\sum_{\ell=1}^{\lfloor n_y/2\rfloor}  
\mathcal{H}_{(m, 2\ell -1),(m,2\ell)} 
+ \frac{1}{4}\sum_{i} \mathcal{H}_i \ .
\end{align}
\end{defn}

On the following pages, we provide explicitly all partitions of the six numerical examples of the main text.

\subsection{{Rel. Sampling complexity improvement lower bound} (Proof of Theorem \ref{thm:improvement})}
\label{app:improvement}
To prove a sampling improvement for measuring in the eigenbases of decoupled subsystems, we express the decoupled Hamiltonians as an orbit of a subgroup of the translations along the $x$- and $ y$-directions of a 2-dimensional lattice. The operators $T_\ell^x$ and $T_\ell^y$ denote a translation by $\ell$ sites into the respective direction. Both directions are connected via a local rotation operation $S$ that maps $T_\ell^y = S T_\ell^x S$. Altogether, we consider subgroups of the symmetry group  
\begin{align}
    \mathcal{T} = \{ T_\ell^x T_m^y | \ell = 1, ..., n_x, m=1,...,n_y \}.
    \label{eq:symmetries}
\end{align}
With this, every $k$-local, translation invariant Hamiltonian can be written as an orbit of $\mathcal{T}$ acting on a unit cell interaction $V$, i.e.
\begin{align}
    H = \sum_{T \in \mathcal{T}} T V T^\dagger + H_{\rm{1-local}}.
    \label{eq:trans_inv_ham}
\end{align}
Here, we focus on geometrically local lattice Hamiltonians as given in Eq.~\eqref{eq: intro_geometrically_local} so that $H_{\rm{1-local}}= \sum_{i, \alpha} h_i^{(\alpha)} o_i^{(\alpha)}$ and $V = \sum_{\alpha, \beta} J_{{i_0},{j_0}}^{(\alpha,\beta)} o_{i_0}^{(\alpha)}o_{j_0}^{(\beta)}$ with arbitrary neighboring lattice vertices $i_0$ and $j_0$.
We will use in the following that any Pauli partitioning $\mathcal B_{\textrm Pauli}$ consists of at least two partitions. Further, at least one of them needs to have translationally invariant 2-local terms. Therefore, assuming all 2-local terms by themselves and respectively all 1-local terms by themselves can be measured simultaneously, and yields a lower bound on the Pauli sampling error:
\begin{align}
    \sigma^2_{\textrm Pauli} = \frac{1}{ M} \left[  \sum_{H_b \in \mathcal B_{\textrm{Pauli}}} \sqrt{ \Var_{\ket{E_i}}  \left( H_b\right) } \right ]^2 
    \geq \frac{2}{ M} \left[ \Var_{\ket{E_i}} \left( H_{\textrm{2-local}} \right) + \Var_{\ket{E_i}} \left( H_{\textrm{1-local}} \right)  \right ]
\end{align}
Further, as we focus on eigenstates here, we have $\Var_{\ket{E_i}} (H_{\rm{2-local}}) = \Var_{\ket{E_i}} (H_{\rm{1-local}})$ due to Corollary.~\ref{corol: Ei_2_partitions}.
We can write the horizontal  interactions as an orbit of the translation group acting on a unit cell Hamiltonian $V$
\begin{align}
    H_{\rm{2-local}}^h &= \sum_{\ell=1}^{n_x} \sum_{m=1}^{n_y} T^y_m T^x_\ell V T^x_{-\ell} T^y_{-m},
\end{align}
and the vertical interactions accordingly
\begin{align}
    H_{\rm{2-local}}^v &= S H_{\rm{2-local}}^h S.
\end{align}
We begin with a proof for 2D partitioning (compare Def.~\ref{def:Explicit geometric partitionings} b.). 
The sample improvement reads
\begin{align}
    &\mathcal G_{\ket{E_i}} \left( \mathcal B_{\rm Pauli}, \mathcal B_{L_x, L_y} \right) 
    \geq \frac{\Var_{\ket{E_i}}(H_{\rm{2-local}})}{\Var_{\ket{E_i}}\left( \frac{1}{2} (H - H_{\rm cut}^x - H_{\rm cut}^y + H_{\rm cut}^{x'} + H_{\rm cut}^{y'}) \right)},
\end{align}
Let us consider the numerator and denominator separately.
The denominator can be simplified further via
    \begin{align}
        \Var_{\ket{E_i}}\left( \frac{1}{2} (H - H_{\rm cut}^x - H_{\rm cut}^y + H_{\rm cut}^{x'} + H_{\rm cut}^{y'}) \right) &= \Var_{\ket{E_i}}\left( \frac{1}{2} (H + H_{\rm cut}^x + H_{\rm cut}^y - H_{\rm cut}^{x'} - H_{\rm cut}^{y'}) \right) \nonumber \\
        &= \frac{1}{4} \Var_{\ket{E_i}}\left( H_{\rm cut}^x + H_{\rm cut}^y - H_{\rm cut}^{x'} - H_{\rm cut}^{y'} \right).
    \end{align}
 First, observe that $\Var_{\ket{E_i}}(H_{\rm{2-local}}) = 2 \Var_{\ket{E_i}}(H_{\rm{2-local}}^h) + 2 \Var_{\ket{E_i}}(S H_{\rm{2-local}}^h)$ from invariance of the ground state. Further, we can express both terms as covariances. We have
\begin{align}
    \Var_{\ket{E_i}}(H_{\rm{2-local}}^h) &= \Var_{\ket{E_i}}\left( \sum_{\ell=1}^{n_x} \sum_{m=1}^{n_y} T^y_m T^x_\ell V T^x_{-\ell} T^y_{-m} \right) = n_x n_y \sum_{\ell=1}^{n_x} \sum_{m=1}^{n_y} \Var_{\ket{E_i}}(T^y_m T^x_\ell V ),
    \label{eq:H2h_c}
\end{align}
where we used the multilinearity of the covariance, the fact that translation operators commute, $\comm{T_\ell^x}{T_m^y} = 0$, and again symmetry of the ground state to identify $n_x n_y$ identical terms. Similarly, we can calculate
\begin{align}
    \Var_{\ket{E_i}}(S H_{\rm{2-local}}^h) &= n_x n_y \sum_{\ell=1}^{n_x} \sum_{m=1}^{n_y} \Var_{\ket{E_i}}(S T^y_m T^x_\ell V ).
\end{align}
Let us separate the qubit number $n = n_x n_y$ from the correlators and write $\Var_{\ket{E_i}}(H_{\rm{2-local}}) =: 2n C_1$. The denominator can be treated similarly. 
\begin{align}
    &\Var_{\ket{E_i}}(H_{\rm cut}^{x} + H_{\rm cut}^{y} - H_{\rm cut}^{x'} - H_{\rm cut}^{y'}) \label{eq:denom_c} \\
    &= 2 \Var_{\ket{E_i}}(H_{\rm cut}^{x}) + 2 \Var_{\ket{E_i}}(H_{\rm cut}^{y}) - \Var_{\ket{E_i}}(T^x_1 H_{\rm cut}^{x}) - \Var_{\ket{E_i}}(T^x_{-1} H_{\rm cut}^{x}) - \Var_{\ket{E_i}}(T^y_1 H_{\rm cut}^{y}) - \Var_{\ket{E_i}}(T^y_{-1} H_{\rm cut}^{y})
    \nonumber
\end{align}
Expressing $H_{\rm cut}^{x}$ in terms of $V$, we calculate
\begin{align}
    \Var_{\ket{E_i}}(H_{\rm cut}^{x}) &= \Var_{\ket{E_i}}\left( \sum_{\ell=1}^{n_x/L_x} \sum_{m=1}^{n_y} T_{\ell L_x}^x T^y_m V T^y_{-m} T_{-\ell L_x}^x \right) = n_y \frac{n_x}{L_x} \sum_{\ell=1}^{n_x/L_x} \sum_{m=1}^{n_y} \Var_{\ket{E_i}}\left( T_{\ell L_x}^x T^y_m V \right).
\end{align}
and similarly
\begin{align}
    &\Var_{\ket{E_i}}(H_{\rm cut}^{y}) = n_x \frac{n_y}{L_y} \sum_{m=1}^{n_y/L_y} \sum_{\ell=1}^{n_x} \Var_{\ket{E_i}}\left( T_{m L_y}^y T^x_\ell SV \right) \\
    &\Var_{\ket{E_i}}( T^x_1 H_{\rm cut}^{x}) = n_y \frac{n_x}{L_x} \sum_{\ell=1}^{n_x/L_x} \sum_{m=1}^{n_y} \Var_{\ket{E_i}}\left( T^x_{\ell L_x + 1} T^y_m V \right) \\
    &\Var_{\ket{E_i}}(T^x_{-1} H_{\rm cut}^{x}) = n_y \frac{n_x}{L_x} \sum_{\ell=1}^{n_x/L_x} \sum_{m=1}^{n_y} \Var_{\ket{E_i}}\left( T^x_{\ell L_x-1} T^y_m V \right) \\
    &\Var_{\ket{E_i}}( T^y_1 H_{\rm cut}^{y}) = n_x \frac{n_y}{L_y} \sum_{m=1}^{n_y/L_y} \sum_{\ell=1}^{n_x} \Var_{\ket{E_i}}\left( T^y_{m L_y + 1} T^x_\ell SV \right) .
\end{align}
Separating qubit number and cluster thickness, we write $\Var_{\ket{E_i}}\left( H_{\rm cut}^x + H_{\rm cut}^y - H_{\rm cut}^{x'} - H_{\rm cut}^{y'} \right) = 2 \left( \frac{n}{L_x} C_2^x + \frac{n}{L_y} C_2^y \right)$ and altogether
\begin{align}
    &\mathcal G_{\ket{E_i}} \left( \mathcal B_{\rm Pauli}, \mathcal B_{L_x, L_y} \right) = 4 L_x L_y \frac{C_1}{L_y C_2^x + L_x C_2^y},    
\end{align}
with the system-dependent constants
\begin{align}
    C_1 &= \sum_{\ell=1}^{n_x} \sum_{m=1}^{n_y} \Var_{\ket{E_i}}(T^y_m T^x_\ell V ) + \sum_{\ell=1}^{n_x} \sum_{m=1}^{n_y} \Var_{\ket{E_i}}( S T^y_{m} T^x_{\ell} V ) \label{eq:c1} \\
    C_2^x &= \sum_{\ell=1}^{n_x/L_x} \sum_{m=1}^{n_y} \left( \Var_{\ket{E_i}}\left( T^x_{\ell L_x} T^y_m V \right) - \frac{1}{2} \Var_{\ket{E_i}}\left( T^x_{\ell L_x + 1} T^y_m V \right) - \frac{1}{2} \Var_{\ket{E_i}}\left( T^x_{\ell L_x - 1} T^y_m V \right) \right) \label{eq:c2x} \\
    C_2^y &= \sum_{m=1}^{n_y/L_y} \sum_{\ell=1}^{n_x} \left( \Var_{\ket{E_i}}\left( T^y_{m L_y} T^x_\ell S V \right) - \frac{1}{2} \Var_{\ket{E_i}}\left( T^y_{m L_y + 1} T^x_\ell S V \right) - \frac{1}{2} \Var_{\ket{E_i}}\left( T^y_{m L_y - 1} T^x_\ell S V \right) \right). \label{eq:c2y}
\end{align}

From $C_2^x \leq C_1$ and $C_2^y \leq C_1$, we already have $\mathcal G_{\ket{E_i}} \left( \mathcal B_{\rm Pauli}, \mathcal B_{L_x, L_y} \right) \geq \frac{4 L_x L_y}{L_x+L_y}$ as a lower bound which can be further improved.
We have, similarly to before: 
\begin{align}
    H_{\rm cut} = T^y_{-1} T^x_{-1} H_{\rm cut}^{'} T^y_{1} T^x_{1} \quad \Rightarrow \quad \Var_{\ket{E_i}}\left( H_{\rm cut}  \right) = \Var_{\ket{E_i}}\left( H_{\rm cut}^{'}  \right)
\end{align}
due to the translational invariance of the eigenstate. Thus, we can write the denominator as
\begin{align}
\Var_{\ket{E_i}}\left( H_{\rm cut}^x + H_{\rm cut}^y - H_{\rm cut}^{x'} - H_{\rm cut}^{y'} \right) 
&= \Var_{\ket{E_i}}\left( H_{\rm cut} - H_{\rm cut}^{'} \right) \\
&= \Var_{\ket{E_i}}\left( H_{\rm cut} \right) + \Var_{\ket{E_i}}\left( H_{\rm cut}^{'} \right) - 2 {\rm CoV}_{\ket{E_i}} \left( H_{\rm cut} , H_{\rm cut}^{'} \right) \\
&= 2 \Var_{\ket{E_i}}\left( H_{\rm cut} \right) \left[ 1 - \frac{{\rm CoV}_{\ket{E_i}} \left( H_{\rm cut} , H_{\rm cut}^{'} \right) }{ \sqrt{\Var_{\ket{E_i}}\left( H_{\rm cut} \right) }  \sqrt{\Var_{\ket{E_i}}\left( H_{\rm cut}^{'} \right) } }\right ] \\
&= 2 \Var_{\ket{E_i}}\left( H_{\rm cut} \right) \left[ 1 - {\rm CoR}_{\ket{E_i}} \left( H_{\rm cut} , H_{\rm cut}^{'} \right)  \right ] 
\end{align}
which we simplified by using the correlation $\operatorname{CoR}_{\ket{E_i}} (H_{\rm cut}, H_{\rm cut}^\prime) = {\rm CoV}_{\ket{E_i}} (H_{\rm cut}, H_{\rm cut}^\prime)/ \sqrt{\Var_{\ket{E_i}} (H_{\rm cut}) \Var_{\ket{E_i}} (H_{\rm cut}^\prime)}$.
Again due to the translational symmetry of $\ket{E_i}$ we have 
\begin{align}
{\rm CoV}_{\ket{E_i}}\left( H_{\rm cut} , H_{\rm cut}^{'} \right) &= {\rm CoV}_{\ket{E_i}} \left( H_{\rm cut} ,T^y_{-1} T^x_{-1} H_{\rm cut} T^y_{1} T^x_{1}  \right) = {\rm CoV}_{\ket{E_i}} \left(  T^x_{-1} T^y_{-1} H_{\rm cut} ,T^y_{-1} T^x_{-1} H_{\rm cut} \right) \\
&= \Var_{\ket{E_i}} \left(  T^x_{-1} T^y_{-1} H_{\rm cut}  \right) \geq 0
\end{align}
From a similar calculation we have $\Var_{\ket{E_i}}\left( H_{\rm cut} \right) = \Var_{\ket{E_i}}\left( H_{\rm cut}^x + H_{\rm cut}^y\right) = \Var_{\ket{E_i}}\left( H_{\rm cut}^x\right)  + \Var_{\ket{E_i}}\left( H_{\rm cut}^y\right) + 2 {\rm CoV}_{\ket{E_i}} \left( H_{\rm cut}^x ,H_{\rm cut}^y\right)  \leq \Var_{\ket{E_i}}\left( H_{\rm cut}^x\right)  + \Var_{\ket{E_i}}\left( H_{\rm cut}^y\right) $
\begin{align}
    \mathcal G_{\ket{E_i}} \left( \mathcal B_{\rm Pauli}, \mathcal B_{L_x, L_y} \right) 
    &\geq 4  \frac{\Var_{\ket{E_i}}(H_{\rm{2-local}})}{2\Var_{\ket{E_i}}\left( H_{\rm cut} \right)} \frac{1}{1- {\rm CoR}_{\ket{E_i}} \left( H_{\rm cut} , H_{\rm cut}^{'} \right) }\\
    &\geq 4  \frac{2n C_1}{2\Var_{\ket{E_i}}\left( H_{\rm cut}^x \right) + 2\Var_{\ket{E_i}}\left( H_{\rm cut}^y \right)} \frac{1}{1- {\rm CoR}_{\ket{E_i}} \left( H_{\rm cut} , H_{\rm cut}^{'} \right) }\\
    &= 4  \frac{2n C_1}{2 \left( \frac{n}{L_x} C_2^x + \frac{n}{L_y} C_2^y \right)} \frac{1}{1- {\rm CoR}_{\ket{E_i}} \left( H_{\rm cut} , H_{\rm cut}^{'} \right) }\\
    &\geq 4 \frac{L_x L_y}{L_x+L_y} \frac{1}{1- {\rm CoR}_{\ket{E_i}} \left( H_{\rm cut} , H_{\rm cut}^{'} \right) }
\end{align}


%
%
For 1D partitioning (compare Def.~\ref{def:Explicit geometric partitionings} a.), we set $L_x = n_x$ and call $L_y =: L$. The sample improvement reads
\begin{align}
    &\mathcal G_{\ket{E_i}} \left( \mathcal B_{\rm Pauli}, \mathcal B_{L_x, L_y} \right) \geq 4 \frac{\Var_{\ket{E_i}}(H_{\rm{2-local}})}{\Var_{\ket{E_i}}\left( H_{\rm cut} - H_{\rm cut}' \right)}.
\end{align}
While the numerator is the same as before, the denominator becomes $\Var_{\ket{E_i}}\left( H_{\rm cut} - H_{\rm cut}' \right) = 2 \left( \Var_{\ket{E_i}}\left( H_{\rm cut} \right) - \Var_{\ket{E_i}}\left( T_1^y H_{\rm cut} \right) \right)$, cf. Eq.~\eqref{eq:denom_c}. We can further simplify Eq.~\eqref{eq:H2h_c} for 1D cuts to
\begin{align}
    \Var_{\ket{E_i}}(H_{\rm{2-local}}^h) = L \sum_{m=1}^{L} \Var_{\ket{E_i}}(T^y_m H_{\rm cut} ).
\end{align}
Altogether we have
\begin{align}
    \mathcal G_{\ket{E_i}} \left( \mathcal B_{\rm Pauli}, \mathcal B_{L_x, L_y} \right) &\geq 4 L \frac{\sum_{m=1}^{L} \Var_{\ket{E_i}}(T^y_m H_{\rm cut} )}{\Var_{\ket{E_i}}\left( H_{\rm cut} \right) - \Var_{\ket{E_i}}\left( T_1^y H_{\rm cut} \right)} \geq 4L \frac{1}{1 - \frac{\Var_{\ket{E_i}}\left( T_1^y H_{\rm cut} \right)}{\Var_{\ket{E_i}}\left( H_{\rm cut} \right)}}
    =4L \frac{1}{1- {\rm CoR}_{\ket{E_i}} \left( H_{\rm cut} , H_{\rm cut}^{'} \right) },
\end{align}
where we dropped the positive term $\Var_{\ket{E_i}}(SH_{\rm{2-local}}^h)$ in the first step.
For $L\geq 2$ we even have: 
\begin{align}
    \mathcal G_{\ket{E_i}} \left( \mathcal B_{\rm Pauli}, \mathcal B_{L_x, L_y} \right) 
    &\geq 4L \frac{1+\frac{\Var_{\ket{E_i}}\left( T_1^y H_{\rm cut} \right)}{\Var_{\ket{E_i}}\left( H_{\rm cut} \right)}}{1 - \frac{\Var_{\ket{E_i}}\left( T_1^y H_{\rm cut} \right)}{\Var_{\ket{E_i}}\left( H_{\rm cut} \right)}}
    =4L \frac{1+{\rm CoR}_{\ket{E_i}} \left( H_{\rm cut} , H_{\rm cut}^{'} \right)}{1- {\rm CoR}_{\ket{E_i}} \left( H_{\rm cut} , H_{\rm cut}^{'} \right) },
\end{align} 

%
%
At last, 2-local partitioning (compare Def.~\ref{def:Explicit geometric partitionings} c.) assumes minimal clusters of $L_x = 1$ and $L_y = 2$ and vice versa. A 2-local Hamiltonian has to be split into four parts, which do not commute in general, $H = H_1 + H_2 + H_3 + H_4$ (as defined in Def.~\ref{def:Explicit geometric partitionings} c.). For higher localities, the following discussion can be straightforwardly generalised. Since the four terms are connected by symmetry under translations and local rotations exchanging $x$ and $y$, $H_1 = T_1^x H_2 T_{-1}^x = S H_3 S = T_1^x S H_4 S T_{-1}^x$, their variances coincide. As a result, the sample improvement again becomes very simple

\begin{align}
    \mathcal G_{\ket{E_i}} \left( \mathcal B_{\rm Pauli}, \mathcal B_{L_x, L_y} \right) &\geq \left[ \frac{ \sqrt{ \Var_{\ket{E_i}} \left(   H_{\rm{1-local}} \right)} + \sqrt{ \Var_{\ket{E_i}} \left(  H_{\rm{2-local}} \right)} }{ \sqrt{ \Var_{\ket{E_i}} \left(  H_1 \right)} + \sqrt{ \Var_{\ket{E_i}} \left(  H_2 \right)} + \sqrt{ \Var_{\ket{E_i}} \left(  H_3 \right)} + \sqrt{ \Var_{\ket{E_i}} \left(  H_4 \right)} }\right]^2 = \frac{1}{4} \frac{\Var_{\ket{E_i}}(H_{\rm{2-local}})}{\Var_{\ket{E_i}}(H_1)}.
\end{align}

We can use $\Var(H_{\rm{2-local}})$ from before and calculate the remaining variance of $H_1 = H_I + \frac{1}{4} H_{\rm{1-local}}  = H_I + \frac{1}{4} (H- H_{\rm{2-local}} )$, with the interactions $H_I = \sum_{\ell=1}^{n_x/2} \sum_{m=1}^{n_y} T_{2\ell}^x T^y_m V T_{-m}^y T_{-2\ell}^x$. The variance of $H_1$ can be calculated via $\Var_{\ket{E_i}}(H_1) = \Var_{\ket{E_i}}(O) + \frac{1}{16} \Var_{\ket{E_i}}(H_{\rm{2-local}}) - \frac{1}{4} ({\rm CoV}_{\ket{E_i}}(O, H_{\rm{2-local}}) + {\rm CoV}_{\ket{E_i}}(H_{\rm{2-local}}, O))$. Using the same methods as before, we calculate
\begin{align}
    \Var_{\ket{E_i}}(H_I) = n_y \frac{n_x}{2} \sum_{\ell=1}^{n_x/2} \sum_{m=1}^{n_y} \Var_{\ket{E_i}}\left( T_{2\ell}^x T^y_m V \right),
    \label{eq:Var_HAx}
\end{align}
which is only the first term of $C_2^x$ from Eq.~\eqref{eq:c2x}. We separate the factor $\frac{n}{2}$ and define the remainder of Eq.~\eqref{eq:Var_HAx} a new constant $C_3$. Similarly, the covariance reads
\begin{align}
    {\rm CoV}_{\ket{E_i}}(O, H_{\rm{2-local}}) &= n_y \frac{n_x}{2} \sum_{\ell=1}^{n_x} \sum_{m=1}^{n_y} {\rm CoV}_{\ket{E_i}}\left( V, T_{\ell}^x T^y_{m} V + T_{\ell}^x T^y_{m} S V \right) = \frac{n}{2} C_1 = {\rm CoV}_{\ket{E_i}}(H_{\rm{2-local}}, O)
\end{align}
Altogether we obtain
\begin{align}
    \mathcal G_{\ket{E_i}} \left( \mathcal B_{\rm Pauli}, \mathcal B_{L_x, L_y} \right) &= \frac{1}{4} \frac{2 n C_1}{\frac{n}{2} C_3 + \frac{n}{8} C_1 - \frac{n}{4} C_1} = \frac{C_1}{C_3 - \frac{1}{4} C_1} \geq \frac{4}{3},
\end{align}
since $C_1 \geq C_3$.  

\section{Sampling Improvements under noise (Proofs of Sec.~\ref{subsec:imperfect eigenstates})}
\label{appendix: For_Sec_II_C}
\subsection{Variance of isotropically perturbed states (Proof of Lemma~\ref{thm:imperfect eigenstates})}
\label{appendix:Lemma_2}

We will use the flip operator defined by 
\begin{align}
    \FF\ket{\psi_1}\ket{\psi_2} 
    \coloneqq 
    \ket{\psi_2}\ket{\psi_1}, 
\end{align}
the so-called swap-trick (e.g. see \cite[Eq.~(7)]{kliesch2021theory})
\begin{align}
    \Tr[AB] = \Tr[(A\otimes B)\FF], 
\end{align}
and that (e.g. see \cite[Eq.~(163)]{kliesch2021theory})
\begin{align}
\EE \,  \left[ \ket\xi \right ] = 0, \qquad \qquad \qquad \qquad \ &\EE\, \left [ \ket\xi\ketbra\xi\xi \right ]= 0, \\
\EE\, \left[ \ketbra\xi\xi \right ]  = \1/d, \qquad \qquad\qquad &\EE\, \left [ \ketbra\xi\xi^{\otimes 2} \right ] = \frac{1}{d(d+1)}  (\1 + \FF). \nonumber
\end{align}

Note that the last equality implies 
\begin{align}
\EE\left[ \sandwich{\xi}{O}{\xi}^2 \right]
&= \EE \left[ \Tr[ \ketbra\xi\xi^{\otimes 2} O^{\otimes 2} \FF ] \right ]
= \frac{1}{d(d+1)} \left( \Tr[ O^{\otimes 2} \FF^2] +  \Tr[ O^{\otimes 2} \FF ]\right) =\frac{\Tr[O]^2 + \norm{O}_{F}^2}{d(d+1)}\, .
\end{align}

Since $\Var(O) = \Var(O+\lambda\1)$ we assume w.l.o.g.\ that $\Tr[O]=0$. 
With these preliminaries, we then verify the claimed identity by a straightforward calculation, 
\begin{align}
\EE \left [ \Var_{\ket{\tilde \psi}}(O) \right]
&= 
\EE\left[ \sandwich{\tilde\psi}{O^2}{\tilde \psi} -\sandwich{\tilde\psi}{O}{\tilde\psi}^2 \right] = 
(1-\epsilon) \sandwich{\psi}{O^2}{\psi} + \epsilon \EE\left[ \sandwich{\xi}{O^2}{\xi} \right]  
- (1-\epsilon)^2\sandwich{\psi}{O}{\psi}^2 \nonumber\\
&\quad - \epsilon^2 \EE\left[ \sandwich{\xi}{O}{\xi}^2 \right]  
 - 4 \epsilon(1-\epsilon) \EE\left[ \sandwich{\psi}{O}{\xi}\sandwich{\xi}{O}{\psi} \right]  
   - 2\epsilon(1-\epsilon) \sandwich{\psi}{O}{\psi} \EE[\sandwich{\xi}{O}{\xi}]   \\
&= 
(1-\epsilon) \sandwich{\psi}{O^2}{\psi} + \epsilon\Tr\left[ O^2\EE\ketbra\xi\xi \right] 
- (1-\epsilon)^2\sandwich{\psi}{O}{\psi}^2 - \epsilon^2 \EE\left[ \sandwich{\xi}{O}{\xi}^2 \right] \nonumber \\
&\quad - 4 \epsilon(1-\epsilon)  \bra{\psi} O \EE[\ketbra\xi\xi] O\ket{\psi} 
 - 2\epsilon(1-\epsilon) \sandwich{\psi}{O}{\psi} \Tr\left[ O\EE\ketbra\xi\xi \right]  \\
&=
(1-\epsilon) \sandwich{\psi}{O^2}{\psi} - (1-\epsilon)^2\sandwich{\psi}{O}{\psi}^2
+ \epsilon\norm{O}_F^2/d
- \frac{4 \epsilon(1-\epsilon)}{d}  \sandwich{\psi}{O^2}{\psi} \
 \nonumber \\&\quad 
- \epsilon^2 \frac{\Tr[O]^2 + \norm{O}_F^2}{d(d+1)}  - \frac{2\epsilon(1-\epsilon)}{d} \sandwich{\psi}{O}{\psi} \Tr\left[ O \right]   
\\
&=
(1-\epsilon) \sandwich{\psi}{O^2}{\psi} - (1-\epsilon)^2\sandwich{\psi}{O}{\psi}^2
+ \epsilon\norm{O}_F^2/d
- \frac{4 \epsilon(1-\epsilon)}{d}  \sandwich{\psi}{O^2}{\psi}  
- \frac{\epsilon^2}{d(d+1)} \norm{O}_F^2 
\\
&=
(1-\epsilon) \Var_{\ket{ \psi}}(O) 
+ \epsilon (1-\epsilon)\sandwich{\psi}{O}{\psi}^2
+ \epsilon \norm{O}^2_F/d
- \frac{1}{d} \left(4 \epsilon (1-\epsilon)\sandwich{\psi}{O}{\psi}^2  + \epsilon^2 \norm{O}^2_\frac{F}{d+1}\right)
\, .
\label{eq: Full RUC}
\end{align}
{
In Eq.~\eqref{eq: Full RUC} the second and fourth, as well as the third and fifth term have the same structure apart from leading factors $4/d$ and $\epsilon/d$. Thus, the last two terms are suppressed with the Hilbert space dimension $d=2^n$, and we can simplify it to:
\begin{align}
\EE \left [ \Var_{\ket{\tilde \psi}}(O) \right ]
&= 
(1-\epsilon) \Var_{\ket{ \psi}}(O) 
+ \epsilon (1-\epsilon)\sandwich{\psi}{O}{\psi}^2
+ \epsilon \norm{O}^2_F/d
+ \mathcal O \left( {1}/{d} \right) \ .
\end{align}
}

Further note that for $O$ being a Hamiltonian $H$, a similar but simpler calculation tells us how the expectation value is gradually shifted from $E_i$ to average energy $\bar E$, $\ket{ \tilde \psi}  = \sqrt{1-\epsilon} \ket{E_i} + \sqrt \epsilon \ket \xi$: 
\begin{align}
\EE\left[ \sandwich{\tilde \psi}{H}{\tilde \psi} \right] = 
(1-\epsilon) E_i + \epsilon \EE\left[ \sandwich{\xi}{H}{\xi} \right] = 
(1-\epsilon) E_i + \epsilon \frac{1}{d} \Tr(H) = 
(1-\epsilon) E_i + \epsilon \overline E \ .
\end{align}
\subsection{{Relative Sampling complexity improvement under global depolarization noise (Proof of Theorem~\ref{thm:2})}}
\label{appendix:Theorem_2}
{
As in the main text, we will concentrate here on bi-partitions $H=H_1+H_2$, which apply for 1D and 2D geometric partitioning. At the same time, partitioning $H$ into more parts, e.g. $H_2 = H_{2a} + H_{2b} $, will always only increase the overall sampling error due to Eq.~\eqref{eq:linear lower bound}, compare also Appendix~\ref{appendix: compare measurement strategies}. Consequently, we will at worst underestimate the Pauli partitioning error. 

As geometric partitioning distributes the couplings equally between $H_1$ and $H_2$, we have,
\begin{align}
\Braket{E_i | H_{L_x, L_y}^{(1)} | E_i} = \Braket{E_i | H_{L_x, L_y}^{(2)} | E_i} = \frac{E_i}{2} &&
\norm{H_{L_x, L_y}^{(1)}}^2_F/d = \norm{H_{L_x, L_y}^{(2)}}^2_F/d \ .
\end{align}
and hence 
 \begin{align}
    \mathbb E \left [\Var_{\ket{\tilde E_i} } \left( H_{L_x, L_y}^{(1)} \right) \right ] = \mathbb E \left [\Var_{\ket{\tilde E_i} } \left( H_{L_x, L_y}^{(2)} \right) \right ] \ 
    \label{eq: appendix_variance_geometric_partitioning}
\end{align}

%
%
\textit{I. Low randomness:} The first statement to show is (Eq.~\eqref{eq:thm2_1}):
\begin{align}
\epsilon \leq \frac{4 \Var_{\ket{E_i}} \left( H_{L_x,L_y}^{(1)} \right)}{E_i^2 + \norm{H}^2_F/d } \quad \Rightarrow \quad 
    \overline{\mathcal G}_{\ket{ E_i}} \left(  \mathcal B_{\rm Pauli}^{\gamma \epsilon}, \mathcal B_{L_x, L_y}^\epsilon\right) \geq \frac{1}{2}
    \mathcal G_{\ket{ E_i}} \left(  \mathcal B_{\rm Pauli}, \mathcal B_{L_x, L_y}\right) \ .
\end{align}
From the requirement that the noise-free state-specific measurement should have lower than average case shot noise ($\norm{O}^2_F/d\geq \Var_{\ket{\psi}} (O)$), if follows that:
\begin{align}
\label{eq: exp_variance_appendix}
     \EE \left [ \Var_{\ket{\tilde \psi (\epsilon)}} (O)\right] &= (1-\epsilon)\Var_{\ket{\psi}} (O) + \epsilon(1-\epsilon)\Braket{\psi | O| \psi}^2 + \epsilon \norm{O}^2_F/d 
     \geq \Var_{\ket{\psi}} (O) + \epsilon \left[ \norm{O}^2_F/d - \Var_{\ket{\psi}} (O) \right ] \\
     &\geq \Var_{\ket{\psi}} (O)
     = \EE \left [ \Var_{\ket{\tilde \psi (\epsilon=0)}} (O)\right] \ .
\end{align}
Next, we need to linearize Eq.~\eqref{eq: average_sampling_error} when plugging in Eq.~\eqref{eq: exp_variance_appendix} from Lemma~\ref{thm:imperfect eigenstates}. The variance standard error of a partitioned observable $H=\sum H_b$ is given by:
\begin{align}
\label{eq: applied_inequality_sqrt_sum}
    \left[ \sum_{H_b \in \mathcal B}  \sqrt{ \EE \left[  \Var_{\ket{E_i(\epsilon)}} \left(  H_b\right) \right ] } \right ]^2 
    = \left[ \sum_{H_b \in \mathcal B}  
    \sqrt{ (1-\epsilon) \Var_{\ket{E_i}}(H_b)
    + \epsilon(1-\epsilon)\sandwich{E_i}{H_b}{E_i}^2 
    + \epsilon  \norm{H_b}^2_F/d  } \right ]^2 + \mathcal{O}(1/d) \ .
\end{align}
By labeling $A_b \coloneqq (1-\epsilon) \Var_{\ket{E_i}}(H_b)$, $B_b \coloneqq \epsilon(1-\epsilon)\sandwich{E_i}{H_b}{E_i}^2 $ and $C_b\coloneqq\epsilon  \norm{H_b}^2_F/d $, we simplify the notation. We will show
\begin{align}
    \left[ \sum_{b}  
    \sqrt{ A_b+B_b} \right ]^2 \geq \left[\sum_{b} \sqrt{ A_b} \right ]^2  + \left[\sum_{b} \sqrt{ B_b} \right ]^2 \ ,
\end{align}
and thus by a nested argument
\begin{align}
    \left[ \sum_{b}  
    \sqrt{ A_b+B_b+C_b} \right ]^2 
    \geq \left[ \sum_{b}\sqrt{ A_b} \right ]^2  +  \left[ \sum_{b} \sqrt{B_b+C_b} \right ]^2 
    \geq \left[ \sum_{b} \sqrt{ A_b} \right ]^2  +  \left[ \sum_{b} \sqrt{B_b} \right ]^2 + \left[ \sum_{b} \sqrt{C_b} \right ]^2 \ .
\end{align}
Note that one is free to relabel $A_b$, $B_b$, and $C_b$ to maximize the lower bound. We have
\begin{align}
    \left[ \sum_{b}  
    \sqrt{ A_b+B_b} \right ]^2 -\left[\sum_{b} \sqrt{ A_b} \right ]^2  - \left[\sum_{b} \sqrt{ B_b} \right ]^2 = \sum_{b\neq b^\prime} \left[ \sqrt{A_b+B_b}\sqrt{A_{b^\prime}+B_{b^\prime}} - \sqrt{A_bA_{b^\prime}} - \sqrt{B_{b^\prime}B_{b^\prime}}  \right] \ .
\end{align}
Next, we will show that each term of the sum on the r.h.s. is larger or equal 0:
\begin{align}
    \left[ \sqrt{A_bB_{b^\prime}}  - \sqrt{B_bA_{b^\prime}}  \right]^2 &\geq 0\\ 
    \Leftrightarrow A_bB_{b^\prime} + B_bA_{b^\prime} - 2 \sqrt{A_bB_{b^\prime}B_bA_{b^\prime}}  &\geq 0 \\
    \Leftrightarrow A_bB_{b^\prime} + B_bA_{b^\prime} + A_bB_b + A_{b^\prime}B_{b^\prime}&\geq 2 \sqrt{A_bB_{b^\prime}B_bA_{b^\prime}} + A_bB_b + A_{b^\prime}B_{b^\prime}\\
    \Leftrightarrow (A_b + B_b)( A_{b^\prime}+B_{b^\prime}) &\geq (\sqrt{A_bA_{b^\prime}} + \sqrt{B_bB_{b^\prime}})^2 \ .
\end{align}
Thus Eq.~\eqref{eq: applied_inequality_sqrt_sum} can be lower bounded by:
\begin{align}
    &\left[ \sum_{H_b \in \mathcal B}  \sqrt{ \EE \left[  \Var_{\ket{E_i(\epsilon)}} \left(  H_b\right) \right ] } \right ]^2 \nonumber \\
    &\geq 
    (1-\epsilon) \left[ \sum_{H_b \in \mathcal B}  
    \sqrt{ \Var_{\ket{E_i}}(H_b) } \right ]^2 
    + \epsilon(1-\epsilon) \left[ \sum_{H_b \in \mathcal B} \sqrt{ \sandwich{E_i}{H_b}{E_i}^2 } \right ]^2
    + \epsilon  \left[ \sum_{H_b \in \mathcal B}\sqrt{  \norm{H_b}^2_F/d  } \right ]^2 + \mathcal{O}(1/d) \ .
    \label{eq: appendix_linarized_variance}
\end{align}
For the proofs of the shot noise improvements under global depolarization noise, we will use Eq.~\eqref{eq: appendix_linarized_variance} extensively. 
Using further Eq.~\eqref{eq: appendix_variance_geometric_partitioning} for the denominator, we get
\begin{align}
\label{eq:appendix_thm2_1}
\overline{\mathcal G}_{\ket{ E_i}} \left(  \mathcal B_{\rm Pauli}^{\gamma \epsilon}, \mathcal B_{L_x, L_y}^\epsilon\right) 
&\geq \overline{\mathcal G}_{\ket{ E_i}} \left(  \mathcal B_{\rm Pauli}^{0 \epsilon}, \mathcal B_{L_x, L_y}^\epsilon\right) 
= \frac{4 \Var_{\ket{E_i}} \left( H_{\rm Pauli}^{(1)} \right)}{4 (1-\epsilon)\Var_{\ket{E_i}} \left( H_{L_x,L_y}^{(1)} \right) + \epsilon (1-\epsilon)E_i^2 + \epsilon (1+\beta)\norm{H}^2_F/d } \nonumber \\
&\geq \frac{4 \Var_{\ket{E_i}} \left( H_{\rm Pauli}^{(1)} \right)}{4 (1-\epsilon)\Var_{\ket{E_i}} \left( H_{L_x,L_y}^{(1)} \right) + \epsilon (1-\epsilon)E_i^2 + \epsilon \norm{H}^2_F/d } ,
\end{align}
where $\beta_{L_x,L_y}  \coloneqq \left| \norm{ H_{L_x,L_y}^{(1)} }^2_F/\norm{ H }^2_F -1/2  \right | \leq 1$ gives the average asymmetry of squared Pauli weights across the partitions. We want to show that the r.h.s. of \eqref{eq:appendix_thm2_1} is bounded from below by $(1/2) \, \mathcal{G}_{\ket{ E_i}} \left(  \mathcal B_{\rm Pauli}, \mathcal B_{L_x, L_y}\right)$. Hence it is sufficient to show that,
\begin{align}
4 (1-\epsilon)\Var_{\ket{E_i}} \left( H_{L_x,L_y}^{(1)} \right) + \epsilon (1-\epsilon)E_i^2 + \epsilon \norm{H}^2_F/d \leq 2 \cdot 4 \Var_{\ket{E_i}} \left( H_{L_x,L_y}^{(1)} \right) \ .
\end{align}
When plugging in $\epsilon \leq \frac{4 \Var_{\ket{E_i}} \left( H_{L_x,L_y}^{(1)} \right)}{E_i^2 + \norm{H}^2_F/d }$, we obtain:
\begin{align}
4 (1-\epsilon)\Var_{\ket{E_i}} \left( H_{L_x,L_y}^{(1)} \right) + \epsilon (1-\epsilon)E_i^2 + \epsilon \norm{H}^2_F/d 
& \leq 4 \Var_{\ket{E_i}} \left( H_{L_x,L_y}^{(1)} \right) \left[2 -\epsilon \frac{2E_i^2+ \norm{H}^2_F/d}{E_i^2 + \norm{H}^2_F/d } \right] \\
& \leq 2 \cdot 4 \Var_{\ket{E_i}} \left( H_{L_x,L_y}^{(1)} \right) \ .
\end{align}

%
%
\textit{II. Intermediate randomness:} Eq.~\eqref{eq:thm2_2a} follows from 
\begin{align}
\overline{\mathcal G}_{\ket{ E_i}} \left(  \mathcal B_{L_x, L_y}^\epsilon,  \mathcal B_{H}^\epsilon\right) 
&=
\frac{4 (1-\epsilon)\Var_{\ket{E_i}} \left( H_{L_x,L_y}^{(1)} \right) + \epsilon (1-\epsilon)E_i^2 + 4 \epsilon \norm{H_{L_x,L_y}^{(1)}}^2_F/d }{\epsilon (1-\epsilon)E_i^2 + \epsilon \norm{H}^2_F/d} \\
&\leq \frac{4 \Var_{\ket{E_i}} \left( H_{L_x,L_y}^{(1)} \right) + \epsilon (1-\epsilon)E_i^2 }{\epsilon (1-\epsilon)E_i^2 + \epsilon \norm{H}^2_F/d} 
\leq 1 + \frac{4 \Var_{\ket{E_i}} \left( H_{L_x,L_y}^{(1)} \right)}{\epsilon (1-\epsilon)E_i^2 + \epsilon \norm{H}^2_F/d} \stackrel{!}{\leq}2 \ .
\label{eq:appendix_thm_2_2a}
\end{align}
This is true for 
\begin{align}
\epsilon \geq \frac{4 \Var_{\ket{E_i}} \left( H_{L_x,L_y}^{(1)} \right)}{(1-\epsilon)E_i^2 + \norm{H}^2_F/d}\geq \frac{4 \Var_{\ket{E_i}} \left( H_{L_x,L_y}^{(1)} \right)}{E_i^2 + \norm{H}^2_F/d} \ .
\end{align}
By a similar calculation one shows Eq.~\eqref{eq:thm2_2b}:
\begin{align}
\overline{\mathcal G}_{\ket{ E_i}} \left(  \mathcal B_{\rm Pauli}^{\gamma \epsilon},  \mathcal B_{H}^\epsilon\right) 
&\geq
\frac{4 (1-\gamma\epsilon)\Var_{\ket{E_i}} \left( H_{\rm Pauli}^{(1)} \right) + \gamma\epsilon (1-\gamma\epsilon)(1+\alpha_{\rm Pauli}^2) E_i^2 + \gamma\epsilon (1+\beta_{\rm Pauli}) \norm{H}^2_F/d }{\epsilon (1-\epsilon)E_i^2 + \epsilon \norm{H}^2_F/d} \\
&\geq
\frac{4 (1-\epsilon)\Var_{\ket{E_i}} \left( H_{\rm Pauli}^{(1)} \right)}{\epsilon (1-\epsilon)E_i^2 + \epsilon \norm{H}^2_F/d} 
+\gamma
\frac{(1-\epsilon) E_i^2 + \norm{H}^2_F/d }{ (1-\epsilon)E_i^2 +  \norm{H}^2_F/d} 
\stackrel{!}{\geq} 2 \ ,
\label{eq:appendix_thm_2_2b}
\end{align}
where we used $\gamma \leq 1$,  $\alpha_{\rm Pauli}  \coloneqq \left| \Braket{ H_{\rm Pauli}^{(1)} }/E_i -1/2 \right | \geq 0$ and $\beta_{\rm Pauli}  \coloneqq \left| \norm{ H_{\rm Pauli}^{(1)} }^2_F/\norm{ H }^2_F -1/2  \right | \geq 0$. Eq.~\eqref{eq:appendix_thm_2_2b} is true for:
\begin{align}
\epsilon \leq \frac{1}{2-\gamma}\frac{4 \Var_{\ket{E_i}} \left( H_{\rm Pauli}^{(1)} \right)}{E_i^2 + \frac{1}{1-\epsilon}\norm{H}^2_F/d}
\leq \frac{1}{2-\gamma} \frac{4 \Var_{\ket{E_i}} \left( H_{\rm Pauli}^{(1)} \right)}{E_i^2 + \norm{H}^2_F/d} \ .
\end{align}
%
%
\textit{III. Random dominated.}
Let $\mathcal B_{\rm bi-part.}^{\gamma \epsilon}$ be a bi-partition of $H$ ($H=H_1+H_2$) and $\overline{\mathcal G}_{\ket{\tilde E_i}} \left(  \mathcal B_{\rm bi-part.}^{\gamma \epsilon}, \mathcal B_{H}^\epsilon\right) \leq C$.
\begin{align}
\overline{\mathcal G}_{\ket{\tilde E_i}} \left(  \mathcal B_{\rm bi-part.}^{\gamma \epsilon}, \mathcal B_{H}^\epsilon\right) \leq \frac{4 (1-\gamma\epsilon)\Var_{\ket{E_i}} \left( H_{\rm bi-part.}^{(1)} \right) + \gamma\epsilon (1-\gamma\epsilon)2 E_i^2 + \gamma\epsilon 2 \norm{H}^2_F/d }{\epsilon (1-\epsilon)E_i^2 + \epsilon \norm{H}^2_F/d} \stackrel{!}{\leq} C \ ,
\label{eq:appendix_thm_2_3}
\end{align}
where we used $\Var_{\ket{E_i}} \left( H_{\rm bi-part.}^{(1)} \right) = \Var_{\ket{E_i}} \left( H_{\rm bi-part.}^{(2)}\right)$, $\alpha_{\rm bi-part.} \leq 1$ and $\beta_{\rm bi-part.} \leq 1$.
Eq.~\eqref{eq:appendix_thm_2_3} holds for:
\begin{align}
C - 2\gamma \geq \frac{4 \Var_{\ket{E_i}} \left( H_{\rm bi-part.}^{(1)} \right) }{\epsilon (1-\epsilon)E_i^2 + \epsilon \norm{H}^2_F/d} \geq \frac{4 \Var_{\ket{E_i}} \left( H_{\rm bi-part.}^{(1)} \right) }{\epsilon E_i^2 + \epsilon \norm{H}^2_F/d} \Rightarrow \epsilon \geq \frac{1}{C-2\gamma}\frac{ 4 \Var_{\ket{E_i}} \left( H_{\rm bi-part.}^{(1)} \right)}{E_i^2 + \norm{H}^2_F/d } \ .
\end{align}

\subsection{Criteria for sampling advantage of geometric partitioning under global depolarization (Proof of Corollary~\ref{lemma:noise-indep sample advantage})}
\label{appendix:Lemma_4}

We have 
\begin{align}
&\overline{\mathcal G}_{\ket{ E_i}} \left(  \mathcal B_{\rm Pauli}^{\gamma \epsilon}, \mathcal B_{L_x, L_y}^\epsilon\right) 
= \frac{\left[ \sum_{H_b \in \mathcal B_{\rm Pauli}}  \sqrt{ \EE \left[  \Var_{\ket{E_i(\gamma\epsilon)}} \left(  H_b\right) \right ] } \right ]^2 }{4\EE \left[  \Var_{\ket{E_i(\epsilon)}} \left(  H_{L_x,L_y}^{(1)}\right) \right ] }\\
&\geq \min \left[ \frac{(1-\gamma\epsilon)}{(1-\epsilon)} \mathcal G_{\ket{ E_i}} \left(  \mathcal B_{\rm Pauli}, \mathcal B_{L_x, L_y}\right)  
, \frac{\gamma \epsilon(1-\gamma\epsilon)}{\epsilon(1-\epsilon)} \frac{ (1+\alpha^2_{\rm Pauli})E_i^2}{E_i^2}
, \frac{\gamma \epsilon}{\epsilon} \left( \frac{\sum_{H_b \in \mathcal B_{\rm Pauli}} \norm{H_b}_F }{\sum_{H_b^\prime \in \mathcal B_{L_x, L_y} } \norm{H_b^\prime}_F }\right)^2 \right] \ ,
\label{eq:appendix_lemma_4}
\end{align}
where we first used Eq.~\eqref{eq: appendix_linarized_variance} on the numerator and then iteratively Eq.~\eqref{eq: frac_inequality} ($\frac{A+A^\prime}{B+B^\prime} \geq min (\frac{A}{B},\frac{A^\prime}{B^\prime})$) to separate into variance-, $E_i^2$- and norm-type fractions. As $ \mathcal G_{\ket{ E_i}} \left(  \mathcal B_{\rm Pauli}, \mathcal B_{L_x, L_y}\right)  > 1$ (Theorem~\ref{thm:improvement}), we get $\overline{\mathcal G}_{\ket{ E_i}} \left(  \mathcal B_{\rm Pauli}^{\gamma \epsilon}, \mathcal B_{L_x, L_y}^\epsilon\right) \geq 1$ for:
\begin{align}
\gamma (1+\alpha^2_{\rm Pauli}) \geq 1 \qquad {\rm and} \qquad \gamma \left( \frac{\sum_{H_b \in \mathcal B_{\rm Pauli}} \norm{H_b}_F }{\sum_{H_b^\prime \in \mathcal B_{L_x, L_y} } \norm{H_b^\prime}_F }\right)^2  \geq 1 \ .
\label{eq:appendix_lemma_4_cond}
\end{align}
}

Alternatively, we can also find a criterion for $\epsilon$ so that
$\overline{\mathcal G}_{\ket{ E_i}} \left(  \mathcal B_{\rm Pauli}^{\gamma \epsilon}, \mathcal B_{L_x, L_y}^\epsilon\right) \geq 1$, similarly to Eq.~\eqref{eq:appendix_thm2_1} we have:
\begin{align}
    &\overline{\mathcal G}_{\ket{ E_i}} \left(  \mathcal B_{\rm Pauli}^{\gamma \epsilon}, \mathcal B_{L_x, L_y}^\epsilon\right) \geq \frac{4 \Var_{\ket{E_i}} \left( H_{\rm Pauli}^{(1)} \right) (1 + \delta - \delta)}{4 (1-\epsilon)\Var_{\ket{E_i}} \left( H_{L_x,L_y}^{(1)} \right) + \epsilon (1-\epsilon)E_i^2 + \epsilon \norm{H}^2_F/d } \ .
\end{align}
Next, we will use Eq.~\eqref{eq: frac_inequality} again: 
\begin{align}
    &\overline{\mathcal G}_{\ket{ E_i}} \left(  \mathcal B_{\rm Pauli}^{\gamma \epsilon}, \mathcal B_{L_x, L_y}^\epsilon\right) \geq 
    \min \left (\frac{\delta \mathcal G_{\ket{ E_i}} \left(  \mathcal B_{\rm Pauli}, \mathcal B_{L_x, L_y}\right)}{1-\epsilon},   
    \frac{4 \Var_{\ket{E_i}} \left( H_{\rm Pauli}^{(1)} \right) (1 - \delta)}{\epsilon (1-\epsilon) E_i^2 + \epsilon \norm{H}^2_F/d }\right) \stackrel{!}{\geq}1 \ ,
    \label{eq:appendix_thm_2_2c}
\end{align}
if we set $\delta = 1/\mathcal G_{\ket{ E_i}} \left(  \mathcal B_{\rm Pauli}, \mathcal B_{L_x, L_y}\right)$ then the first term will always be larger than 1 as $\frac{1}{1-\epsilon} \geq 1$, while the second term will be larger 1 for
\begin{align}
\frac{1}{\epsilon}
\frac{ 4 \Var_{\ket{E_i}} \left( H_{\rm Pauli}^{(1)} \right) (1-\delta)}{(1-\epsilon) E_i^2 + \norm{H}^2_F/d } \geq \frac{1- \delta}{\epsilon} 
\frac{4 \Var_{\ket{E_i}} \left( H_{\rm Pauli}^{(1)} \right)}{E_i^2 + \norm{H}^2_F/d }  \stackrel{!}{\geq}1
\quad \Rightarrow \quad 
\epsilon \leq \left(1 - \frac{1}{\mathcal G_{\ket{ E_i}} \left(  \mathcal B_{\rm Pauli}, \mathcal B_{L_x, L_y}\right)} \right ) \frac{4 \Var_{\ket{E_i}} \left( H_{\rm Pauli}^{(1)} \right)}{E_i^2 + \norm{H}^2_F/d}
\ .
\end{align}

\subsection{Gate budget for noisy geometric partitioning (Proof of Corollary~\ref{corol: gate_budget})}
\label{appendix:Corollary_4}
{
Here, we use Theorem~\ref{thm:2} to derive a criterion that allows a sample improvement of at least a factor $G$:
\begin{align}
    \overline{\mathcal G}_{\ket{ E_i}} \left(  \mathcal B_{\rm Pauli}^{\gamma \epsilon}, \mathcal B_{L_x, L_y}^\epsilon\right) \geq G
\end{align}
For this, we replace the bound in Theorem~\ref{thm:2} in Eq.~\eqref{eq:appendix_thm_2_2c} (main text Eq.~\eqref{eq:thm2_2c}) "$\geq 1$" with "$\geq G$". This gives:
\begin{align}
\epsilon \leq \left(1 - \frac{G}{\mathcal G_{\ket{ E_i}} \left(  \mathcal B_{\rm Pauli}, \mathcal B_{L_x, L_y}\right)} \right) \epsilon_{II \leftrightarrow III}
\end{align}
Further, we keep Eq.~\eqref{eq:appendix_thm_2_2a} (main text Eq.~\eqref{eq:thm2_2a}) and replace the "2" at the r.h.s. with $2G$. This gives the alternative inequality
\begin{align}
\epsilon \leq \frac{1}{2G - \gamma} \epsilon_{II \leftrightarrow III}
\end{align}
By choosing the maximum of both, $c_{\epsilon} = \max(\frac{1}{2G-\gamma},  \,  \left[ 1 - {G}/{\mathcal G_{\ket{ E_i}} \left(  \mathcal B_{\rm Pauli}, \mathcal B_{L_x, L_y}\right)} \right ])$, we get
\begin{align}
    &1-(1-\epsilon_{\rm gate})^{\mathcal{N}_{2q}(U_{\ket{E_i}}) + \mathcal{N}_{2q}\left( U_{L_x,L_y} \right)} \approx \epsilon \leq c_{\epsilon} \epsilon_{II \leftrightarrow III} 
     & \Leftrightarrow \quad 
     {\mathcal{N}_{2q}(U_{\ket{E_i}}) + \mathcal{N}_{2q}\left( U_{L_x,L_y} \right)} \lessapprox 
     \frac{\log\left( 1 - c_\epsilon \epsilon_{II \leftrightarrow III} \right)}{\log(1- \epsilon_{\rm gate})} 
\end{align}
}
\section{Additional material for Numerical tests}
\label{appendix:TFIM}
%
%

\subsection{Proof of Lemma~\ref{lemma:TFIM improvements} "Disordered phase" $h \gg J$}
We normalize the Hamiltonian to obtain (this does not change the variance ratio):
\begin{align}
H = - \sum_i X_i -\left( \frac{J}{h} \right) \sum_{\left< i,j\right >} Z_i Z_j =  - \sum_i X_i -\lambda \sum_{\left< i,j\right >} Z_i Z_j  = H_0 + \lambda V \ .
 \end{align}

 The ground state $\ket{E_0}$ is given up to second order in $\lambda$ by:
\begin{align}
 \ket{E_0^{(0)}} &= \ket{+}^{\otimes n} \\
 \ket{E_0^{(1)}} &= \sum_{\ket{b} \ne \ket{+}^{\otimes n}} \frac{\Braket{ b | \sum_{\left< i,j\right >} Z_i Z_j | +^{\otimes n} }}{-n - (n-2\cdot2)} \ket{b}
 = \frac{1}{4} \sum_{\left< i,j\right >} Z_i Z_j \ket{+}^{\otimes n}\\
\ket{E_0^{(2)}} &= \sum_{\ket{b_1} \ne \ket{+}^{\otimes n}}\sum_{\ket{b_2} \ne \ket{+}^{\otimes n}} \ket{b_1}\frac{\Braket{b_1 | \sum_{\left< i,j\right >} Z_i Z_j  | b_2} \Braket{b_2 | \sum_{\left< i,j\right >} Z_i Z_j | +^{\otimes n}}}{\left (E_{\ket{+}^{\otimes n}}^{(0)}-E_{\ket{b_1}}^{(0)}\right ) \left (E_{\ket{+}^{\otimes n}}^{(0)}-E_{\ket{b_2}}^{(0)} \right )} - \frac{1}{2}\ket{+}^{\otimes n} \frac{(2n)}{4^2}\\
  &=\left[ \frac{2}{4\cdot 8} \sum_{\substack{\left< i,j\right > ,  \left< k,\ell \right > \\ i,j \neq k,\ell}} (Z_i Z_j) (Z_k Z_\ell)+ \frac{4}{16} \sum_{\left< \left< i,j\right > \right >_{\text{diagonal}}} Z_i Z_j + \frac{2}{16} \sum_{\left< \left< i,j\right > \right >_{\text{axial}}} Z_i Z_j  - \frac{n}{16} \mathds{1} \right] \ket{+}^{\otimes n}\\
 \ket{E_0} &= \left[ \mathds{1} + \frac{\lambda}{4} \sum_{\left< i,j\right >} Z_i Z_j   + \frac{\lambda^2}{16} \left(  \sum_{\substack{\left< i,j\right > ,  \left< k,\ell \right > \\ i,j \neq k,\ell}} (Z_i Z_j) (Z_k Z_\ell)+ 4 \sum_{\left< \left< i,j\right > \right >_{\text{diagonal}}} Z_i Z_j + 2 \sum_{\left< \left< i,j\right > \right >_{\text{axial}}} Z_i Z_j - n \mathds{1} \right) \right] \ket{+}^{\otimes n}  + \mathcal O (\lambda^3)\\
 &=P_X^{(2)}\ket{+}^{\otimes n}  + \mathcal O (\lambda^3)
 \end{align}
Note that $\sum_{\left< \left< i,j\right > \right >_{\text{axial}}} $ obtains a factor 2 if $n_x, n_y=4$ and does not exist for $n_x, n_y=3$.  Let's start with calculating the variance of $H_0$ up to second order
 \begin{align}
\Var(H_0) &= \Braket{(H_0)^2} - \Braket{H_0}^2 \\
\Braket{H_0} &= \Bra{+}^{\otimes n} 
 \left[ \mathds{1} + \frac{\lambda}{4} \sum_{\left< i,j\right >} Z_i Z_j   + \frac{\lambda^2}{16} \left(  \sum_{\substack{\left< i,j\right > ,  \left< k,\ell \right > \\ i,j \neq k,\ell}} (Z_i Z_j) (Z_k Z_\ell)+ 4 \sum_{\substack{ \left< \left< i,j\right > \right > \\ \text{diagonal}}} Z_i Z_j + 2 \sum_{\substack{ \left< \left< i,j\right > \right > \\ \text{axial}}} Z_i Z_j - n \mathds{1} \right) \right] \\
&\cdot \sum_i X_i 
 \left[ \mathds{1} + \frac{\lambda}{4} \sum_{\left< i,j\right >} Z_i Z_j   + \frac{\lambda^2}{16} \left(  \sum_{\substack{\left< i,j\right > ,  \left< k,\ell \right > \\ i,j \neq k,\ell}} (Z_i Z_j) (Z_k Z_\ell)+ 4 \sum_{\substack{ \left< \left< i,j\right > \right > \\ \text{diagonal}}} Z_i Z_j + 2 \sum_{\substack{ \left< \left< i,j\right > \right > \\ \text{axial}}} Z_i Z_j - n \mathds{1} \right) \right]
\Ket{+}^{\otimes n} + \mathcal O (\lambda^3)\\
&= n + \lambda^2 \left( \frac{1}{4^2} (2n (n-2 -2 )) - 2\frac{n^2}{16} \right) + \mathcal O (\lambda^3) 
=n - \frac{\lambda^2}{2} n   + \mathcal O (\lambda^3) 
  \end{align}
  
   \begin{align}
\Braket{(H_0)^2}  &= \Bra{+}^{\otimes n} 
 \left[ \mathds{1} + \frac{\lambda}{4} \sum_{\left< i,j\right >} Z_i Z_j   + \frac{\lambda^2}{16} \left(  \sum_{\substack{\left< i,j\right > ,  \left< k,\ell \right > \\ i,j \neq k,\ell}} (Z_i Z_j) (Z_k Z_\ell)+ 4 \sum_{\substack{ \left< \left< i,j\right > \right > \\ \text{diagonal}}} Z_i Z_j + 2 \sum_{\substack{ \left< \left< i,j\right > \right > \\ \text{axial}}} Z_i Z_j - n \mathds{1} \right) \right]
\left[ n \mathds{1} + \sum_{i \neq j} X_i X_j \right]\\
 &\cdot \left[ \mathds{1} + \frac{\lambda}{4} \sum_{\left< i,j\right >} Z_i Z_j   + \frac{\lambda^2}{16} \left(  \sum_{\substack{\left< i,j\right > ,  \left< k,\ell \right > \\ i,j \neq k,\ell}} (Z_i Z_j) (Z_k Z_\ell)+ 4 \sum_{\substack{ \left< \left< i,j\right > \right > \\ \text{diagonal}}} Z_i Z_j + 2 \sum_{\substack{ \left< \left< i,j\right > \right > \\ \text{axial}}} Z_i Z_j - n \mathds{1} \right) \right]
\Ket{+}^{\otimes n} + \mathcal O (\lambda^3)\\
&=n^2 + \lambda^2 \left(\frac{2n[n+(n-2)(n-3) +2 -4 (n-2)]}{16} -\frac{2n^3}{16}\right)+ \mathcal O (\lambda^3) 
=n^2 + \frac{\lambda^2}{16} \left(-16n^2 +32n \right)+ \mathcal O (\lambda^3) 
  \end{align}

   \begin{align}
\Var(H_0) &= \Braket{(H_0)^2} - \Braket{H_0}^2= n^2 + \lambda^2 \left(2n - n^2 \right)- \left[ n^2 - n^2 \lambda^2 \right] + \mathcal O (\lambda^3) = 2n \lambda^2  + \mathcal O (\lambda^3)
\end{align}

Next lets calculate (For $L>1$):
\begin{align}
\Var( H_l - H_r) &= \Braket{(H_l - H_r)^2} - \Braket{H_l - H_r}^2 = 2 \Braket{(H_l)^2} - 2\Braket{H_l H_r} \\
H_l &= \lambda \sum_{i=1}^{n_x/L} \sum_{j=1}^{n_y} Z_{Li, j}Z_{Li+1, j} \qquad \qquad
H_r =\lambda  \sum_{i=1}^{n_x/L} \sum_{j=1}^{n_y} Z_{Li+1, j}Z_{Li+2, j}
\end{align}
This calculation is also true for $L=3$ as there the axial next nearest neighbors would be neighbors, but the term in the expansion does not exist then.  
\begin{align}
&\left( \bra{E_0} H_l \right) \left(H_l  \ket{E_0} \right)= \\
 &\Bra{+}^{\otimes n} 
 \left[ \mathds{1} + \frac{\lambda}{4} \sum_{\left< i,j\right >} Z_i Z_j   + \frac{\lambda^2}{16} \left(  \sum_{\substack{\left< i,j\right > ,  \left< k,\ell \right > \\ i,j \neq k,\ell}} (Z_i Z_j) (Z_k Z_\ell)+ 4 \sum_{\substack{ \left< \left< i,j\right > \right > \\ \text{diagonal}}} Z_i Z_j + 2 \sum_{\substack{ \left< \left< i,j\right > \right > \\ \text{axial}}} Z_i Z_j - n \mathds{1} \right) \right]\\
&\cdot \lambda^2 \left[ \frac{n}{L} \mathds{1} + \sum_{i=1}^{n_x/L} \sum_{j=1}^{n_y} \sum_{k=1}^{n_x/L} \sum_{\ell=1, (i,j)\neq (k,\ell)}^{n_y}Z_{Li, j}Z_{Li+1, j} Z_{Lk, \ell}Z_{Lk+1, \ell} \right]\\
 & \cdot \left[ \mathds{1} + \frac{\lambda}{4} \sum_{\left< i,j\right >} Z_i Z_j   + \frac{\lambda^2}{16} \left(  \sum_{\substack{\left< i,j\right > ,  \left< k,\ell \right > \\ i,j \neq k,\ell}} (Z_i Z_j) (Z_k Z_\ell)+ 4 \sum_{\substack{ \left< \left< i,j\right > \right > \\ \text{diagonal}}} Z_i Z_j + 2 \sum_{\substack{ \left< \left< i,j\right > \right > \\ \text{axial}}} Z_i Z_j - n \mathds{1} \right) \right]
\Ket{+}^{\otimes n} + \mathcal O (\lambda^3)\\
&= \frac{n}{L} \lambda^2 + \frac{\lambda^4}{16} \left(\frac{n}{L} 2 \cdot 2n + 2\cdot  \frac{n}{L}  \cdot \frac{n-1}{L} - 2\cdot  \frac{n^2}{L} + 2 \cdot 2 \cdot  \frac{n}{L}  \cdot \frac{n-1}{L} \right) + \mathcal O (\lambda^5)\\
&= \frac{n}{L} \lambda^2 + \frac{\lambda^4}{16} \frac{n^2}{L}  \left(4 +  \frac{2}{L}  - \frac{2}{nL}  - 2 +  \frac{4}{L}  - \frac{4}{nL} \right) + \mathcal O (\lambda^5)
= \frac{n}{L}\lambda^2 + \frac{\lambda^4}{16} \frac{n^2}{L}  \left(2 +  \frac{6}{L} \left( 1  - \frac{1}{n} \right)  \right) + \mathcal O (\lambda^5)
\end{align}

\begin{align}
&\left( \bra{E_0} H_l \right) \left(H_r  \ket{E_0} \right)= \\
 &\Bra{+}^{\otimes n} 
 \left[ \mathds{1} + \frac{\lambda}{4} \sum_{\left< i,j\right >} Z_i Z_j   + \frac{\lambda^2}{16} \left(  \sum_{\substack{\left< i,j\right > ,  \left< k,\ell \right > \\ i,j \neq k,\ell}} (Z_i Z_j) (Z_k Z_\ell)+ 4 \sum_{\substack{ \left< \left< i,j\right > \right > \\ \text{diagonal}}} Z_i Z_j + 2 \sum_{\substack{ \left< \left< i,j\right > \right > \\ \text{axial}}} Z_i Z_j - n \mathds{1} \right) \right]\\
&\cdot \left[ \lambda \sum_{i=1}^{n_x/L} \sum_{j=1}^{n_y} Z_{Li, j}Z_{Li+1, j} \right]\left[\lambda \sum_{i=1}^{n_x/L} \sum_{j=1}^{n_y} Z_{Li+1, j}Z_{Li+2, j} \right]\\
 &\cdot \left[ \mathds{1} + \frac{\lambda}{4} \sum_{\left< i,j\right >} Z_i Z_j   + \frac{\lambda^2}{16} \left(  \sum_{\substack{\left< i,j\right > ,  \left< k,\ell \right > \\ i,j \neq k,\ell}} (Z_i Z_j) (Z_k Z_\ell)+ 4 \sum_{\substack{ \left< \left< i,j\right > \right > \\ \text{diagonal}}} Z_i Z_j + 2 \sum_{\substack{ \left< \left< i,j\right > \right > \\ \text{axial}}} Z_i Z_j - n \mathds{1} \right) \right]
\Ket{+}^{\otimes n} + \mathcal O (\lambda^3 \lambda^2) \\
&= \mathcal O (\lambda^4) 
\end{align}

Thus we have:
\begin{align}
\Var(H_0) &= 2n \lambda^2 + \mathcal O (\lambda^3) \\
\Var( H_l - H_r) &= 2\cdot\frac{n}{L}\lambda^2 + \mathcal O (\lambda^4) \\
\frac{\Var(H_0)}{\Var(\frac{1}{2} (H- H_l + H_r))} & = 4 \frac{\Var(H_1)}{\Var( H_l - H_r)} = 4 \frac{2n \lambda^2}{2\cdot\frac{n}{L}\lambda^2} + \mathcal O (\lambda) = 4L + \mathcal O (\lambda) 
\end{align}
\subsection{"Ordered phase" $h << J$}
We normalize the Hamiltonian to obtain (this does not change the variance ratio):
\begin{align}
H = -\sum_{\left< i,j\right >} Z_i Z_j - \left( \frac{h}{J} \right) \sum_i X_i =  -\sum_{\left< i,j\right >} Z_i Z_j - \lambda \sum_i X_i = H_0 + \lambda V
\end{align}

Then we can calculate the perturbed state order by order:
\begin{align}
\frac{\Var(H_2)}{\Var(\frac{1}{2} (H- H_l + H_r))} = 4 \frac{\Var(H_1)}{\Var( H_l - H_r)}
 \end{align}

We now want to calculate 
\begin{align}
 \ket{E_0^{(0)}} &= \ket{0}^{\otimes n} \\
 \ket{E_0^{(1)}} &= \sum_{\ket{b} \ne \ket{0}^{\otimes n}} \frac{\Braket{ b | \sum_i X_i | 0^{\otimes n} }}{-2n - E_k^{(0)}} \ket{b}
 = \frac{1}{8} \sum_i X_i \ket{0}^{\otimes n}\\
 \ket{E_0^{(2)}} &=  \sum_{\ket{b_1} \ne \ket{0}^{\otimes n}}\sum_{\ket{b_2} \ne \ket{0}^{\otimes n}} \ket{b_1}\frac{\Braket{b_1 | \sum_i X_i  | b_2} \Braket{b_2 | \sum_i X_i | 0^{\otimes n}}}{\left (E_n^{(0)}-E_k^{(0)}\right ) \left (E_n^{(0)}-E_\ell^{(0)} \right )} \\
& -\sum_{k\neq n}\left |k^{(0)}\right\rangle \frac{\left \langle k^{(0)} \right |V\left |n^{(0)} \right \rangle \left \langle  0^{\otimes n} | \sum_i X_i | 0^{\otimes n}  \right \rangle}{\left (E_n^{(0)}-E_k^{(0)} \right )^2} 
- \frac{1}{2}\ket{0}^{\otimes n} \frac{n}{8^2} \\
&=  \left( \frac{2}{8\cdot2\cdot6} \sum_{\left <i, j \right>} X_iX_j + \frac{1}{8\cdot2\cdot8} \sum_{i \neq j \backslash  \left <i, j \right>} X_iX_j - \frac{1}{128} n \mathds{1} \right)\ket{0}^{\otimes n} \\ 
 \ket{E_0} &= \left[ \mathds{1} + \frac{\lambda}{8} \sum_i X_i + \frac{\lambda^2}{16} \left( \frac{1}{3} \sum_{\left <i, j \right>} X_iX_j + \frac{1}{8} \sum_{i \neq j \backslash  \left <i, j \right>} X_iX_j - \frac{1}{8} n \mathds{1}\right) \right] \ket{0}^{\otimes n}  + \mathcal O (\lambda^3)
 \end{align}
Note here that the $\sum_{\left <i, j \right>}$ receives a factor 2 as then there is no double counting of lattice sides any more, while $\sum_{i \neq j \backslash  \left <i, j \right>} $ still has every term twice. 

Further note that we defined $H_r = T_{-1} H_l T_1$. In the $L=1$ case we use instead $H_r = S H_l S$. Further, we could have taken $H_r = T_{-\lceil L/2 \rceil} H_l T_{\lceil L/2 \rceil}$. Indeed, it turns out that the last definition performs a factor 2 worse than the first one in the perturbative limit. 

Let's start with calculating the variance of $H_1 = \lambda \sum_i X_i$ up to second order
 \begin{align}
\Var(H_1) &= \Braket{(H_1)^2} - \Braket{H_1}^2 \\
\Braket{(H_1)^2} &= \lambda^2n +\lambda^2 \Bra{0}^{\otimes n}
\left[ \mathds{1} + \frac{\lambda}{8} \sum_i X_i + \frac{\lambda^2}{16} \left( \frac{1}{3} \sum_{\left <i, j \right>} X_iX_j + \frac{1}{8} \sum_{i \neq j \backslash  \left <i, j \right>} X_iX_j - \frac{1}{8} n \mathds{1}\right) \right] \left[ \sum_{i \neq j} X_i X_j \right]\\
&\cdot \left[ \mathds{1} + \frac{\lambda}{8} \sum_i X_i + \frac{\lambda^2}{16} \left( \frac{1}{3} \sum_{\left <i, j \right>} X_iX_j + \frac{1}{8} \sum_{i \neq j \backslash  \left <i, j \right>} X_iX_j - \frac{1}{8} n \mathds{1}\right) \right] 
\Ket{0}^{\otimes n} + \mathcal O (\lambda^3\lambda^2 )\\
&=\lambda^2 n  + \lambda^4 \left( \frac{n\cdot 2 (n-1) }{8^2} +2 \frac{2n \cdot 2}{3\cdot 16} +2 \frac{[n(n-1) -2n] \cdot 2}{8\cdot 16} \right) + \mathcal O (\lambda^3\lambda^2 ) \\
&=\lambda^2 n  + \lambda^4  \left( n^2 \left( \frac{1}{32} + \frac{1}{32} \right) +n \left( -\frac{1}{32} + \frac{1}{6} - \frac{3}{32} \right) \right) + \mathcal O (\lambda^5 ) = \lambda^2 n  + \frac{\lambda^4 }{16} \left( n^2 -\frac{4}{3}n\right)+ \mathcal O (\lambda^5 )
\end{align}
\begin{align}
\Braket{H_1} &= \lambda \Bra{0}^{\otimes n}
\left[ \mathds{1} + \frac{\lambda}{8} \sum_i X_i + \frac{\lambda^2}{16} \left( \frac{1}{3} \sum_{\left <i, j \right>} X_iX_j + \frac{1}{8} \sum_{i \neq j \backslash  \left <i, j \right>} X_iX_j - \frac{1}{8} n \mathds{1}\right) \right] \left[ \sum_{i } X_i \right]\\
&\cdot \left[ \mathds{1} + \frac{\lambda}{8} \sum_i X_i + \frac{\lambda^2}{16} \left( \frac{1}{3} \sum_{\left <i, j \right>} X_iX_j + \frac{1}{8} \sum_{i \neq j \backslash  \left <i, j \right>} X_iX_j - \frac{1}{8} n \mathds{1}\right) \right] 
\Ket{0}^{\otimes n} + \mathcal O (\lambda^3\lambda^1 ) \\
&= \lambda^2 \left(  \frac{2}{8} n \right)  + \mathcal O (\lambda^4 ) \\
\Var(H_1) &= n\lambda^2 - \frac{1}{12} n \lambda^4  + \mathcal O (\lambda^5 )
  \end{align}

  Let's calculate the variance of $H_l - H_r$ up to second order. Let's consider a general $H_r$ which is translated with respect to $H_l$ by $i_r \leq L/2$.
   \begin{align}
  \Var(\frac{1}{2} (H- H_l + H_r)) &= \frac{1}{4} \Var( H_l - H_r) = \frac{1}{2}\left( \Braket{(H_l)^2} - \Braket{H_l H_r} \right)\\
  H_l &= \sum_{i=1}^{n_x/L} \sum_{j=1}^{n_y} Z_{Li, j}Z_{Li+1, j}
\qquad \qquad H_r =  \sum_{i=1}^{n_x/L} \sum_{j=1}^{n_y} Z_{Li+i_r, j}Z_{Li+i_r+1, j}
\end{align}

\begin{align}
\Braket{(H_l)^2} &=  \frac{n}{L} +  \Bra{0}^{\otimes n}
\left[ \mathds{1} + \frac{\lambda}{8} \sum_i X_i + \frac{\lambda^2}{16} \left( \frac{1}{3} \sum_{\left <i, j \right>} X_iX_j + \frac{1}{8} \sum_{i \neq j \backslash  \left <i, j \right>} X_iX_j - \frac{1}{8} n \mathds{1}\right) \right] \\
&\cdot \left[  \sum_{i, k = 1}^{n_x/L} \sum_{\substack{j, \ell = 1\\ (i,j) \neq (k,\ell)}}^{n_y}  (Z_{Li, j}Z_{Li+1, j}) (Z_{Lk, \ell}Z_{Lk+1, \ell}) \right]\\
&\cdot \left[ \mathds{1} + \frac{\lambda}{8} \sum_i X_i + \frac{\lambda^2}{16} \left( \frac{1}{3} \sum_{\left <i, j \right>} X_iX_j + \frac{1}{8} \sum_{i \neq j \backslash  \left <i, j \right>} X_iX_j - \frac{1}{8} n \mathds{1}\right) \right] 
\Ket{0}^{\otimes n} + \mathcal O (\lambda^4 ) \\
&= \left(\frac{n}{L} \right)^2  + \frac{\lambda^2}{64} \left(  \frac{n}{L} \left( \frac{n}{L} -1 \right) (n-2\cdot 4) - 2   \frac{n}{2} \frac{n}{L} \left( \frac{n}{L} -1 \right) \right) + \mathcal O (\lambda^4 ) = \left(\frac{n}{L} \right)^2  - \frac{\lambda^2}{8} \frac{n}{L} \left( \frac{n}{L} -1 \right) + \mathcal O (\lambda^4 ) 
 \end{align}
\begin{align}
\Braket{H_l H_r} &= \Bra{0}^{\otimes n}
\left[ \mathds{1} + \frac{\lambda}{8} \sum_i X_i + \frac{\lambda^2}{16} \left( \frac{1}{3} \sum_{\left <i, j \right>} X_iX_j + \frac{1}{8} \sum_{i \neq j \backslash  \left <i, j \right>} X_iX_j - \frac{1}{8} n \mathds{1}\right) \right] \\
&\cdot \left[ \delta_{L,2}\delta_{i_r,1}  \sum_{i=1}^{n_x/L} \sum_{j=1}^{n_y} Z_{Li-1, j}Z_{Li+1, j} + \delta_{i_r,1} \sum_{i=1}^{n_x/L} \sum_{j=1}^{n_y} Z_{Li, j}Z_{Li+2, j} \right. \\
& \left. +  
\sum_{i, k = 1}^{n_x/L} \sum_{\substack{j, \ell = 1\\ (Li,j) \neq   (Lk + i_r +\{0 \pm 1\},\ell)}}^{n_y}  
(Z_{Li, j}Z_{Li+1, j}) (Z_{Lk+i_r, \ell}Z_{Lk+i_r+1, \ell})  \right]\\
&\cdot \left[ \mathds{1} + \frac{\lambda}{8} \sum_i X_i + \frac{\lambda^2}{16} \left( \frac{1}{3} \sum_{\left <i, j \right>} X_iX_j + \frac{1}{8} \sum_{i \neq j \backslash  \left <i, j \right>} X_iX_j - \frac{1}{8} n \mathds{1}\right) \right] 
\Ket{0}^{\otimes n} + \mathcal O (\lambda^3 ) \\
&=  \left(\frac{n}{L} \right)^2 
+ \frac{\lambda^2}{64} \left( (\delta_{L,2} + \delta_{i_r,1} ) n \cdot \frac{n-2 \cdot 2}{L} 
    + (n-2\cdot4) \frac{n}{L} \left(\frac{n}{L} - \delta_{L,2} - \delta_{i_r,1}\right)  - 2 \frac{n^3}{2L^2} \right ) +O (\lambda^4 ) \\
&=\left(\frac{n}{L} \right)^2 
+ \frac{\lambda^2}{16} \frac{n^2}{L}\left( \left( \frac{n-8}{L} + (\delta_{L,2} + \delta_{i_r,1} ) \frac{4}{n} \right)- \frac{n}{4L} \right ) +O (\lambda^4 )
  \end{align}
  note that $L=2$ implies $i_r=1$. We have $  O (\lambda^4 )$, not only $  O (\lambda^3 )$ as the third perturbation order contains single and triple excitations and thus the product of zeroth and third order terms vanish as well. The third sum has $\frac{n^2}{L^2} -  (\delta_{L,2} + \delta_{i_r,1} ) \frac{n}{L} $ elements. Whenever $XZX$ act on the same qubits, we obtain $-Z$ and we obtain $+Z$ if they act on different qubits. For $L=2$, we have
\begin{align}
    H_l &= \sum_{i=1}^{n_x/L} \sum_{j=1}^{n_y} Z_{2i, j}Z_{2i+1, j}
\qquad \qquad H_r =  \sum_{i=1}^{n_x/L} \sum_{j=1}^{n_y} Z_{2i+1, j}Z_{2i+2, j} \\
H_l H_r &= \sum_{i=1}^{n_x/L} \sum_{j=1}^{n_y} \sum_{k=1}^{n_x/L} \sum_{l=1}^{n_y} Z_{2i, j}Z_{2i+1, j} Z_{2k+1, l}Z_{2k+2, l}
\end{align}
So
\begin{align}
\Braket{H_l H_r} &= \Bra{0}^{\otimes n}
\left[ \mathds{1} + \frac{\lambda}{8} \sum_i X_i + \frac{\lambda^2}{16} \left( \frac{1}{3} \sum_{\left <i, j \right>} X_iX_j + \frac{1}{8} \sum_{i \neq j \backslash  \left <i, j \right>} X_iX_j - \frac{1}{8} n \mathds{1}\right) \right] \\
&\cdot \sum_{i=1}^{n_x/L} \sum_{j=1}^{n_y} \sum_{k=1}^{n_x/L} \sum_{l=1}^{n_y} Z_{2i, j}Z_{2i+1, j} Z_{2k+1, l}Z_{2k+2, l} \\
&\cdot \left[ \mathds{1} + \frac{\lambda}{8} \sum_i X_i + \frac{\lambda^2}{16} \left( \frac{1}{3} \sum_{\left <i, j \right>} X_iX_j + \frac{1}{8} \sum_{i \neq j \backslash  \left <i, j \right>} X_iX_j - \frac{1}{8} n \mathds{1}\right) \right] 
\Ket{0}^{\otimes n} + \mathcal O (\lambda^3 ) \\
&=  \left(\frac{n}{L} \right)^2 
+ \frac{\lambda^2}{64} \left( 
 2 \frac{n}{L} (n-4) 
+ ( (\frac{n}{L})^2 - 2 \frac{n}{L} )(n - 8) 
- \frac{n^3}{L^2}
\right ) + O (\lambda^4 ) \\
&= \left(\frac{n}{L} \right)^2 + \frac{\lambda^2}{8} \frac{n}{L} \left( 1 - \frac{n}{L} \right) + O (\lambda^4 )
  \end{align}

Together, this gives:
\begin{align}
    \left( \Braket{(H_l)^2} - \Braket{H_l H_r} \right) &= \lambda^2 \left( - \frac{1}{8} \frac{n}{L} \left( \frac{n}{L} -1 \right) + \frac{1}{8} \frac{n}{L} \left( \frac{n}{L} - 1 \right) \right) + O (\lambda^4 ) = O (\lambda^4)
\end{align}

For $L>2$ we have:
\begin{align}
\Braket{H_l H_r} &= \Bra{0}^{\otimes n}
\left[ \mathds{1} + \frac{\lambda}{8} \sum_i X_i + \frac{\lambda^2}{16} \left( \frac{1}{3} \sum_{\left <i, j \right>} X_iX_j + \frac{1}{8} \sum_{i \neq j \backslash  \left <i, j \right>} X_iX_j - \frac{1}{8} n \mathds{1}\right) \right] \\ 
&\cdot \sum_{i=1}^{n_x/L} \sum_{j=1}^{n_y} \sum_{k=1}^{n_x/L} \sum_{l=1}^{n_y} Z_{Li, j} Z_{Li+1, j} Z_{Lk+1, l}Z_{Lk+2, l} \\
&\cdot \left[ \mathds{1} + \frac{\lambda}{8} \sum_i X_i + \frac{\lambda^2}{16} \left( \frac{1}{3} \sum_{\left <i, j \right>} X_iX_j + \frac{1}{8} \sum_{i \neq j \backslash  \left <i, j \right>} X_iX_j - \frac{1}{8} n \mathds{1}\right) \right] 
\Ket{0}^{\otimes n} + \mathcal O (\lambda^3 ) \\
&= \left(\frac{n}{L} \right)^2 
+ \frac{\lambda^2}{64} \left( 
(\frac{n}{L}) (n - 4) 
+ ((\frac{n}{L})^2 - \frac{n}{L}) (n - 8) 
- \frac{n^3}{L^2}
\right ) + O ( \lambda^4 ) \\
&= \left(\frac{n}{L} \right)^2 
- \frac{\lambda^2}{8} \left( \left( \frac{n}{L} \right)^2 - \frac{1}{2} \frac{n}{L} \right) + O ( \lambda^4 )
\end{align}

Together, this gives:
\begin{align}
    \left( \Braket{(H_l)^2} - \Braket{H_l H_r} \right) &= \frac{\lambda^2}{8} \left( - \frac{n}{L} \left( \frac{n}{L} -1 \right) + \left( \frac{n}{L} \right)^2 - \frac{1}{2} \frac{n}{L} \right) + O (\lambda^4 ) = \frac{\lambda^2}{16} \frac{n}{L} + O (\lambda^4)
\end{align}

Finally
\begin{align}
    \frac{\Var(H_2)}{\Var(\frac{1}{2} (H- H_l + H_r))} = 32 L + O(\lambda^2)
\end{align}

For $L>2$ und $1 < i_r \leq L/2$
\begin{align}
\Braket{H_l H_r} &= \Bra{0}^{\otimes n}
\left[ \mathds{1} + \frac{\lambda}{8} \sum_i X_i + \frac{\lambda^2}{16} \left( \frac{1}{3} \sum_{\left <i, j \right>} X_iX_j + \frac{1}{8} \sum_{i \neq j \backslash  \left <i, j \right>} X_iX_j - \frac{1}{8} n \mathds{1}\right) \right] \\
&\cdot \sum_{i=1}^{n_x/L} \sum_{j=1}^{n_y} \sum_{k=1}^{n_x/L} \sum_{l=1}^{n_y} Z_{Li, j} Z_{Li+1, j} Z_{Lk+i_r, l}Z_{Lk+i_r+1, l} \\
&\cdot \left[ \mathds{1} + \frac{\lambda}{8} \sum_i X_i + \frac{\lambda^2}{16} \left( \frac{1}{3} \sum_{\left <i, j \right>} X_iX_j + \frac{1}{8} \sum_{i \neq j \backslash  \left <i, j \right>} X_iX_j - \frac{1}{8} n \mathds{1}\right) \right] 
\Ket{0}^{\otimes n} + \mathcal O (\lambda^3 ) \\
&= \left(\frac{n}{L} \right)^2 
+ \frac{\lambda^2}{64} \left( 
(\frac{n}{L})^2 (n - 8) 
- \frac{n^3}{L^2}
\right ) + O ( \lambda^4 ) = \left(\frac{n}{L} \right)^2 
- \frac{\lambda^2}{8} \left( \frac{n}{L} \right)^2 + O ( \lambda^4 )
\end{align}

Together, this gives:
\begin{align}
    \left( \Braket{(H_l)^2} - \Braket{H_l H_r} \right) &= \frac{\lambda^2}{8} \left( - \frac{n}{L} \left( \frac{n}{L} -1 \right) + \left( \frac{n}{L} \right)^2 \right) + O (\lambda^4 ) = \frac{\lambda^2}{8} \frac{n}{L} + O (\lambda^4)
\end{align}
Finally
\begin{align}
    \frac{\Var(H_2)}{\Var(\frac{1}{2} (H- H_l + H_r))} = 16 L + O(\lambda^2)
\end{align}
{This means that it is beneficial to choose $H_{\text cut}$ and $H_{\text cut}^\prime$ neighboring each other ($i_r =1$). As this is possible in general, we only report the $i_r=1$ results in Lemma~\ref{lemma:TFIM improvements} in the main text.}
For $L= 1$ we need to define:
\begin{align}
    H_l &= \sum_{i=1}^{n_x} \sum_{j=1}^{n_y} Z_{i, j}Z_{i+1, j}
\qquad \qquad H_r = S H_l S =   \sum_{i=1}^{n_x} \sum_{j=1}^{n_y} Z_{i, j}Z_{i, j+1} \\
H_l H_r &= \sum_{i=1}^{n_x} \sum_{j=1}^{n_y} \sum_{k=1}^{n_x} \sum_{l=1}^{n_y} Z_{i, j}Z_{i+1, j} Z_{k, l}Z_{k+1, l}
\end{align}

\begin{align}
\Braket{H_l H_r} &= \Bra{0}^{\otimes n}
\left[ \mathds{1} + \frac{\lambda}{8} \sum_i X_i + \frac{\lambda^2}{16} \left( \frac{1}{3} \sum_{\left <i, j \right>} X_iX_j + \frac{1}{8} \sum_{i \neq j \backslash  \left <i, j \right>} X_iX_j - \frac{1}{8} n \mathds{1}\right) \right] \\
&\cdot \sum_{i=1}^{n_x} \sum_{j=1}^{n_y} \sum_{k=1}^{n_x} \sum_{l=1}^{n_y} Z_{i, j}Z_{i+1, j} Z_{k, l}Z_{k+1, l} \\
&\cdot \left[ \mathds{1} + \frac{\lambda}{8} \sum_i X_i + \frac{\lambda^2}{16} \left( \frac{1}{3} \sum_{\left <i, j \right>} X_iX_j + \frac{1}{8} \sum_{i \neq j \backslash  \left <i, j \right>} X_iX_j - \frac{1}{8} n \mathds{1}\right) \right] 
\Ket{0}^{\otimes n} + \mathcal O (\lambda^3 ) \\
&= n^2 
+ \frac{\lambda^2}{64} \left( 
 2 n (n-4) 
+ ( (n^2 - 2 n )(n - 8) 
- n^3
\right ) + O (\lambda^4 ) \\
&= n^2 + \frac{\lambda^2}{8} n \left( 1 - n \right) + O (\lambda^4 )
  \end{align}

Thus:
\begin{align}
    \left( \Braket{(H_l)^2} - \Braket{H_l H_r} \right) &= \lambda^2 \left( - \frac{n}{8} \left( n -1 \right) + \frac{n}{8} \left( n - 1 \right) \right) + O (\lambda^4 ) = O (\lambda^4)
\end{align}

\newpage 
\subsection{List of all used partitions}
\label{appendix:list_all_partitions}
%
%
\begin{table*}[h!]
        \centering
        \begin{tabularx}{\linewidth}{p{15mm} || p{10mm} |  >{\centering\arraybackslash} p{150mm} }
            \centering Label
            & \centering Color
            & {\centering Partitioning} 
            \\ \hline \hline 
           \vspace{0.2cm}
           \centering Pauli 
            &   
            \centering {\color{FAUred} \rule{3 mm}{21 mm}}
            & \vspace{-0.6cm}
            \begin{equation}
            \begin{aligned}
               H_1 &= -\sum_{i=1}^{n_x = 4} \sum_{j=1}^{n_x = 6} J_X \left( X_{i,j} X_{i+1,j} + X_{i,j} X_{i,j+1}\right) \qquad H_3 = -\sum_{i=1}^{n_x = 4} \sum_{j=1}^{n_x = 6} h_Z Z_{i,j} \\
               H_2 &= -\sum_{i=1}^{n_x = 4} \sum_{j=1}^{n_x = 6} J_Y \left( Y_{i,j} Y_{i+1,j} + Y_{i,j} Y_{i,j+1}\right)
            \end{aligned}
            \end{equation}
            \vspace{-0.4cm}
            \\ \hline
            \vspace{0.2cm}
           \centering $L=1$, $(L_x,L_y) = (n_x,1)$ 
            &   
            \centering {\color{FAUlightgreen} \rule{3 mm}{21 mm}}
            & \vspace{-0.7cm}
            \begin{equation}
            \begin{aligned}
               H_1 &= -\sum_{i=1}^{n_x = 4} \sum_{j=1}^{n_x = 6} \left( J_X X_{i,j} X_{i+1,j} + J_Y Y_{i,j} Y_{i+1,j} \right)
               - \sum_{i=1}^{n_x = 4} \sum_{j=1}^{n_x = 6}   \frac{h_Z}{2} Z_{i,j} \\ 
               H_2 &= -\sum_{i=1}^{n_x = 4} \sum_{j=1}^{n_x = 6} \left( J_X X_{i,j} X_{i,j+1} + J_Y Y_{i,j} Y_{i,j+1} \right)
               - \sum_{i=1}^{n_x = 4} \sum_{j=1}^{n_x = 6}   \frac{h_Z}{2} Z_{i,j} 
            \end{aligned}
            \end{equation}
            \vspace{-0.4cm}
            \\ \hline
            \vspace{0.2cm}
           \centering $L=2$, $(L_x,L_y) = (n_x,2)$ 
            &   
            \centering {\color{FAUcyan} \rule{3 mm}{48 mm}}
            & \vspace{-0.5cm}
            \begin{equation}
            \begin{aligned}
               H_1 = &-\sum_{i=1}^{n_x = 4} \sum_{j=1}^{n_y/L = 3}  \left( J_X X_{i,2j} X_{i,2j+1} + J_Y Y_{i,2j} Y_{i,2j+1}\right)\\
               &-\sum_{i=1}^{n_x = 4} \sum_{j=1}^{n_y = 6} \left( \frac{J_X}{2} X_{i,j} X_{i+1,j} + \frac{J_Y}{2} Y_{i,j} Y_{i+1,j}\right)
               - \sum_{i=1}^{n_x = 4} \sum_{j=1}^{n_y = 6}   \frac{h_Z}{2} Z_{i,j} \\
               H_2 = &-\sum_{i=1}^{n_x = 4} \sum_{j=1}^{n_y/L = 3} \left( J_X X_{i,2j-1} X_{i,2j} - J_Y Y_{i,2j-1} Y_{i,2j} \right) \\
               &-\sum_{i=1}^{n_x = 4} \sum_{j=1}^{n_y = 6} \left( \frac{J_X}{2} X_{i,j} X_{i+1,j} + \frac{J_Y}{2} Y_{i,j} Y_{i+1,j} \right) 
               - \sum_{i=1}^{n_x = 4} \sum_{j=1}^{n_y = 6}   \frac{h_Z}{2} Z_{i,j} 
            \end{aligned}
            \end{equation}
            \vspace{-0.3cm}
                        \\ \hline
            \vspace{0.2cm}
           \centering $L=3$, $(L_x,L_y) = (n_x,3)$ 
            &   
            \centering {\color{FAUblue} \rule{3 mm}{56 mm}}
            & \vspace{-0.5cm}
            \begin{equation}
            \begin{aligned}
               H_1 = &-\sum_{i=1}^{n_x = 4} \sum_{j=1}^{n_y/L = 2} 
               \left[ J_X \left( X_{i,3j} X_{i,3j+1} + \frac{1}{2} X_{i,3j+1} X_{i,3j+2} \right ) 
               + J_Y \left( Y_{i,3j} Y_{i,3j+1} + \frac{1}{2} Y_{i,3j+1} Y_{i,3j+2} \right ) \right]\\
               &-\sum_{i=1}^{n_x = 4} \sum_{j=1}^{n_y = 6} \left ( \frac{J_X}{2} X_{i,j} X_{i+1,j} \frac{J_Y}{2} Y_{i,j} Y_{i+1,j} \right)
               - \sum_{i=1}^{n_x = 4} \sum_{j=1}^{n_y = 6}   \frac{h_Z}{2} Z_{i,j} \\
               H_2 = &-\sum_{i=1}^{n_x = 4} \sum_{j=1}^{n_y/L = 2} 
               \left[ J_X \left( X_{i,3j-1} X_{i,3j} + \frac{1}{2} X_{i,3j+1} X_{i,3j+2} \right ) 
               + J_Y \left( Y_{i,3j-1} Y_{i,3j} + \frac{1}{2} Y_{i,3j+1} Y_{i,3j+2} \right ) \right]\\
               &-\sum_{i=1}^{n_x = 4} \sum_{j=1}^{n_y = 6} \left ( \frac{J_X}{2} X_{i,j} X_{i+1,j} \frac{J_Y}{2} Y_{i,j} Y_{i+1,j} \right)
               - \sum_{i=1}^{n_x = 4} \sum_{j=1}^{n_y = 6}   \frac{h_Z}{2} Z_{i,j}
            \end{aligned}
            \end{equation}
            \\ \hline \hline
        \end{tabularx}
        \\ \ \\
        \hfill 
        Table continues on the next page.
\end{table*}

%
%
\begin{table*}[t!]
        \centering
        \begin{tabularx}{\linewidth}{p{15mm} || p{10mm} |  >{\centering\arraybackslash} p{150mm} }
            \centering Label
            & \centering Color
            & {\centering Partitioning} 
            \\ \hline \hline 
            \vspace{0.2cm}
           \centering $(L_x,L_y) =$ 2-local 
            &   \vspace{-4 mm}
            \centering {\rotatebox{270}{
            {\color{FAUyellow} \rule{3 mm}{3 mm}}{\color{white} \rule{2 mm}{3 mm}}
            {\color{FAUyellow} \rule{3 mm}{3 mm}}{\color{white} \rule{2 mm}{3 mm}}
            {\color{FAUyellow} \rule{3 mm}{3 mm}}{\color{white} \rule{2 mm}{3 mm}}
            {\color{FAUyellow} \rule{3 mm}{3 mm}}{\color{white} \rule{2 mm}{3 mm}}
            {\color{FAUyellow} \rule{3 mm}{3 mm}}{\color{white} \rule{2 mm}{3 mm}}
            {\color{FAUyellow} \rule{3 mm}{3 mm}}{\color{white} \rule{2 mm}{3 mm}}
            {\color{FAUyellow} \rule{3 mm}{3 mm}}{\color{white} \rule{2 mm}{3 mm}}
            {\color{FAUyellow} \rule{3 mm}{3 mm}}{\color{white} \rule{2 mm}{3 mm}}
            {\color{FAUyellow} \rule{3 mm}{3 mm}}{\color{white} \rule{2 mm}{3 mm}}
            {\color{FAUyellow} \rule{3 mm}{3 mm}}
            }}
            & 
            \begin{equation}
            \begin{aligned}
               H_1 &= -\sum_{i=1}^{n_x/2 = 2} \sum_{j=1}^{n_y = 6} \left( J_X X_{2i,j} X_{2i+1,j} + J_Y Y_{2i,j} Y_{2i+1,j} \right) 
               - \sum_{i=1}^{n_x = 4} \sum_{j=1}^{n_y = 6}   \frac{h_Z}{4} Z_{i,j}\\
               H_2 &=  -\sum_{i=1}^{n_x/2 = 2} \sum_{j=1}^{n_y = 6} \left( J_X X_{2i-1,j} X_{2i,j} + J_Y Y_{2i-1,j} Y_{2i,j} \right)  
               - \sum_{i=1}^{n_x = 4} \sum_{j=1}^{n_y = 6}   \frac{h}{4} X_{i,j}\\
                H_3 &= -\sum_{i=1}^{n_x/2 = 2} \sum_{j=1}^{n_y = 6} \left( J_X X_{i,2j} X_{i,2j+1} + J_Y Y_{i,2j} Y_{i,2j+1} \right) 
               - \sum_{i=1}^{n_x = 4} \sum_{j=1}^{n_y = 6}   \frac{h_Z}{4} Z_{i,j}\\
               H_4 &=  -\sum_{i=1}^{n_x/2 = 2} \sum_{j=1}^{n_y = 6} \left( J_X X_{,2j-1} X_{i,2j} + J_Y Y_{i,2j-1} Y_{i,2j} \right)  
               - \sum_{i=1}^{n_x = 4} \sum_{j=1}^{n_y = 6}   \frac{h}{4} X_{i,j}
            \end{aligned}
            \end{equation}
            \\ \hline
            \vspace{0.2cm}
           \centering $(L_x,L_y) = (2,2)$ 
            &    \vspace{-4 mm}
            \centering {\rotatebox{270}{
            {\color{FAUlightgreen} \rule{3 mm}{3 mm}}{\color{white} \rule{2 mm}{3 mm}}
            {\color{FAUlightgreen} \rule{3 mm}{3 mm}}{\color{white} \rule{2 mm}{3 mm}}
            {\color{FAUlightgreen} \rule{3 mm}{3 mm}}{\color{white} \rule{2 mm}{3 mm}}
            {\color{FAUlightgreen} \rule{3 mm}{3 mm}}{\color{white} \rule{2 mm}{3 mm}}
            {\color{FAUlightgreen} \rule{3 mm}{3 mm}}{\color{white} \rule{2 mm}{3 mm}}
            {\color{FAUlightgreen} \rule{3 mm}{3 mm}}{\color{white} \rule{2 mm}{3 mm}}
            {\color{FAUlightgreen} \rule{3 mm}{3 mm}}{\color{white} \rule{2 mm}{3 mm}}
            {\color{FAUlightgreen} \rule{3 mm}{3 mm}}{\color{white} \rule{2 mm}{3 mm}}
            {\color{FAUlightgreen} \rule{3 mm}{3 mm}}{\color{white} \rule{2 mm}{3 mm}}
            {\color{FAUlightgreen} \rule{3 mm}{3 mm}}
            }}
            & 
            \begin{equation}
            \begin{aligned}
               H_1 = &-\sum_{i=1}^{n_x = 4} \sum_{j=1}^{n_y/L_y = 3} \left( J_X X_{i,2j} X_{i,2j+1} + ( J_Y Y_{i,2j} Y_{i,2j+1} \right)\\
               &- \sum_{i=1}^{n_x/L_x = 2} \sum_{j=1}^{n_y = 6} \left(  J_X X_{2i,j} X_{2i+1,j} + J_Y Y_{2i,j} Y_{2i+1,j} \right)
               - \sum_{i=1}^{n_x = 4} \sum_{j=1}^{n_y = 6}   \frac{h_Z}{2} Z_{i,j} \\
               H_2 &= -\sum_{i=1}^{n_x = 4} \sum_{j=1}^{n_y/L_y = 3} \left( J_X X_{i,2j-1} X_{i,2j} + ( J_Y Y_{i,2j-1} Y_{i,2j} \right)\\
               &- \sum_{i=1}^{n_x/L_x = 2} \sum_{j=1}^{n_y = 6} \left(  J_X X_{2i-1,j} X_{2i,j} + J_Y Y_{2i-1,j} Y_{2i,j} \right)
               - \sum_{i=1}^{n_x = 4} \sum_{j=1}^{n_y = 6}   \frac{h_Z}{2} Z_{i,j} 
            \end{aligned}
            \end{equation}
                        \\ \hline
            \vspace{0.2cm}
           \centering $(L_x,L_y) = (2,3)$ 
            &   \vspace{-4 mm}
            \centering {\rotatebox{270}{
            {\color{FAUdarkgreen} \rule{3 mm}{3 mm}}{\color{white} \rule{2 mm}{3 mm}}
            {\color{FAUdarkgreen} \rule{3 mm}{3 mm}}{\color{white} \rule{2 mm}{3 mm}}
            {\color{FAUdarkgreen} \rule{3 mm}{3 mm}}{\color{white} \rule{2 mm}{3 mm}}
            {\color{FAUdarkgreen} \rule{3 mm}{3 mm}}{\color{white} \rule{2 mm}{3 mm}}
            {\color{FAUdarkgreen} \rule{3 mm}{3 mm}}{\color{white} \rule{2 mm}{3 mm}}
            {\color{FAUdarkgreen} \rule{3 mm}{3 mm}}{\color{white} \rule{2 mm}{3 mm}}
            {\color{FAUdarkgreen} \rule{3 mm}{3 mm}}{\color{white} \rule{2 mm}{3 mm}}
            {\color{FAUdarkgreen} \rule{3 mm}{3 mm}}{\color{white} \rule{2 mm}{3 mm}}
            {\color{FAUdarkgreen} \rule{3 mm}{3 mm}}{\color{white} \rule{2 mm}{3 mm}}
            {\color{FAUdarkgreen} \rule{3 mm}{3 mm}}
            }}
            &
             \begin{equation}
            \begin{aligned}
               H_1 = &-\sum_{i=1}^{n_x = 4} \sum_{j=1}^{n_y/L = 2} 
               \left[ J_X \left( X_{i,3j} X_{i,3j+1} + \frac{1}{2} X_{i,3j+1} X_{i,3j+2} \right ) 
               + J_Y \left( Y_{i,3j} Y_{i,3j+1} + \frac{1}{2} Y_{i,3j+1} Y_{i,3j+2} \right ) \right]\\
               &-\sum_{i=1}^{n_x/L_x = 2} \sum_{j=1}^{n_y = 6} \left ( \frac{J_X}{2} X_{2i,j} X_{2i+1,j} \frac{J_Y}{2} Y_{2i,j} Y_{2i+1,j} \right)
               - \sum_{i=1}^{n_x = 4} \sum_{j=1}^{n_y = 6}   \frac{h_Z}{2} Z_{i,j} \\
               H_2 = &-\sum_{i=1}^{n_x = 4} \sum_{j=1}^{n_y/L = 2} 
               \left[ J_X \left( X_{i,3j-1} X_{i,3j} + \frac{1}{2} X_{i,3j+1} X_{i,3j+2} \right ) 
               + J_Y \left( Y_{i,3j-1} Y_{i,3j} + \frac{1}{2} Y_{i,3j+1} Y_{i,3j+2} \right ) \right]\\
               &-\sum_{i=1}^{n_x/L_x = 2} \sum_{j=1}^{n_y = 6} \left ( \frac{J_X}{2} X_{2i-1,j} X_{2i,j} \frac{J_Y}{2} Y_{2i-1,j} Y_{2i,j} \right)
               - \sum_{i=1}^{n_x = 4} \sum_{j=1}^{n_y = 6}   \frac{h_Z}{2} Z_{i,j}
            \end{aligned}
            \end{equation}
        \end{tabularx}
         \caption{\textbf{Explicit partitions of the TFXYM example, Fig.~\ref{fig: Examples} (a), and Hard-core Bose-Hubbard model example,  Fig.~\ref{fig: Examples1b} (a)}
        To recall, all examples apart from the Spinful Hubbard model, are on a $(n_x,n_y)=(4,6)$ rectangular lattice with periodic boundary conditions, thus all indices are modulo $n_x$ and respectively $n_y$. For the geometric partitionings we show in Fig.~\ref{fig: Examples} (a) the sampling improvements $\mathcal G_{\ket{E_0}}(\mathcal B_{\rm Pauli}, \mathcal B_{L_x,L_y})$, and for Pauli partitioning the combined variance of the partition 
        for Pauli measurements $[  \sum_{H_b \in \mathcal B_{\rm Pauli}} \sqrt{ \Var_{\ket \psi} ( H_b) } ]^2 $ is given.
        For the Transverse Field XY Model, Eq.~\eqref{eq: results_examples_XY_Hamiltonian}, $J_X = \frac{1}{2} (1+ \eta)$,  $J_Y = \frac{1}{2} (1- \eta)$,  and $h_Z = h$. Similarly, for the Bose-Hubbard model, when allowing maximally one Boson per lattice side, Eq.~\eqref{eq: Fock_op_HCBH}, one can map it to a spin basis Eq.~\eqref{eq: Spin_HCBH}, where it has TFXYM form with $J_X = J_Y =  \frac{1}{2} J$, and $h_Z = -\frac{1}{2} h$.
        }
        \label{table: explicit_partitions_TFXYM}
\end{table*}

%
%
\begin{table*}[t!]
        \centering
        \begin{tabularx}{\linewidth}{p{15mm} || p{10mm} |  >{\centering\arraybackslash} p{150mm} }
            \centering Label
            & \centering Color
            & {\centering Partitioning} 
            \\ \hline \hline 
           \vspace{0.2cm}
           \centering Pauli 
            &   
            \centering {\color{FAUred} \rule{3 mm}{15 mm}}
            & \vspace{-0.5cm}
            \begin{equation}
            \begin{aligned}
               H_1 = -\sum_{i=1}^{n_x = 4} \sum_{j=1}^{n_x = 6} J \left( Z_{i,j} Z_{i+1,j} + Z_{i,j} Z_{i,j+1}\right) \qquad && \qquad H_2 = -\sum_{i=1}^{n_x = 4} \sum_{j=1}^{n_x = 6} h X_{i,j}
            \end{aligned}
            \end{equation}
            \\ \hline
            \vspace{0.2cm}
           \centering $L=1$, $(L_x,L_y) = (n_x,1)$ 
            &   
            \centering {\color{FAUlightgreen} \rule{3 mm}{15 mm}}
            & \vspace{-0.5cm}
            \begin{align}
               H_1 = -\sum_{i=1}^{n_x = 4} \sum_{j=1}^{n_x = 6} J Z_{i,j} Z_{i+1,j} 
               - \sum_{i=1}^{n_x = 4} \sum_{j=1}^{n_x = 6}   \frac{h}{2} X_{i,j} \qquad 
               H_2 = -\sum_{i=1}^{n_x = 4} \sum_{j=1}^{n_x = 6} J Z_{i,j} Z_{i,j+1} 
               - \sum_{i=1}^{n_x = 4} \sum_{j=1}^{n_x = 6}   \frac{h}{2} X_{i,j} 
            \end{align}
            \\ \hline
            \vspace{0.2cm}
           \centering $L=2$, $(L_x,L_y) = (n_x,2)$ 
            &   
            \centering {\color{FAUcyan} \rule{3 mm}{24 mm}}
            & \vspace{-0.5cm}
            \begin{equation}
            \begin{aligned}
               H_1 &= -\sum_{i=1}^{n_x = 4} \sum_{j=1}^{n_y = 6} \frac{J}{2} Z_{i,j} Z_{i+1,j} 
               - \sum_{i=1}^{n_x = 4} \sum_{j=1}^{n_y/L = 3} J Z_{i,2j} Z_{i,2j+1} 
               - \sum_{i=1}^{n_x = 4} \sum_{j=1}^{n_y = 6}   \frac{h}{2} X_{i,j} \\
               H_2 &= -\sum_{i=1}^{n_x = 4} \sum_{j=1}^{n_y = 6} \frac{J}{2} Z_{i,j} Z_{i+1,j} 
               - \sum_{i=1}^{n_x = 4} \sum_{j=1}^{n_y/L = 3} J Z_{i,2j-1} Z_{i,2j} 
               - \sum_{i=1}^{n_x = 4} \sum_{j=1}^{n_y = 6}   \frac{h}{2} X_{i,j} 
            \end{aligned}
            \end{equation}
            \vspace{-0.3cm}
                        \\ \hline
            \vspace{0.2cm}
           \centering $L=3$, $(L_x,L_y) = (n_x,3)$ 
            &   
            \centering {\color{FAUblue} \rule{3 mm}{24 mm}}
            & \vspace{-0.5cm}
            \begin{equation}
            \begin{aligned}
               H_1 &= -\sum_{i=1}^{n_x = 4} \sum_{j=1}^{n_y = 6} \frac{J}{2} Z_{i,j} Z_{i+1,j} 
               - \sum_{i=1}^{n_x = 4} \sum_{j=1}^{n_y/L = 2} J \left( Z_{i,3j} Z_{i,3j+1} + \frac{1}{2} Z_{i,3j+1} Z_{i,3j+2} \right )
               - \sum_{i=1}^{n_x = 4} \sum_{j=1}^{n_y = 6}   \frac{h}{2} X_{i,j} \\
               H_2 &= -\sum_{i=1}^{n_x = 4} \sum_{j=1}^{n_y = 6} \frac{J}{2} Z_{i,j} Z_{i+1,j} 
              - \sum_{i=1}^{n_x = 4} \sum_{j=1}^{n_y/L = 2} J \left( Z_{i,3j-1} Z_{i,3j} + \frac{1}{2} Z_{i,3j+1} Z_{i,3j+2} \right )
               - \sum_{i=1}^{n_x = 4} \sum_{j=1}^{n_y = 6}   \frac{h}{2} X_{i,j} 
            \end{aligned}
            \end{equation}
            \vspace{-0.3cm}
            \\ \hline
            \vspace{0.2cm}
           \centering $(L_x,L_y) =$ 2-local 
            &   \vspace{-4 mm}
            \centering {\rotatebox{270}{
            {\color{FAUyellow} \rule{3 mm}{3 mm}}{\color{white} \rule{2 mm}{3 mm}}
            {\color{FAUyellow} \rule{3 mm}{3 mm}}{\color{white} \rule{2 mm}{3 mm}}
            {\color{FAUyellow} \rule{3 mm}{3 mm}}{\color{white} \rule{2 mm}{3 mm}}
            {\color{FAUyellow} \rule{3 mm}{3 mm}}{\color{white} \rule{2 mm}{3 mm}}
            {\color{FAUyellow} \rule{3 mm}{3 mm}}
            }}
            & \vspace{-0.5cm}
            \begin{equation}
            \begin{aligned}
               H_1 = -\sum_{i=1}^{n_x/2 = 2} \sum_{j=1}^{n_y = 6} J Z_{2i,j} Z_{2i+1,j} 
               - \sum_{i=1}^{n_x = 4} \sum_{j=1}^{n_y = 6}   \frac{h}{4} X_{i,j}
               && H_2 = -\sum_{i=1}^{n_x/2 = 2} \sum_{j=1}^{n_y = 6} J Z_{2i-1,j} Z_{2i,j} 
               - \sum_{i=1}^{n_x = 4} \sum_{j=1}^{n_y = 6}   \frac{h}{4} X_{i,j}\\
               H_3 = -\sum_{i=1}^{n_x/2 = 2} \sum_{j=1}^{n_y = 6} J Z_{i,2j} Z_{i,2j+1} 
               - \sum_{i=1}^{n_x = 4} \sum_{j=1}^{n_y = 6}   \frac{h}{4} X_{i,j}
               && H_4 = -\sum_{i=1}^{n_x/2 = 2} \sum_{j=1}^{n_y = 6} J Z_{i,2j-1} Z_{i,2j} 
               - \sum_{i=1}^{n_x = 4} \sum_{j=1}^{n_y = 6}   \frac{h}{4} X_{i,j}
            \end{aligned}
            \end{equation}
            \vspace{-0.3cm}
            \\ \hline
            \vspace{0.2cm}
           \centering $(L_x,L_y) = (2,2)$ 
            &    \vspace{-4 mm}
            \centering {\rotatebox{270}{
            {\color{FAUlightgreen} \rule{3 mm}{3 mm}}{\color{white} \rule{2 mm}{3 mm}}
            {\color{FAUlightgreen} \rule{3 mm}{3 mm}}{\color{white} \rule{2 mm}{3 mm}}
            {\color{FAUlightgreen} \rule{3 mm}{3 mm}}{\color{white} \rule{2 mm}{3 mm}}
            {\color{FAUlightgreen} \rule{3 mm}{3 mm}}{\color{white} \rule{2 mm}{3 mm}}
            {\color{FAUlightgreen} \rule{3 mm}{3 mm}}
            }}
            & \vspace{-0.5cm}
            \begin{equation}
            \begin{aligned}
               H_1 &= -\sum_{i=1}^{n_x = 4} \sum_{j=1}^{n_y/L_y = 3} J Z_{i,2j} Z_{i,2j+1} 
               - \sum_{i=1}^{n_x/L_x = 2} \sum_{j=1}^{n_y = 6} J Z_{2i,j} Z_{2i+1,j} 
               - \sum_{i=1}^{n_x = 4} \sum_{j=1}^{n_y = 6}   \frac{h}{2} X_{i,j} \\
               H_2 &= -\sum_{i=1}^{n_x = 4} \sum_{j=1}^{n_y/L_y = 3} J Z_{i,2j-1} Z_{i,2j} 
               - \sum_{i=1}^{n_x/L_x = 2} \sum_{j=1}^{n_y = 6} J Z_{2i-1,j} Z_{2i,j} 
               - \sum_{i=1}^{n_x = 4} \sum_{j=1}^{n_y = 6}   \frac{h}{2} X_{i,j} 
            \end{aligned}
            \end{equation}
            \vspace{-0.3cm}
                        \\ \hline
            \vspace{0.2cm}
           \centering $(L_x,L_y) = (2,3)$ 
            &   \vspace{-4 mm}
            \centering {\rotatebox{270}{
            {\color{FAUdarkgreen} \rule{3 mm}{3 mm}}{\color{white} \rule{2 mm}{3 mm}}
            {\color{FAUdarkgreen} \rule{3 mm}{3 mm}}{\color{white} \rule{2 mm}{3 mm}}
            {\color{FAUdarkgreen} \rule{3 mm}{3 mm}}{\color{white} \rule{2 mm}{3 mm}}
            {\color{FAUdarkgreen} \rule{3 mm}{3 mm}}{\color{white} \rule{2 mm}{3 mm}}
            {\color{FAUdarkgreen} \rule{3 mm}{3 mm}}
            }}
            & \vspace{-0.5cm}
            \begin{equation}
            \begin{aligned}
                H_1 &= -\sum_{i=1}^{n_x/L_x = 2} \sum_{j=1}^{n_y = 6} J Z_{2i,j} Z_{2i+1,j} 
               - \sum_{i=1}^{n_x = 4} \sum_{j=1}^{n_y/L = 2} J \left( Z_{i,3j} Z_{i,3j+1} + \frac{1}{2} Z_{i,3j+1} Z_{i,3j+2} \right )
               - \sum_{i=1}^{n_x = 4} \sum_{j=1}^{n_y = 6}   \frac{h}{2} X_{i,j} \\
               H_2 &= -\sum_{i=1}^{n_x/L_x = 2} \sum_{j=1}^{n_y = 6} J Z_{2i-1,j} Z_{2i,j} 
              - \sum_{i=1}^{n_x = 4} \sum_{j=1}^{n_y/L = 2} J \left( Z_{i,3j-1} Z_{i,3j} + \frac{1}{2} Z_{i,3j+1} Z_{i,3j+2} \right )
               - \sum_{i=1}^{n_x = 4} \sum_{j=1}^{n_y = 6}   \frac{h}{2} X_{i,j} 
            \end{aligned}
            \end{equation}
            \vspace{-0.3cm}
        \end{tabularx}
        \caption{\textbf{Explicit partitions of the TFIM example, Fig.~\ref{fig: Examples} (b)} 
        To recall, all examples apart from the Spinful Hubbard model, are on a $(n_x,n_y)=(4,6)$ rectangular lattice with periodic boundary conditions, thus all indices are modulo $n_x$ and respectively $n_y$. For the geometric partitionings we show in Fig.~\ref{fig: Examples} (b) the sampling improvements $\mathcal G_{\ket{E_0}}(\mathcal B_{\rm Pauli}, \mathcal B_{L_x,L_y})$, and for Pauli partitioning the combined variance of the partition 
        for Pauli measurements $[  \sum_{H_b \in \mathcal B_{\rm Pauli}} \sqrt{ \Var_{\ket \psi} ( H_b) } ]^2 $ is given.
        \vspace{-0.5 cm} 
        }
        \label{table: explicit_partitions_TFIM}
\end{table*}
%
%
\begin{table*}[t!]
        \centering
        \begin{tabularx}{\linewidth}{p{15mm} || p{10mm} |  >{\centering\arraybackslash} p{150mm} }
            \centering Label
            & \centering Color
            & {\centering Partitioning} 
            \\ \hline \hline 
           \vspace{0.2cm}
           \centering Pauli 
            &   
            \centering {\color{FAUred} \rule{3 mm}{15 mm}}
            & \vspace{-0.5cm}
            \begin{equation}
            \begin{aligned}
               H_1 = -\sum_{i=1}^{n_x = 4} \sum_{j=1}^{n_x = 6} J \left[ \left( Z_{i,j} Z_{i+1,j} + Z_{i,j} Z_{i,j+1}\right) 
               -\kappa J \left( Z_{i,j} Z_{i+2,j} + Z_{i,j} Z_{i,j+2}\right)\right] \ && \
               H_2 = -\sum_{i=1}^{n_x = 4} \sum_{j=1}^{n_x = 6} h X_{i,j}
            \end{aligned}
            \end{equation}
            \\ \hline
            \vspace{0.2cm}
           \centering $L=1$, $(L_x,L_y) = (n_x,1)$ 
            &   
            \centering {\color{FAUlightgreen} \rule{3 mm}{24 mm}}
            & \vspace{-0.5cm}
             \begin{equation}
            \begin{aligned}
               H_1 &= -\sum_{i=1}^{n_x = 4} \sum_{j=1}^{n_x = 6} J \left( Z_{i,j} Z_{i+1,j}  
               -\kappa  Z_{i,j} Z_{i+2,j} \right)
               - \sum_{i=1}^{n_x = 4} \sum_{j=1}^{n_x = 6}   \frac{h}{2} X_{i,j} \\
               H_2 &= -\sum_{i=1}^{n_x = 4} \sum_{j=1}^{n_x = 6} J \left( Z_{i,j} Z_{i,j+1}  
               -\kappa  Z_{i,j} Z_{i,j+2} \right)
               - \sum_{i=1}^{n_x = 4} \sum_{j=1}^{n_x = 6}   \frac{h}{2} X_{i,j} 
            \end{aligned}
            \end{equation}
            \\ \hline
            \vspace{0.2cm}
           \centering $L=3$, $(L_x,L_y) = (n_x,3)$ 
            &   
            \centering {\color{FAUblue} \rule{3 mm}{75 mm}}
            & \vspace{-0.5cm}
            \begin{equation}
            \begin{aligned}
               H_1 = 
               &- \sum_{i=1}^{n_x = 4} \sum_{j=1}^{n_y/L = 2} J \left(\frac{1}{2} Z_{i,3j} Z_{i,3j+1} + \frac{1}{2} Z_{i,3j+1}  Z_{i,3j+2} -\kappa Z_{i,3j} Z_{i,3j+2} \right )\\
               &-\sum_{i=1}^{n_x = 4} \sum_{j=1}^{n_y = 6} \frac{J}{3} \left( Z_{i,j} Z_{i+1,j}  
               -\kappa  Z_{i,j} Z_{i+2,j} \right)
               - \sum_{i=1}^{n_x = 4} \sum_{j=1}^{n_y = 6}   \frac{h}{3} X_{i,j} \\
               H_2 &- \sum_{i=1}^{n_x = 4} \sum_{j=1}^{n_y/L = 2} J \left(\frac{1}{2} Z_{i,3j+1} Z_{i,3j+2} + \frac{1}{2} Z_{i,3j+2}  Z_{i,3j+3} -\kappa Z_{i,3j+1} Z_{i,3j+3} \right )\\
               &-\sum_{i=1}^{n_x = 4} \sum_{j=1}^{n_y = 6} \frac{J}{3} \left( Z_{i,j} Z_{i+1,j}  
               -\kappa  Z_{i,j} Z_{i+2,j} \right)
               - \sum_{i=1}^{n_x = 4} \sum_{j=1}^{n_y = 6}   \frac{h}{3} X_{i,j}  \\
               H_3 &- \sum_{i=1}^{n_x = 4} \sum_{j=1}^{n_y/L = 2} J \left(\frac{1}{2} Z_{i,3j-1} Z_{i,3j} + \frac{1}{2} Z_{i,3j}  Z_{i,3j+1} -\kappa Z_{i,3j-1} Z_{i,3j+1} \right )\\
               &-\sum_{i=1}^{n_x = 4} \sum_{j=1}^{n_y = 6} \frac{J}{3} \left( Z_{i,j} Z_{i+1,j}  
               -\kappa  Z_{i,j} Z_{i+2,j} \right)
               - \sum_{i=1}^{n_x = 4} \sum_{j=1}^{n_y = 6}   \frac{h}{3} X_{i,j} 
            \end{aligned}
            \end{equation}
            \\ \hline
            \vspace{0.2cm}
           \centering $(L_x,L_y) =$ 3-local 
            &   \vspace{-4 mm}
            \centering {\rotatebox{270}{
            {\color{FAUyellow} \rule{3 mm}{3 mm}}{\color{white} \rule{2 mm}{3 mm}}
            {\color{FAUyellow} \rule{3 mm}{3 mm}}{\color{white} \rule{2 mm}{3 mm}}
            {\color{FAUyellow} \rule{3 mm}{3 mm}}{\color{white} \rule{2 mm}{3 mm}}
            {\color{FAUyellow} \rule{3 mm}{3 mm}}{\color{white} \rule{2 mm}{3 mm}}
            {\color{FAUyellow} \rule{3 mm}{3 mm}}{\color{white} \rule{2 mm}{3 mm}}
            {\color{FAUyellow} \rule{3 mm}{3 mm}}{\color{white} \rule{2 mm}{3 mm}}
            {\color{FAUyellow} \rule{3 mm}{3 mm}}{\color{white} \rule{2 mm}{3 mm}}
            {\color{FAUyellow} \rule{3 mm}{3 mm}}{\color{white} \rule{2 mm}{3 mm}}
            {\color{FAUyellow} \rule{3 mm}{3 mm}}{\color{white} \rule{2 mm}{3 mm}}
            {\color{FAUyellow} \rule{3 mm}{3 mm}}
            }}
            & \vspace{-0.5cm}
            \begin{equation}
            \begin{aligned}
               H_1 &= 
               -\sum_{i=1}^{n_x = 4} \sum_{j=1}^{n_y/3 = 2} J \left(\frac{1}{2} Z_{i,3j} Z_{i,3j+1} + \frac{1}{2} Z_{i,3j+1}  Z_{i,3j+2} -\kappa Z_{i,3j} Z_{i,3j+2} \right )
               - \sum_{i=1}^{n_x = 4} \sum_{j=1}^{n_y = 6}   \frac{h}{6} X_{i,j}\\
               H_2 &= 
               -\sum_{i=1}^{n_x = 4} \sum_{j=1}^{n_y/3 = 2} J \left(\frac{1}{2} Z_{i,3j+1} Z_{i,3j+2} + \frac{1}{2} Z_{i,3j+2}  Z_{i,3j+3} -\kappa Z_{i,3j+1} Z_{i,3j+3} \right )
               - \sum_{i=1}^{n_x = 4} \sum_{j=1}^{n_y = 6}   \frac{h}{6} X_{i,j}\\
               H_{2+k} &= 
               - \sum_{j=1}^{n_y/3 = 2} J \left(\frac{1}{2} Z_{k,3j-1} Z_{1,3j} + \frac{1}{2} Z_{k,3j}  Z_{k,3j+1} -\kappa Z_{k,3j-1} Z_{k,3j+1} \right )\\
               &-\sum_{j=1}^{n_y = 6} J \left(\frac{1}{2} Z_{k+1,j} Z_{k+2,j} + \frac{1}{2} Z_{k+2,j}  Z_{k+3,j} -\kappa Z_{k+1,j} Z_{k+3,j} \right )
               - \sum_{i=1}^{n_x = 4} \sum_{j=1}^{n_y = 6}   \frac{h}{6} X_{i,j}\\
               H_3 &= H_{2+k=1} \qquad
               H_4 = H_{2+k=2} \qquad
               H_5 = H_{2+k=3} \qquad
               H_6 = H_{2+k=4}
            \end{aligned}
            \end{equation}
            \vspace{-0.3cm}
        \end{tabularx}
        \caption{\textbf{Explicit partitions of the TF-BNNNI example, Fig.~\ref{fig: Examples} (c)} 
        To recall, all examples apart from the Spinful Hubbard model, are on a $(n_x,n_y)=(4,6)$ rectangular lattice with periodic boundary conditions, thus all indices are modulo $n_x$ and respectively $n_y$. For the geometric partitionings we show in Fig.~\ref{fig: Examples} (c) the sampling improvements $\mathcal G_{\ket{E_0}}(\mathcal B_{\rm Pauli}, \mathcal B_{L_x,L_y})$, and for Pauli partitioning the combined variance of the partition 
        for Pauli measurements $[  \sum_{H_b \in \mathcal B_{\rm Pauli}} \sqrt{ \Var_{\ket \psi} ( H_b) } ]^2 $ is given.
        Further, for the TF-BNNNI model the unit cell will be axial 3-local, and hence partitions $L=2$, $(L_x,L_y) = (2,2)$, and $(L_x,L_y) = (2,3)$ are not applicable. Similarly, a bi-partition in $H=H_1+H_2$ will again be possible for $L_x, L_y \geq 5$, where then the unit cell is shifted by 2 due to the next-nearest neighbor coupling.  
        \vspace{-0.5 cm} 
        }
        \label{table: explicit_partitions_TFBNNNI}
\end{table*}
%
%
\begin{table*}[t!]
        \centering
        \begin{tabularx}{\linewidth}{p{15mm} || p{10mm} |  >{\centering\arraybackslash} p{150mm} }
            \centering Label
            & \centering Color
            & {\centering Partitioning} 
            \\ \hline \hline 
           \vspace{0.2cm}
           \centering Pauli
            &   
            \centering {\color{FAUred} \rule{3 mm}{36 mm}}
            & \vspace{-0.6 cm} 
            \begin{equation}
            \begin{aligned}
               H_1 &= -\sum_{i=1}^{n_x/2 = 2} \sum_{j=1}^{n_y/2 = 3} t \left( c_{i,j}^\dagger c_{i+1,j} + c_{i+1,j}^\dagger c_{i,j} \right) \qquad
               H_2 = -\sum_{i=1}^{n_x/2 = 2} \sum_{j=1}^{n_y/2 = 3} t \left( c_{i,j}^\dagger c_{i+1,j} + c_{i+1,j}^\dagger c_{i,j} \right) \\
               H_3 &= -\sum_{i=1}^{n_x/2 = 2} \sum_{j=1}^{n_y/2 = 3} t \left( c_{i,j}^\dagger c_{i+1,j} + c_{i+1,j}^\dagger c_{i,j} \right) \qquad
               H_4 = -\sum_{i=1}^{n_x/2 = 2} \sum_{j=1}^{n_y/2 = 3} t \left( c_{i,j}^\dagger c_{i+1,j} + c_{i+1,j}^\dagger c_{i,j} \right)\\
               H_5 &= +\sum_{i=1}^{n_x = 4} \sum_{j=1}^{n_x = 6} U (n_{i,j}n_{i,j+1}+n_{i,j}n_{i+1,j})
               - \sum_{i=1}^{n_x = 4} \sum_{j=1}^{n_y = 6}   \mu n_{i,j}
            \end{aligned}
            \end{equation}
            \vspace{-0.3 cm} 
            \\ \hline
            \vspace{0.2cm}
           \centering $L=1$, $(L_x,L_y) = (n_x,1)$ 
            &   
            \centering {\color{FAUlightgreen} \rule{3 mm}{24 mm}}
            & \vspace{-0.6 cm} 
            \begin{equation}
            \begin{aligned}
               H_1 &= -\sum_{i=1}^{n_x = 4} \sum_{j=1}^{n_x = 6} 
               \left( t  c_{i,j}^\dagger c_{i+1,j} + tc_{i+1,j}^\dagger c_{i,j} - U n_{i,j}n_{i+1,j}\right)
               - \sum_{i=1}^{n_x = 4} \sum_{j=1}^{n_y = 6}   \frac{\mu}{2} n_{i,j}
               \\
               H_2 &= -\sum_{i=1}^{n_x = 4} \sum_{j=1}^{n_x = 6} 
               \left( t  c_{i,j}^\dagger c_{i,j+1} + tc_{i,j+1}^\dagger c_{i,j} - U n_{i,j}n_{i,j+1}\right)
               - \sum_{i=1}^{n_x = 4} \sum_{j=1}^{n_y = 6}   \frac{\mu}{2} n_{i,j}
            \end{aligned}
            \end{equation}
            \vspace{-0.3 cm} 
            \\ \hline
            \vspace{0.2cm}
           \centering $L=2$, $(L_x,L_y) = (n_x,2)$ 
            &   
            \centering {\color{FAUcyan} \rule{3 mm}{48 mm}}
            & \vspace{-0.6 cm} 
            \begin{equation}
            \begin{aligned}
               H_1 = &-\sum_{i=1}^{n_x = 4} \sum_{j=1}^{n_x = 6} 
               \left( \frac{t}{2}  c_{i,j}^\dagger c_{i+1,j} + \frac{t}{2} c_{i+1,j}^\dagger c_{i,j} - \frac{U}{2} n_{i,j}n_{i+1,j}\right)\\
               &- \sum_{i=1}^{n_x = 4} \sum_{j=1}^{n_y/L = 3}
                \left( t  c_{i,2j}^\dagger c_{i,2j+1} + tc_{i,2j+1}^\dagger c_{i,2j} - U n_{i,2j}n_{i,2j+1}\right)
                - \sum_{i=1}^{n_x = 4} \sum_{j=1}^{n_y = 6}   \frac{\mu}{2} n_{i,j}
                \\
                H_2 = &-\sum_{i=1}^{n_x = 4} \sum_{j=1}^{n_x = 6} 
                   \left( \frac{t}{2}  c_{i,j}^\dagger c_{i+1,j} + \frac{t}{2} c_{i+1,j}^\dagger c_{i,j} - \frac{U}{2} n_{i,j}n_{i+1,j}\right)
                   \\
               &- \sum_{i=1}^{n_x = 4} \sum_{j=1}^{n_y/L = 3}
                \left( t  c_{i,2j-1}^\dagger c_{i,2j} + tc_{i,2j}^\dagger c_{i,2j-1} - U n_{i,2j-1}n_{i,2j}\right)
                - \sum_{i=1}^{n_x = 4} \sum_{j=1}^{n_y = 6}   \frac{\mu}{2} n_{i,j}\\
            \end{aligned}
            \end{equation}
            \vspace{-0.3 cm} 
                        \\ \hline
            \vspace{0.2cm}
           \centering $L=3$, $(L_x,L_y) = (n_x,3)$ 
            &   
            \centering {\color{FAUblue} \rule{3 mm}{72 mm}}
            & \vspace{-0.6cm}
            \begin{equation}
            \begin{aligned}
               H_1 = &-\sum_{i=1}^{n_x = 4} \sum_{j=1}^{n_x = 6} 
               \left( \frac{t}{2}  c_{i,j}^\dagger c_{i+1,j} + \frac{t}{2} c_{i+1,j}^\dagger c_{i,j} - \frac{U}{2} n_{i,j}n_{i+1,j}\right)\\
               &- \sum_{i=1}^{n_x = 4} \sum_{j=1}^{n_y/L = 2} 
               \left( t  c_{i,3j}^\dagger c_{i,3j+1} + tc_{i,3j+1}^\dagger c_{i,3j} - U n_{i,3j}n_{i,3j+1} \right)
               \\
               &- \sum_{i=1}^{n_x = 4} \sum_{j=1}^{n_y/L = 2} 
               \frac{1}{2} \left(  t  c_{i,3j+1}^\dagger c_{i,3j+2} + tc_{i,3j+2}^\dagger c_{i,3j+1} - U n_{i,3j+1}n_{i,3j+2} \right) - \sum_{i=1}^{n_x = 4} \sum_{j=1}^{n_y = 6}   \frac{\mu}{2} n_{i,j}
                \\
               H_2 &-\sum_{i=1}^{n_x = 4} \sum_{j=1}^{n_x = 6} 
               \left( \frac{t}{2}  c_{i,j}^\dagger c_{i+1,j} + \frac{t}{2} c_{i+1,j}^\dagger c_{i,j} - \frac{U}{2} n_{i,j}n_{i+1,j}\right)\\
               &- \sum_{i=1}^{n_x = 4} \sum_{j=1}^{n_y/L = 2} 
               \left( t  c_{i,3j-1}^\dagger c_{i,3j} + tc_{i,3j}^\dagger c_{i,3j-1} - U n_{i,3j-1}n_{i,3j} \right)
               \\
               &- \sum_{i=1}^{n_x = 4} \sum_{j=1}^{n_y/L = 2} 
               \frac{1}{2} \left(  t  c_{i,3j-1}^\dagger c_{i,3j-2} + tc_{i,3j-2}^\dagger c_{i,3j-1} - U n_{i,3j-1}n_{i,3j-2} \right) - \sum_{i=1}^{n_x = 4} \sum_{j=1}^{n_y = 6}   \frac{\mu}{2} n_{i,j}
            \end{aligned}
            \end{equation}
            \vspace{-0.3 cm} 
        \\ \hline \hline
        \end{tabularx}
        \\ \ \\
        \hfill 
        Table continues on the next page.
\end{table*}
%
%
\begin{table*}[t!]
        \centering
        \begin{tabularx}{\linewidth}{p{15mm} || p{10mm} |  >{\centering\arraybackslash} p{150mm} }
            \centering Label
            & \centering Color
            & {\centering Partitioning} 
            \\ \hline \hline 
            \vspace{0.2cm}
           \centering $(L_x,L_y) =$ 2-local 
            &   \vspace{-4 mm}
            \centering {\rotatebox{270}{
            {\color{FAUyellow} \rule{3 mm}{3 mm}}{\color{white} \rule{2 mm}{3 mm}}
            {\color{FAUyellow} \rule{3 mm}{3 mm}}{\color{white} \rule{2 mm}{3 mm}}
            {\color{FAUyellow} \rule{3 mm}{3 mm}}{\color{white} \rule{2 mm}{3 mm}}
            {\color{FAUyellow} \rule{3 mm}{3 mm}}{\color{white} \rule{2 mm}{3 mm}}
            {\color{FAUyellow} \rule{3 mm}{3 mm}}{\color{white} \rule{2 mm}{3 mm}}
            {\color{FAUyellow} \rule{3 mm}{3 mm}}{\color{white} \rule{2 mm}{3 mm}}
            {\color{FAUyellow} \rule{3 mm}{3 mm}}{\color{white} \rule{2 mm}{3 mm}}
            {\color{FAUyellow} \rule{3 mm}{3 mm}}{\color{white} \rule{2 mm}{3 mm}}
            {\color{FAUyellow} \rule{3 mm}{3 mm}}{\color{white} \rule{2 mm}{3 mm}}
            {\color{FAUyellow} \rule{3 mm}{3 mm}}
            }}
            &
            \begin{equation}
            \begin{aligned}
               H_1 = &-\sum_{i=1}^{n_x/2 = 2} \sum_{j=1}^{n_y = 6} 
               \left( t  c_{2i,j}^\dagger c_{2i+1,j} + t c_{2i+1,j}^\dagger c_{2i,j} - U n_{2i,j}n_{2i+1,j}\right)
               - \sum_{i=1}^{n_x = 4} \sum_{j=1}^{n_y = 6}   \frac{\mu}{4} n_{i,j} \\
               H_2 = &-\sum_{i=1}^{n_x/2 = 2} \sum_{j=1}^{n_y = 6} 
               \left( t  c_{2i-1,j}^\dagger c_{2i,j} + t c_{2i,j}^\dagger c_{2i-1,j} - U n_{2i-1,j}n_{2i,j}\right)
               - \sum_{i=1}^{n_x = 4} \sum_{j=1}^{n_y = 6}   \frac{\mu}{4} n_{i,j}\\
               H_3 = &-\sum_{i=1}^{n_x = 4} \sum_{j=1}^{n_y/2 = 3} 
               \left( t  c_{i,2j}^\dagger c_{i,2j+1} + t c_{i,2j+1}^\dagger c_{i,2j} - U n_{i,2j}n_{i,2j+1}\right)
               - \sum_{i=1}^{n_x = 4} \sum_{j=1}^{n_y = 6}   \frac{\mu}{4} n_{i,j} \\
               H_4 = &-\sum_{i=1}^{n_x = 4} \sum_{j=1}^{n_y/2 = 3} 
               \left( t  c_{i,2j-1}^\dagger c_{i,2j} + t c_{i,2j}^\dagger c_{i,2j-1} - U n_{i,2j-1}n_{i,2j}\right)
               - \sum_{i=1}^{n_x = 4} \sum_{j=1}^{n_y = 6}   \frac{\mu}{4} n_{i,j}\\
            \end{aligned}
            \end{equation}
            \\ \hline
            \vspace{0.2cm}
           \centering $(L_x,L_y) = (2,2)$ 
            &    \vspace{-4 mm}
            \centering {\rotatebox{270}{
            {\color{FAUlightgreen} \rule{3 mm}{3 mm}}{\color{white} \rule{2 mm}{3 mm}}
            {\color{FAUlightgreen} \rule{3 mm}{3 mm}}{\color{white} \rule{2 mm}{3 mm}}
            {\color{FAUlightgreen} \rule{3 mm}{3 mm}}{\color{white} \rule{2 mm}{3 mm}}
            {\color{FAUlightgreen} \rule{3 mm}{3 mm}}{\color{white} \rule{2 mm}{3 mm}}
            {\color{FAUlightgreen} \rule{3 mm}{3 mm}}{\color{white} \rule{2 mm}{3 mm}}
            {\color{FAUlightgreen} \rule{3 mm}{3 mm}}{\color{white} \rule{2 mm}{3 mm}}
            {\color{FAUlightgreen} \rule{3 mm}{3 mm}}{\color{white} \rule{2 mm}{3 mm}}
            {\color{FAUlightgreen} \rule{3 mm}{3 mm}}{\color{white} \rule{2 mm}{3 mm}}
            {\color{FAUlightgreen} \rule{3 mm}{3 mm}}{\color{white} \rule{2 mm}{3 mm}}
            {\color{FAUlightgreen} \rule{3 mm}{3 mm}}
            }}
            & 
            \begin{equation}
            \begin{aligned}
               H_1 = &-\sum_{i=1}^{n_x/L_x = 2} \sum_{j=1}^{n_x = 6} 
               \left( t  c_{2i,j}^\dagger c_{2i+1,j} + t c_{2i+1,j}^\dagger c_{2i,j} - U n_{2i,j}n_{2i+1,j}\right)\\
               &- \sum_{i=1}^{n_x = 4} \sum_{j=1}^{n_y/L_y = 3}
                \left( t  c_{i,2j}^\dagger c_{i,2j+1} + tc_{i,2j+1}^\dagger c_{i,2j} - U n_{i,2j}n_{i,2j+1}\right)
                - \sum_{i=1}^{n_x = 4} \sum_{j=1}^{n_y = 6}   \frac{\mu}{2} n_{i,j}
                \\
                H_2 = &-\sum_{i=1}^{n_x/L_x = 2} \sum_{j=1}^{n_x = 6} 
                   \left( t c_{2i-1,j}^\dagger c_{2i,j} + t c_{2i,j}^\dagger c_{2i-1,j} - U n_{2i-1,j}n_{2i,j}\right)
                   \\
               &- \sum_{i=1}^{n_x = 4} \sum_{j=1}^{n_y/L = 3}
                \left( t  c_{i,2j-1}^\dagger c_{i,2j} + tc_{i,2j}^\dagger c_{i,2j-1} - U n_{i,2j-1}n_{i,2j}\right)
                - \sum_{i=1}^{n_x = 4} \sum_{j=1}^{n_y = 6}   \frac{\mu}{2} n_{i,j}\\
            \end{aligned}
            \end{equation}
                        \\ \hline
            \vspace{0.2cm}
           \centering $(L_x,L_y) = (2,3)$ 
            &   \vspace{-4 mm}
            \centering {\rotatebox{270}{
            {\color{FAUdarkgreen} \rule{3 mm}{3 mm}}{\color{white} \rule{2 mm}{3 mm}}
            {\color{FAUdarkgreen} \rule{3 mm}{3 mm}}{\color{white} \rule{2 mm}{3 mm}}
            {\color{FAUdarkgreen} \rule{3 mm}{3 mm}}{\color{white} \rule{2 mm}{3 mm}}
            {\color{FAUdarkgreen} \rule{3 mm}{3 mm}}{\color{white} \rule{2 mm}{3 mm}}
            {\color{FAUdarkgreen} \rule{3 mm}{3 mm}}{\color{white} \rule{2 mm}{3 mm}}
            {\color{FAUdarkgreen} \rule{3 mm}{3 mm}}{\color{white} \rule{2 mm}{3 mm}}
            {\color{FAUdarkgreen} \rule{3 mm}{3 mm}}{\color{white} \rule{2 mm}{3 mm}}
            {\color{FAUdarkgreen} \rule{3 mm}{3 mm}}{\color{white} \rule{2 mm}{3 mm}}
            {\color{FAUdarkgreen} \rule{3 mm}{3 mm}}{\color{white} \rule{2 mm}{3 mm}}
            {\color{FAUdarkgreen} \rule{3 mm}{3 mm}}{\color{white} \rule{2 mm}{3 mm}}
            {\color{FAUdarkgreen} \rule{3 mm}{3 mm}}{\color{white} \rule{2 mm}{3 mm}}
            {\color{FAUdarkgreen} \rule{3 mm}{3 mm}}{\color{white} \rule{2 mm}{3 mm}}
            {\color{FAUdarkgreen} \rule{3 mm}{3 mm}}{\color{white} \rule{2 mm}{3 mm}}
            {\color{FAUdarkgreen} \rule{3 mm}{3 mm}}
            }}
            & 
            \begin{equation}
            \begin{aligned}
               H_1 = &-\sum_{i=1}^{n_x/L_x = 2} \sum_{j=1}^{n_x = 6} 
               \left( t  c_{2i,j}^\dagger c_{2i+1,j} + t c_{2i+1,j}^\dagger c_{2i,j} - U n_{2i,j}n_{2i+1,j}\right)\\
               &- \sum_{i=1}^{n_x = 4} \sum_{j=1}^{n_y/L_y = 2} 
               \left( t  c_{i,3j}^\dagger c_{i,3j+1} + tc_{i,3j+1}^\dagger c_{i,3j} - U n_{i,3j}n_{i,3j+1} \right)
               \\
               &- \sum_{i=1}^{n_x = 4} \sum_{j=1}^{n_y/L_y = 2} 
               \frac{1}{2} \left(  t  c_{i,3j+1}^\dagger c_{i,3j+2} + tc_{i,3j+2}^\dagger c_{i,3j+1} - U n_{i,3j+1}n_{i,3j+2} \right) - \sum_{i=1}^{n_x = 4} \sum_{j=1}^{n_y = 6}   \frac{\mu}{2} n_{i,j}
                \\
               H_2 = &-\sum_{i=1}^{n_x/L_x = 2} \sum_{j=1}^{n_x = 6} 
               \left( t  c_{2i-1,j}^\dagger c_{2i,j} + t c_{2i,j}^\dagger c_{2i-1,j} - U n_{2i-1,j}n_{2i,j}\right)\\
               &- \sum_{i=1}^{n_x = 4} \sum_{j=1}^{n_y/L = 2} 
               \left( t  c_{i,3j-1}^\dagger c_{i,3j} + tc_{i,3j}^\dagger c_{i,3j-1} - U n_{i,3j-1}n_{i,3j} \right)
               \\
               &- \sum_{i=1}^{n_x = 4} \sum_{j=1}^{n_y/L = 2} 
               \frac{1}{2} \left(  t  c_{i,3j-1}^\dagger c_{i,3j-2} + tc_{i,3j-2}^\dagger c_{i,3j-1} - U n_{i,3j-1}n_{i,3j-2} \right) - \sum_{i=1}^{n_x = 4} \sum_{j=1}^{n_y = 6}   \frac{\mu}{2} n_{i,j}
            \end{aligned}
            \end{equation}
        \end{tabularx}
        \caption{\textbf{Explicit partitions of the spinless Hubbard example, Fig.~\ref{fig: Examples1b} (b)} 
        To recall, all examples apart from the Spinful Hubbard model, are on a $(n_x,n_y)=(4,6)$ rectangular lattice with periodic boundary conditions, thus all indices are modulo $n_x$ and respectively $n_y$. For the geometric partitionings we show in Fig.~\ref{fig: Examples1b} (b) the sampling improvements $\mathcal G_{\ket{E_0}}(\mathcal B_{\rm Pauli}, \mathcal B_{L_x,L_y})$, and for Pauli partitioning the combined variance of the partition 
        for Pauli measurements $[  \sum_{H_b \in \mathcal B_{\rm Pauli}} \sqrt{ \Var_{\ket \psi} ( H_b) } ]^2 $ is given.
        \vspace{-0.5 cm} 
        }
        \label{table: explicit_partitions_spinless_Hubbard}
\end{table*}
%
%
%
\begin{table*}[t!]
        \centering
        \begin{tabularx}{\linewidth}{p{20mm} || p{10mm} |  >{\centering\arraybackslash} p{145mm} }
            \centering Label
            & \centering Color
            & {\centering Partitioning} 
            \\ \hline \hline 
           \vspace{0.2cm}
           \centering Pauli
            &   
            \centering {\color{FAUred} \rule{3 mm}{65 mm}}
            &\vspace{-0.3 cm} 
            \begin{equation}
            \begin{aligned}
               H_1 &=  -\sum_{i=1}^{n_x/2 = 2} \sum_{\sigma \in \{\uparrow, \downarrow\}} 
               t \left( c_{2i,1, \sigma}^\dagger c_{2i+1,1, \sigma} + c_{2i+1,1}^\dagger c_{2i,1} \right) 
               -\sum_{i=1}^{n_x = 4} \sum_{\sigma \in \{\uparrow, \downarrow\}} 
               t \left( c_{i,2, \sigma}^\dagger c_{i,3, \sigma} + c_{i,3}^\dagger c_{i,2} \right) \\
                H_2 &=  -\sum_{i=1}^{n_x/2 = 2} \sum_{\sigma \in \{\uparrow, \downarrow\}} 
               t \left( c_{2i,2, \sigma}^\dagger c_{2i+1,2, \sigma} + c_{2i+1,2}^\dagger c_{2i,2} \right) 
               -\sum_{i=1}^{n_x = 4} \sum_{\sigma \in \{\uparrow, \downarrow\}} 
               t \left( c_{i,3, \sigma}^\dagger c_{i,1, \sigma} + c_{i,1}^\dagger c_{i,3} \right) \\
                H_3 &=  -\sum_{i=1}^{n_x/2 = 2} \sum_{\sigma \in \{\uparrow, \downarrow\}} 
               t \left( c_{2i,3, \sigma}^\dagger c_{2i+1,3, \sigma} + c_{2i+1,3}^\dagger c_{2i,3} \right) 
               -\sum_{i=1}^{n_x = 4} \sum_{\sigma \in \{\uparrow, \downarrow\}} 
               t \left( c_{i,1, \sigma}^\dagger c_{i,2, \sigma} + c_{i,2}^\dagger c_{i,1} \right) \\
               H_4 &= -\sum_{i=1}^{n_x/2 = 2} \sum_{j=1}^{n_y = 3} \sum_{\sigma \in \{\uparrow, \downarrow\}} 
               t \left( c_{2i-1,j}^\dagger c_{2i,j} + c_{2i,j}^\dagger c_{2i-1,j} \right)\\
               H_5 &=  +\sum_{i=1}^{n_x = 4} \sum_{j=1}^{n_x = 3} U n_{i,j, \uparrow}n_{i,j, \downarrow}
               - \mu (n_{i,j, \uparrow} + n_{i,j, \downarrow})
            \end{aligned}
            \end{equation}
            \vspace{-0.3 cm} 
            \\ \hline
            \vspace{0.2cm}
           \centering $(L_x,L_y,L_{\uparrow,\downarrow})$ $= (n_x, n_y, 1)$, \& $(1,1,n_{\uparrow,\downarrow})$ 
            &   
            \centering {\color{FAUlightgreen} \rule{3 mm}{25 mm}}
            & \vspace{-0.3 cm} 
            \begin{equation}
            \begin{aligned}
               H_1 &= - \sum_{\sigma \in \{\uparrow, \downarrow\}} 
               \left[ 
               \sum_{i=1}^{n_x = 4} \sum_{j=1}^{n_y = 3} 
               \left( t  c_{i,j}^\dagger c_{i+1,j} + tc_{i+1,j}^\dagger c_{i,j} 
               + t  c_{i,j}^\dagger c_{i,j+1} + tc_{i,j+1}^\dagger c_{i,j} \right)
               +  \frac{\mu}{2} n_{i,j, \sigma}
               \right ]
               \\
              H_2 &= +\sum_{i=1}^{n_x = 4} \sum_{j=1}^{n_x = 3} U n_{i,j, \uparrow}n_{i,j, \downarrow}
               - \frac{\mu}{2} (n_{i,j, \uparrow} + n_{i,j, \downarrow})
            \end{aligned}
            \end{equation}
            \vspace{-0.3 cm} 
            \\ \hline
            \vspace{0.2cm}
           \centering $(L_x,L_y,L_{\uparrow,\downarrow})$ $= (2,n_y,n_{\uparrow,\downarrow})$ 
            &   
            \centering {\color{FAUcyan} \rule{3 mm}{55 mm}}
            & \vspace{-0.3 cm} 
            \begin{equation}
            \begin{aligned}
               H_1 = &\sum_{i=1}^{n_x = 4} \sum_{j=1}^{n_y = 3} \left[ \sum_{\sigma \in \{\uparrow, \downarrow\}}   
                - \left( \frac{t}{2}  c_{i,j, \sigma}^\dagger c_{i,j+1, \sigma} + \frac{t}{2} c_{i,j+1,\sigma}^\dagger c_{i,j, \sigma} \right)
               - \frac{\mu}{2} n_{i,j, \sigma} \right]
               + \frac{U}{2}   n_{i,j, \uparrow} n_{i,j, \downarrow}\\
               &-\sum_{i=1}^{n_x/L_x = 2} \sum_{j=1}^{n_y = 3}
                \sum_{\sigma \in \{\uparrow, \downarrow\}}   
                t c_{2i,j, \sigma}^\dagger c_{2i+1,j, \sigma} + t c_{2i+1,j,\sigma}^\dagger c_{2i,j, \sigma} \\
                H_2 = &\sum_{i=1}^{n_x = 4} \sum_{j=1}^{n_y = 3} \left[ \sum_{\sigma \in \{\uparrow, \downarrow\}}   
                - \left( \frac{t}{2}  c_{i,j, \sigma}^\dagger c_{i,j+1, \sigma} + \frac{t}{2} c_{i,j+1,\sigma}^\dagger c_{i,j, \sigma} \right)
               - \frac{\mu}{2} n_{i,j, \sigma} \right]
               + \frac{U}{2}   n_{i,j, \uparrow} n_{i,j, \downarrow}\\
               &-\sum_{i=1}^{n_x/L_x = 2} \sum_{j=1}^{n_y = 3}
                \sum_{\sigma \in \{\uparrow, \downarrow\}}   
                t c_{2i-1,j, \sigma}^\dagger c_{2i,j, \sigma} + t c_{2i,j,\sigma}^\dagger c_{2i-1,j, \sigma} 
            \end{aligned}
            \end{equation}
           \vspace{-0.3 cm} 
             \\ \hline
            \vspace{0.2cm}
           \centering $(L_x,L_y,L_{\uparrow,\downarrow})$ $= (n_x,1,1)$, $(1,n_y,1)$, \& $(1,1,n_{\uparrow,\downarrow})$
            &   
            \centering {\color{FAUyellow} \rule{3 mm}{40 mm}}
            & \vspace{-0.3 cm} 
            \begin{equation}
            \begin{aligned}
               H_1 = &-\sum_{i=1}^{n_x = 4} \sum_{j=1}^{n_y = 3}  \sum_{\sigma \in \{\uparrow, \downarrow\}}   
               \left( t  c_{i,j, \sigma}^\dagger c_{i+1,j, \sigma} + t c_{i+1,j,\sigma}^\dagger c_{i,j, \sigma} \right)
               - \frac{\mu}{3} n_{i,j, \sigma} 
               \\
               H_2 = &-\sum_{i=1}^{n_x = 4} \sum_{j=1}^{n_y = 3}  \sum_{\sigma \in \{\uparrow, \downarrow\}}   
               \left( t  c_{i,j, \sigma}^\dagger c_{i,j+1, \sigma} + t c_{i,j+1,\sigma}^\dagger c_{i,j, \sigma} \right)
               - \frac{\mu}{3} n_{i,j, \sigma} \\
               H_3 = &+\sum_{i=1}^{n_x = 4} \sum_{j=1}^{n_y = 3}  
               U   n_{i,j, \uparrow} n_{i,j, \downarrow}
               - \frac{\mu}{3} \left( n_{i,j, \uparrow} +  n_{i,j, \downarrow}\right)
            \end{aligned}
            \end{equation}
            \vspace{-0.3 cm} 
        \\ \hline \hline
        \end{tabularx}
        \\ \ \\
        \hfill 
        Table continues on the next page.
\end{table*}
%
%
\begin{table*}[t!]
        \centering
        \begin{tabularx}{\linewidth}{p{20mm} || p{10mm} |  >{\centering\arraybackslash} p{145mm} }
            \centering Label
            & \centering Color
            & {\centering Partitioning} 
            \\ \hline \hline 
            \vspace{0.2cm}
           \centering  \centering $(L_x,L_y,L_{\uparrow,\downarrow})$ $=(1,n_y,n_{\uparrow,\downarrow})$, \& $(n_x,1,1)$
            &   
            \centering {\color{FAUdarkgreen} \rule{3 mm}{28 mm}}
            &  \vspace{-0.3 cm} 
            \begin{equation}
            \begin{aligned}
                H_1 = &\sum_{i=1}^{n_x = 4} \sum_{j=1}^{n_y = 3} 
                \left[ \sum_{\sigma \in \{\uparrow, \downarrow\}}
                - \left( t  c_{i,j, \sigma}^\dagger c_{i,j+1, \sigma} + t c_{i,j+1,\sigma}^\dagger c_{i,j, \sigma} \right)
               - \frac{\mu}{2} n_{i,j, \sigma} \right]
               + U   n_{i,j, \uparrow} n_{i,j, \downarrow}
                \\
                H_2 = &\sum_{i=1}^{n_x = 4} \sum_{j=1}^{n_y = 3}  \sum_{\sigma \in \{\uparrow, \downarrow\}}   
               - \left( t  c_{i,j, \sigma}^\dagger c_{i+1,j, \sigma} + t c_{i+1,j,\sigma}^\dagger c_{i,j, \sigma} \right)
               - \frac{\mu}{2} n_{i,j, \sigma} 
            \end{aligned}
            \end{equation}
             \vspace{-0.3 cm} 
                        \\ \hline
            \vspace{0.2cm}
           \centering \centering  \centering $(L_x,L_y,L_{\uparrow,\downarrow})$ $=(n_x,1,n_{\uparrow,\downarrow})$, \& $(1,n_y,1)$ 
            &   \vspace{-4 mm}
            \centering {\rotatebox{270}{
            {\color{FAUdarkgreen} \rule{3 mm}{3 mm}}{\color{white} \rule{2 mm}{3 mm}}
            {\color{FAUdarkgreen} \rule{3 mm}{3 mm}}{\color{white} \rule{2 mm}{3 mm}}
            {\color{FAUdarkgreen} \rule{3 mm}{3 mm}}{\color{white} \rule{2 mm}{3 mm}}
            {\color{FAUdarkgreen} \rule{3 mm}{3 mm}}{\color{white} \rule{2 mm}{3 mm}}
            {\color{FAUdarkgreen} \rule{3 mm}{3 mm}}
            }}
            &  \vspace{-0.5 cm} 
            \begin{equation}
            \begin{aligned}
                 H_1 = &\sum_{i=1}^{n_x = 4} \sum_{j=1}^{n_y = 3}
                \left[ \sum_{\sigma \in \{\uparrow, \downarrow\}}   
                - \left( t  c_{i,j, \sigma}^\dagger c_{i+1,j, \sigma} + t c_{i+1,j,\sigma}^\dagger c_{i,j, \sigma} \right)
               - \frac{\mu}{2} n_{i,j, \sigma} \right]
               + U   n_{i,j, \uparrow} n_{i,j, \downarrow}
                \\
                H_2 = &\sum_{i=1}^{n_x = 4} \sum_{j=1}^{n_y = 3}  \sum_{\sigma \in \{\uparrow, \downarrow\}}
                - \left( t  c_{i,j, \sigma}^\dagger c_{i,j+1, \sigma} + t c_{i,j+1,\sigma}^\dagger c_{i,j, \sigma} \right)
               - \frac{\mu}{2} n_{i,j, \sigma} 
            \end{aligned}
            \end{equation}
             \vspace{-0.3 cm} 
             \\ \hline
            \vspace{0.2cm}
           \centering \centering  \centering $(L_x,L_y,L_{\uparrow,\downarrow})$ $=(1,n_y,n_{\uparrow,\downarrow})$, \& $(n_x,1,n_{\uparrow,\downarrow})$ 
            &   \vspace{-4 mm}
            \centering {\rotatebox{270}{
            {\color{FAUblue} \rule{3 mm}{3 mm}}{\color{white} \rule{2 mm}{3 mm}}
            {\color{FAUblue} \rule{3 mm}{3 mm}}{\color{white} \rule{2 mm}{3 mm}}
            {\color{FAUblue} \rule{3 mm}{3 mm}}{\color{white} \rule{2 mm}{3 mm}}
            {\color{FAUblue} \rule{3 mm}{3 mm}}{\color{white} \rule{2 mm}{3 mm}}
            {\color{FAUblue} \rule{3 mm}{3 mm}}
            }}
            &  \vspace{-0.5 cm} 
            \begin{equation}
            \begin{aligned}
             H_1 = &\sum_{i=1}^{n_x = 4} \sum_{j=1}^{n_y = 3} 
                \left[ \sum_{\sigma \in \{\uparrow, \downarrow\}}
                - \left( t  c_{i,j, \sigma}^\dagger c_{i,j+1, \sigma} + t c_{i,j+1,\sigma}^\dagger c_{i,j, \sigma} \right)
               - \frac{\mu}{2} n_{i,j, \sigma} \right]
               + \frac{U}{2}   n_{i,j, \uparrow} n_{i,j, \downarrow}
               \\
                H_2 = &\sum_{i=1}^{n_x = 4} \sum_{j=1}^{n_y = 3}
                \left[ \sum_{\sigma \in \{\uparrow, \downarrow\}}   
                - \left( t  c_{i,j, \sigma}^\dagger c_{i+1,j, \sigma} + t c_{i+1,j,\sigma}^\dagger c_{i,j, \sigma} \right)
               - \frac{\mu}{2} n_{i,j, \sigma} \right]
               + \frac{U}{2}   n_{i,j, \uparrow} n_{i,j, \downarrow}
            \end{aligned}
            \end{equation}
             \vspace{-0.3 cm} 
        \end{tabularx}
        \caption{\textbf{Explicit partitions of the spinful Hubbard example, Fig.~\ref{fig: Examples1b} (c) and Fig.~\ref{fig: Examples2b}} 
        The spinful Hubbard model example uses a bi-layer $(n_x,n_y)=(4,3,2)$ rectangular lattice with periodic boundary conditions in $n_x$ and $n_y$ direction, thus all indices in the first two indices are modulo $n_x$ and respectively $n_y$. For the geometric partitionings we show in Fig.~\ref{fig: Examples1b} (c) the sampling improvements $\mathcal G_{\ket{E_0}}(\mathcal B_{\rm Pauli}, \mathcal B_{L_x,L_y})$, and for Pauli partitioning the combined variance of the partition 
        for Pauli measurements $[  \sum_{H_b \in \mathcal B_{\rm Pauli}} \sqrt{ \Var_{\ket \psi} ( H_b) } ]^2 $ is given.
        }
        \label{table: explicit_partitions_spinfull_Hubbard}
\end{table*}
\newpage

\end{document}